\documentclass{ws-rv975x65}

\usepackage{subfigure}     
\usepackage{ws-rv-thm}     
\usepackage{ws-rv-van}     
\makeindex
\usepackage{cite}

\newcommand{\be}{\begin{eqnarray}}
\newcommand{\ee}{\end{eqnarray}}
\newcommand{\non}{\nonumber\\}
\newcommand{\EQ}{{\,=\,}}

\newcommand{\bfx}{{\bf x}}
\newcommand{\bfu}{{\bf u}}
\newcommand{\bfv}{{\bf v}}
\newcommand{\bfp}{{\bf p}}
\newcommand{\bfq}{{\bf q}}
\newcommand{\bfk}{{\bf k}}
\newcommand{\bfr}{{\bf r}}
\newcommand{\bfj}{{\bf j}}
\newcommand{\bfJ}{{\bf J}}
\newcommand{\bfV}{{\bf V}}
\newcommand{\bfP}{{\bf P}}

\newcommand{\hatH}{{\hat{H}}}
\newcommand{\hatA}{{\hat{A}}}
\newcommand{\hatO}{{\hat{O}}}
\newcommand{\hatD}{{\hat{D}}}
\newcommand{\hatJ}{{\hat{J}}}
\newcommand{\hatQ}{{\hat{Q}}}
\newcommand{\hatT}{{\hat{T}}}
\newcommand{\hatP}{{\hat{P}}}
\newcommand{\hatd}{{\hat{d}}}
\newcommand{\hatk}{{\hat{k}}}
\newcommand{\hatu}{{\hat{u}}}
\newcommand{\hatrho}{{\hat{\rho}}}

\newcommand{\barG}{{\bar{G}}}
\newcommand{\barE}{{\bar{E}}}
\newcommand{\baru}{{\bar{u}}}

\newcommand{\tildeq}{{\tilde{q}}}
\newcommand{\tildej}{{\tilde{j}}}
\newcommand{\tildeu}{{\tilde{u}}}
\newcommand{\tildeT}{{\tilde{T}}}

\newcommand{\calD}{{\cal D}}
\newcommand{\calL}{{\cal L}}
\newcommand{\calT}{{\cal T}}
\newcommand{\calS}{{\cal S}}

\newcommand{\deltaJ}{{\delta J}}
\newcommand{\deltan}{{\delta n}}
\newcommand{\deltaf}{{\delta f}}

\newcommand{\ave}[1]{{\langle {#1} \rangle}}
\newcommand{\dotpr}[2]{{\bf #1}{\cdot}{\bf #2}}
\newcommand{\bra}[1]{\langle {#1} |}
\newcommand{\ket}[1]{| {#1} \rangle}

\newcommand{\tw}{\textwidth}

\newcommand{\bmat}[1]{\left(\begin{array}{#1}}
\newcommand{\emat}{\end{array}\right)}

\newcommand{\bcent}{\begin{center}}
\newcommand{\ecent}{\end{center}}

\def\gsim{ \,\, \vcenter{\hbox{$\buildrel{\displaystyle >}\over\sim$}} \,\,}
\begin{document}

\chapter[Introduction to Hydrodynamics]{Introduction to
Hydrodynamics}\label{ra_ch1}

\author[S. Jeon and U. Heinz]{Sangyong Jeon$^1$ and Ulrich Heinz$^2$}

\address{
$^1$Department of Physics, McGill University,
Montreal, QC, Canada\\
$^2$Department of Physics, The Ohio State University,
Columbus, Ohio, USA
}

\begin{abstract}
\end{abstract}
\body

\section{Introduction}

Application of hydrodynamics in high-energy physics has a long 
and illustrious history starting from L.D.~Landau's
seminal paper\cite{Landau:1953gs}. In its history of more than half a
century,
many papers have been written on a broad spectrum of topics, too numerous
to list them
all here. In this review, our emphasis will be more on the basics of the
theory 
of hydrodynamics than to report on the current phenomenological status,
of which several excellent reviews already 
exist (for instance, see
Refs.\cite{Kolb:2003dz,Huovinen:2003fa,Gale:2013da}).

Recent ultra-relativistic heavy ion collision
experiments at the Super Proton Synchrotron (SPS), Relativistic Heavy Ion
Collider (RHIC)
and the Large Hadron Collider (LHC) have demonstrated beyond any doubt that
Quark-Gluon Plasma (QGP)
is being created in these collisions. Unfortunately, direct access to 
the QGP properties such 
as the temperature, equation of state, transport coefficients, etc.~is not
very feasible. 
The only experimentally accessible information is contained in the
spectra of the final state particles.
To connect them to the QGP properties such as above, one must use
theoretical models.
It would be wonderful to have an analytic or a numerical method that can
calculate 
evolution of heavy ion collisions from first principles. 
But this microscopic, non-equilibrium, many-body QCD problem is currently
intractable.  
What is tractable is the coarse-grained collective motion of the system as
a fluid
after the local thermal equilibrium is established. 
Since the properties of QGP we are after are mostly (local) equilibrium
properties, 
it is natural that the dynamics of collective motion -- hydrodynamics --
is an integral part of the theoretical modelling.

What has been exciting and interesting in QGP research is the close
discourse
between the experiment and theory. In elementary particle
experiments, perturbative QCD is being tested with amazing successes.
More and more precise perturbative QCD calculations prove to describe
experimental
data more and more accurately.
In contrast, QGP research is much more dynamic. For instance,
before the discovery of QGP, theoretical expectation was that QGP would be
a weakly coupled plasma of quarks and gluons based partially on the fact
that the QGP properties seem to approach about 80\,\% of the Stefan-Boltzmann 
limit rather quickly on lattice, around $2T_c$ \cite{Karsch:2001vs}.
But almost from the day-1 of RHIC operation,
strong elliptic flow quickly proved this initial expectation not very
viable.
QGP around the transition temperature turned out to be {\em the} most
strongly
coupled many-body system ever observed. Soon after, 
the authors of Ref.\cite{Kovtun:2004de}
used string theory techniques to calculate the infinite coupling limit of
the shear
viscosity and came up with a surprising result that the limit is small,
but has a non-zero lower bound.
Subsequent
Hydrodynamic calculations
then demonstrated 
the importance of small but finite shear viscosity
in understanding the RHIC flow data
\cite{Song:2007fn,Song:2008hj,Schenke:2010rr}.
Comparing the ensuing
LHC predictions with the LHC data now confirmed the expectation that as the
temperature
increases, shear viscosity of QGP should also 
increase~\cite{Schenke:2011tv,Song:2011qa,Petersen:2011sb,Shen:2011eg}.

All these connections between exciting theoretical developments and
experiments
cannot be made without hydrodynamics. More recently, the systems created
in
the highest multiplicity proton-proton collisions and proton-nucleus
collisions 
were also seen to exhibit strong collective 
behavior~\cite{Aad:2012gla,CMS:2012qk,Abelev:2012ola,Adare:2013piz}. 
This is deeply puzzling as the size
of the system ought to be too small to behave collectively. 
It is hoped that more thorough investigation
of the possible origin of the collectivity in such small systems can
illuminate the 
inner workings of the QGP formation greatly~\cite{Kozlov:2014hya}.

As mentioned in the beginning, the aim of this review is the introduction
of the theory of the hydrodynamics in ultra-relativistic heavy ion collisions.
This actually entails a large
number of disciplines in addition to the relativistic fluid dynamics.
Our plan for this review is as follows. Section 2 contains the basic
concepts of hydrodynamics and their definitions. 
In sections 3, second order viscous hydrodynamics is derived from 
a very general linearly response theory of conserved currents.
Section 4 discusses how coarse-graining of kinetic
theory can result in more general form of viscous hydrodynamics. 
In section 5, various numerical techniques needed to implement
relativistic viscous hydrodynamics in ultra-relativistic heavy ion collisions
are discussed. We conclude in section 6.

\section{Hydrodynamic Form of the Stress Energy Tensor and the Net Baryon
Number Current}

Hydrodynamics is all about flow of conserved quantities.
In this review, we strictly deal with relativistic
hydrodynamics. Therefore, unlike the non-relativistic case,
mass is a part of the energy budget.
In the Minkowski coordinates,
conservation laws in their local form are
\be
\partial_\mu T^{\mu\nu} &=& 0
\non
\partial_\mu J^\mu_i &=& 0
\label{eq:1}
\ee
where $T^{\mu\nu}$ is the stress-energy tensor
and the roman letter $i$ on the current $J_i^\mu$ labels 
any other conserved charges
such as the net baryon number, net electric charge, etc.
For the bulk evolution in relativistic heavy ion collisions, 
usually only the net baryon number current, $J_B^\mu$, is considered.
If needed, additional electric current and the strangeness current
can be easily accommodated. 

Using the divergence theorem,
the integral form of the conservation laws read
\be
{d\over dt} \int_V d^3x\, T^{0\nu}
&=& -\int_{\partial V} dS_i  T^{i\nu}
\non
{d\over dt} \int_V d^3x\, J_B^{0}
&=& -\int_{\partial V} dS_i J_B^{i}
\label{eq:conservation1}
\ee
where in the right hand side the integration is over the boundary of the
volume $V$
assuming that the size and the shape of the volume is independent of time.

This form admits a very physical
interpretation that the rate of change of the conserved quantity 
in a fixed volume equals the net current entering the volume.
Hence, the dynamics of conserved quantities 
are governed by the dynamics that governs the currents. 
In essence, hydrodynamics is all about the dynamics of the currents.

Hydrodynamics is useful because it is a coarse-grained theory. 
When a system contains too many particles, 
it becomes difficult to follow microscopic details of the system.
When the system contains sufficiently many particles,
the system again starts to admit analytic studies because thermodynamic
concepts start to apply, in particular the static equilibrium.
A system in static equilibrium is characterized by only few
quantities such as the temperature, collective velocity
and chemical potential. These control
the energy density, momentum density and charge density, respectively.
The price to pay for this simplification is that questions on
short time scale phenomena or short length scale phenomena
cannot be answered any longer.

The systems we would like to study, the ultra-relativistic heavy ion
collisions, 
do contain a large number of particles, but they certainly are not 
static. In fact, they will never actually reach the state of static
equilibrium.
Nevertheless, if one is interested only in the coarse-grained
collective motion of the system,
the concept of {\em local equilibrium} may still apply
provided that the expansion rate is much slower than the microscopic
interaction rate.
If one considers a macroscopically small but microscopically large 
fluid cell around a position $\bfx$ at a given time $t$, 
then within a macroscopically short but microscopically long time scale,
the time averages should approach the static equilibrium values
according to the ergodicity hypothesis of statistical mechanics.
More details on the length and the time scale analysis can be found
in Section \ref{sec:kinetic}.

When the local equilibrium is reached, it becomes meaningful to
describe the system with the local temperature $T(t,\bfx)$, the
collective velocity $u^\mu(t,\bfx)$ of the fluid cell
and the local chemical potential $\mu_B(t,\bfx)$.
One can then study dynamics of only those few thermodynamic quantities.
Since $T, u^\mu, \mu_B$ are basically the Lagrange multipliers to fix 
the average energy, momentum and net charge, it is natural that we turn to
the conservation laws for their dynamics.
The goal of hydrodynamics is then to study collective motion of a system
using the conservation laws with only the {\em statistical} inputs 
from the underlying microscopic theory.

In an dynamically evolving fluid, 
the concept of the local rest frame is essential in order
to apply the concept of local equilibrium.
However defining the collective velocity of a fluid cell turned out
to be quite intricate.
This is not so simple even for the simplest
system composed of a single kind of non-interacting particles in the
non-relativistic setting. 
One kind of collective velocity comes from the average momentum 
\be
\bfu_p = {\sum_i m\bfv_i\over \sum_i m}
\ee
or from the mass current.
Here the sum is over all particles in the given fluid cell.
Another comes from the energy-weighted average
\be
\bfu_E = {\sum_i (m\bfv_i^2/2) \bfv_i\over \sum_i (m\bfv_i^2/2)}
\ee
or from the energy current.
If the particles in the system have an additional conserved charge, $B$,
and
the net charge is non-zero,
then one can define yet another collective velocity by
performing the charge-weighted average
\be
\bfu_B = {\sum_{i} B_i \bfv_i\over \sum_{i} B_i}
\ee
Here $B_i$ is the conserved charge of the $i$-th particle.

For a state in static equilibrium, all three collective velocities above
coincide because they must all vanish. 
However, 
they do not necessarily coincide
to an observer moving with a uniform speed $-\bfu_O$. 
The mass weighted average velocity
and the charge weighted average velocity
are both $\bfu_O$. But the energy weighted average velocity
\be
\bfu'_E = {\sum_i (m(\bfv_i + \bfu_O)^2/2) (\bfv_i + \bfu_O) 
\over 
\sum_i m(\bfv_i + \bfu_O)^2/2} \ne \bfu_O
\ee
clearly does not coincide with $\bfu_O$.
There is no unambiguous choice of the flow velocity even in this simple case.
One must choose among these options what will be
regarded
as the flow velocity.

In the relativistic setting, mass is a part of the energy. Hence, 
there are only two options for choosing the flow velocity:
The energy current or the net baryon current. 
One must choose one of these velocity options
in order to decompose $T^{\mu\nu}$ and $J_B^\mu$ into a useful hydrodynamic form. 
The net baryon number is relatively small
in ultra-relativistic heavy ion collisions. 
Furthermore, the flow observables we are interested in are mostly 
patterns in energy and momentum distributions. 
Therefore,
there is no real benefit to choose the net baryon number collective velocity
as long as the heavy ion analysis is concerned. 

Choosing to follow the energy current,
\footnote{
This choice of frame is often referred to as
the Landau-Lifshitz frame. If one choose to follow the charge current,
it is referred to as the Eckart frame.}
the flow velocity for the energy current 
is defined by the eigenvalue problem
\be
\label{EVeq}
T^{\mu}_{\ \nu} u^\nu = \varepsilon u^\mu
\ee
where $T^\mu_{\ \nu} = T^{\mu\alpha}g_{\alpha\nu}$
and the flow vector is normalized to $u^\mu u_\mu = g_{\mu\nu}u^\mu u^\nu =
1$.
We use $g_{\mu\nu} = {\rm diag}(1,-1,-1,-1)$ throughout this review.
Note that while $T^{\mu\nu}$ is a symmetric matrix,
$T^{\mu}_{\ \nu}$ is no longer a symmetric matrix. Therefore,
there
is no guarantee that the eigenvalues are real. But any physically
consistent
system should admit a positive eigenvalue $\varepsilon$ and the associated
time-like real eigenvector $u^\mu$.

Decomposing $T^{\mu\nu}$ using $\varepsilon$ and $u^\mu$, one gets
\be
T^{\mu\nu} =
\varepsilon u^\mu u^\nu 
+ {1\over 3} (T^\alpha_{\ \alpha} - \varepsilon) \Delta^{\mu\nu} +
\pi^{\mu\nu}
\label{eq:Tmunu1}
\ee
where the local 3-metric $\Delta^{\mu\nu}$ is given by
\be
\label{eq:Deltamunu}
\Delta^{\mu\nu} = g^{\mu\nu} - u^\mu u^\nu
\ee
The residual shear tensor $\pi^{\mu\nu}$ is symmetric
$\pi^{\mu\nu}=\pi^{\nu\mu}$,
transverse $ \pi^{\mu\nu}u_\nu = 0 $ and traceless $ \pi^{\mu\nu}g_{\mu\nu}
= 0 $.
Hence altogether the expression (\ref{eq:Tmunu1}) 
has the required 10 independent degrees of freedom;
the local energy density $\varepsilon$,
the local fluid velocity $\bfu$, the trace $T^{\alpha}_{\ \alpha}$ 
and the residual shear tensor $\pi^{\mu\nu}$.
The net-baryon current is 
\be
J_B^\mu = J^\mu_{B,\rm id} + V_B^\mu
\label{eq:JB1}
\ee
where 
$J^\mu_{B,\rm id} = \rho_B u^\mu$ is the ideal fluid part of the current
and $V_B^\mu$ is a space-like vector satisfying the transversality
condition $u_\mu V^\mu_{B} = 0$. It has the required 4 independent
degrees of freedom; the local net baryon density $\rho_B$ and the residual
vector $\bfV_B$.

So far we did not use any thermodynamic information.
We will do so now to re-write the trace $T^\alpha_{\ \alpha}$ 
in a more physical form.
A static medium at rest has 
$T^{\mu\nu}_{\rm eq}
= {\rm diag}(\varepsilon, P, P, P)$ where $P$ is the pressure.
Therefore, the trace should contain the equilibrium piece
$g_{\mu\nu}T^{\mu\nu}_{\rm eq} = \varepsilon - 3P$.
Furthermore, the thermodynamic identities
\be
dP &=& sdT + \rho_B d\mu_B
\\
ds &=& {1\over T}d\varepsilon - {\mu_B\over T} d\rho_B 
\ee
where $s$ is the entropy density, indicate that the pressure
$P$ is a function of the
temperature $T$ and the baryon chemical potential $\mu_B$, and they are in turn
function of the energy density and the net baryon density.
Hence we must be able to find $P$ as a function of $\varepsilon$ and $\rho_B$:
\be
P = P(\varepsilon, \rho_B)
\ee
This relationship is known as the equation of state.
Writing $T^{\alpha}_{\ \alpha} = \varepsilon - 3(P + \Pi)$,
the stress-energy tensor then becomes
\be
T^{\mu\nu} =
T_{\rm id}^{\mu\nu}
- \Pi \Delta^{\mu\nu} + \pi^{\mu\nu}
\label{eq:Tmunu2}
\ee
where
$T_{\rm id}^{\mu\nu} =
\varepsilon u^\mu u^\nu - P(\varepsilon, \rho_B) \Delta^{\mu\nu}$
is the ideal fluid stress-energy tensor.

From the arguments presented above, it should be clear that
the residual scalar term $\Pi$ and the tensor term $\pi^{\mu\nu}$
as well as the vector $V_B^\mu$ in the baryon current must vanish
in the static equilibrium limit.
As these quantities represent deviation from equilibrium,
the size of these terms will depend on how fast the local
equilibrium is achieved.
If local equilibration is instantaneous in the macroscopic time scale,
then fluid cells will always be in strict local equilibrium and hence
$\Pi = \pi^{\mu\nu} = V_B^\mu = 0$.
Microscopically, this would happen if scattering cross-section is large 
so that the mean free path is much shorter than
any macroscopic time scale or length scale (c.f.~section \ref{sec:kinetic}).
If this is the case, the number of unknowns $(\varepsilon, \bfu, \rho_B)$
matches the 5 conservation laws and one has the ideal hydrodynamics
whose dynamics is completely specified by 
\be
\partial_\mu \left(\varepsilon u^\mu u^\nu - P(\varepsilon,
\rho_B)\Delta^{\mu\nu}\right) =
0
\ee
and
\be
\partial_\mu (\rho_B u^\mu) = 0
\ee
The information on the underlying system is only in the equation of state
$P(\varepsilon, \rho_B)$.

\section{Hydrodynamics from Linear Response Theory}
\label{sec:linear_response}
\subsection{Linear Response Theory}

In realistic systems, the approach to the local equilibrium is never going
to
be infinitely fast. Therefore, $\Pi$, $\pi^{\mu\nu}$ and $V_B^\mu$ cannot
simply
be set to vanish, although if the system is conformal one strictly has
$\Pi = 0$. This is because 
the trace must vanish $T^{\alpha}_{\ \alpha} = 0$ in a conformal system.
When any of these quantities are non-zero, the system is out of
equilibrium. 
Therefore, local entropy must increase. 
The evolution equations
of these quantities are then necessarily of the dissipative type.
In this and following sections, we use linear response theory to obtain 
such equations for dissipative hydrodynamics. 
In section \ref{sec:kinetic_hydro}, kinetic theory approaches that can 
go beyond the near-equilibrium restriction of 
the linear response theory is discussed.

To gain more insights on the behavior of the dissipative quantities, we start
with
a system very slightly out of equilibrium at $t = 0$.
We then consider how the system approaches the equilibrium. 
This is the realm of the linear response theory. 
 Full analysis of the quantum linear response theory 
 can be found in any number of standard text books (for example, see
 Ref.\cite{Kapusta:2006pm}).
 Here we will only go over the main ideas.
 
 Suppose that at the remote past, $t = -T$,
 the density operator had the equilibrium form
 $\hat\rho_0 = e^{-\beta \hatH_0}/Z_0$ where $\hatH_0$ is the 
 system Hamiltonian and $Z_0 = {\rm Tr} e^{-\beta \hatH_0}$ is the partition function.
 Then a force term is adiabatically
 turned on $f(t,x) = \theta(-t)f(x)e^{\epsilon t}$
 at an infinitesimally slow rate $\epsilon$.
 At $t=0$, the force term is turned off and at this point the system is out of equilibrium. 
 The full time-dependent Hamiltonian for this process is
 \be
 \hat{H}(t) = \hatH_0 
 - \int d^3x\, \hatA(x)f(t,\bfx)
 \label{eq:fullH}
 \ee
 Here $\hat{A}$ 
 represents a set of Hermitian operators for the 
 conserved quantities
 we are interested in.
 Namely, $T^{0\mu}$ and $\rho_B$. 
 Hence, the expressions presented below are in general matrix expressions.

 Treating $\delta\hatH = -\int d^3x\, f(t,\bfx)\hatA(\bfx)$
 as a perturbation, the formal first order solution 
 of the quantum Liouville equation 
 $i\partial_t\hatrho(t) = [\hatH(t), \hatrho(t)]$
 is given by
 \be
 \delta\hat\rho_H(t)
 =
 -i\int_{-T}^t dt'\, [\delta\hat{H}_H(t'), \hat\rho_0(0)]
 \ee
 where $\delta\hatrho(t) = \hatrho(t) - \hatrho_0$ and
 the subscript $H$ denotes Heisenberg picture operators
 \be
 \hatO_H(t) = e^{i\hatH_0 t}\hatO e^{-i\hatH_0 t}
 \ee
 Using $\delta\hatH$ from Eq.(\ref{eq:fullH}),
 the deviation of an observable $A$ from the equilibrium value is 
 found to be
 \be
 \delta\langle \hat{A}(t,\bfx) \rangle
 & = &
 \int d^4x'\,
 G_R^{AA}(t-t',\bfx-\bfx') \theta(-t')e^{\epsilon t'}f(\bfx')
 \label{eq:Aexpr}
 \ee
 where $t > 0$ and
 \be
 G_R^{AA}(t-t',\bfx-\bfx')
 =
 i\theta(t-t')
 {\rm Tr}\left( \hat\rho_0 [ \hat{A}_H(t,\bfx), \hat{A}_H(t',\bfx')
] \right)
 \ee
 is the retarded response function.
 We also took the $-T\to -\infty$ limit.
 
 Suppose that one can find an operator $\hatD_A$
 for which $G_R^{AA}$ is the generalized retarded Green function, 
 \be
 \hatD_A G_R^{AA}(t-t',\bfx-\bfx') = \hatd_A\delta(t-t')\delta(\bfx-\bfx')
 \ee
 where $\hatd_A$ can contain a finite number of derivatives.
 For $t>0$, $t$ and $t'$ can never be the same in Eq.(\ref{eq:Aexpr}).
 Hence $\delta\ave{\hatA}$ satisfies the evolution equation
 \be
 \hatD_A \delta\ave{\hat{A}(t,\bfx)} = 0
 \ee
 for $t> 0$.
 Therefore finding the pole structure of the response function is
equivalent to finding the evolution 
equation\cite{forster1975,Kadanoff1963}.
We will use this to find hydrodynamic equations in the following sections.

 For further analysis, it is useful to define the spectral density
 \be
 \rho^{AA}(\omega,\bfk)
 & = &
 \int d^4x\, e^{i\omega t - i\dotpr{k}{x}}\,
 \ave{[\hatA_H(t,\bfx), \hatA_H^\dagger(0)]}
 \ee
 Using the thermal average 
 $\ave{(\cdots)} = {\rm Tr}\hatrho_0(\cdots)$,
 it is not hard to show 
 (for instance, see Ref.\cite{Kapusta:2006pm})
 \be
 \rho^{AA}(\omega,\bfk)
 & = &
 {1\over Z_0}\sum_{m,n}\left( e^{-\beta E_n} - e^{-\beta E_m}\right)
 (2\pi)^4\delta(k - p_m + p_n) \left| \bra{n}\hatA\ket{m} \right|^2
 \ee
 where $k = (\omega, \bfk)$ and 
$\ket{m}$ is the simultaneous eigenstate of 
the system Hamiltonian $\hatH_0$ and the total momentum $\hat\bfP$ 
of the system with the eigenvalue $p_{m} = (E_{m}, \bfp_{m})$.
From this expression, we can derive
 \be
 \rho^{AA}(-\omega,-\bfk) = -\rho^{AA}(\omega,\bfk)
 \label{eq:odd_rhoAA}
 \ee
 by exchanging $m$ and $n$.
 When the underlying equilibrium system is isotropic, 
 then $\rho^{AA}(\omega,\bfk)$ must be a function of $|\bfk|$ only.
 Hence the spectral density  is an odd function of $\omega$,
 $ \rho^{AA}(-\omega,\bfk) = -\rho^{AA}(\omega,\bfk)$.
 We will use this property often in later sections to parametrize
 the analytic structure of the response functions.
 
 In terms of $\rho^{AA}$, the retarded correlator is 
 \be
 G_R^{AA}(\omega, \bfk)
 & = &
 \int {d\omega'\over 2\pi}\,
 {\rho^{AA}(\omega',\bfk)\over \omega' - \omega - i\epsilon }
 \label{eq:GRAA}
 \ee
 The imaginary part of the retarded correlator 
 directly gives the spectral density
 \be
 {\rm Im}\,G_R^{AA}(\omega,\bfk)
 = {1\over 2}\rho^{AA}(\omega,\bfk)
 \ee
 It is also useful to know that the Euclidean correlator is given by
 (for derivations, see, for instance, Ref.\cite{Kapusta:2006pm})
 \be
 G_E^{AA}(\omega_n, \bfk)
 & = &
 \int {d\omega'\over 2\pi}\,
 {\rho^{AA}(\omega',\bfq)\over \omega' - i\omega_n }
 \label{eq:GEAA}
 \ee
 where $\omega_n = 2\pi nT$ is the Matsubara frequency.
 One important fact we will often use in the following sections is
 that the $\omega\to 0$ limit
 \be
 G_R^{AA}(0,\bfk)
 &=&
 G_E^{AA}(0,\bfk) 
 \non
 & = &
 \int {d\omega'\over 2\pi}{\rho^{AA}(\omega,\bfk) \over \omega'}
 \non
 & = &
 {1\over Z_0}
 \sum_{m,n}\left( e^{-\beta E_n} - e^{-\beta E_m}\right)
 {1\over E_m - E_n}
 (2\pi)^3\delta(\bfk - \bfp_m + \bfp_n)
 \left| \bra{n}\hatA\ket{m} \right|^2
 \non
 \label{eq:GR0}
 \ee
 is real and positive and function only of the magnitude of 
 $\bfk$.\footnote{To see this, just exchange $m$ and $n$.}
 
 \subsection{Baryon Density Diffusion}
 
 From the previous section, it is clear that the analytic structure of
 the response function determines the evolution of small disturbances.
 In this and the following sections, we show that this fact combined with
 the conservation laws is powerful enough to produce dissipative
 hydrodynamic equations.
 
 The analysis in this section and those up to section \ref{sec:sound_bulk} 
 closely follow the unpublished note by Laurence G.~Yaffe 
 (private communication, see also Refs.\cite{Herzog:2009xv,Kovtun:2012rj})
 in a simplified form.
 Additional discussion on the 2nd-order formulation of dissipative 
 hydrodynamics is given in section \ref{sec:2nd_order_visc}.
 
 We start with the net baryon number conservation which is the simplest to
examine.
 Suppose we set up a system where only the net baryon density
 $\ave{\rho_B(0,\bfx)}$ is non-uniform at $t = 0$.
 The perturbing Hamiltonian is
 \be
 \delta\hat{H}(t) = -\int d^3x\,\hatrho_B(\bfx)\, e^{\epsilon t}\,
\mu_B(\bfx) 
 \ee
 which results in the following
 linear response in the mixed space of $t$ and the wavevector $\bfk$,
 \be
 \delta\langle \hatrho_B(t,\bfk) \rangle
 & = &
 \mu_B(\bfk)
 \int_{-\infty}^{\infty} dt'\,
 \theta(-t') e^{\epsilon t'}
 G_R^{00}(t-t',\bfk) 
 \ee
 for $t > 0$.
 
 Applying the current conservation, $\partial_t \rho_B =
-\nabla{\cdot}\bfJ_B$,
 to the retarded correlation functions
$ G_R^{\mu\nu}(t,\bfx) = 
i\theta(t)\ave{[\hatJ_B^\mu (t,\bfx), \hatJ_B^\nu(0)]} $
 results in the following relationships between them in the
frequency-wavevector
 space
 \be
 \omega G_R^{00}(\omega,\bfk) = k_i G^{0i}_R(\omega, \bfk)
 \label{eq:GR001}
 \\
 \omega G_R^{0j}(\omega,\bfk) = k_i G^{ij}_R(\omega, \bfk)
 \label{eq:GR002}
 \ee
 provided that 
 $[J^0_B(0,\bfx), J^\nu_B(0,\bfx')] = 0$.
 The underlying isotropic equilibrium permits 
 decomposition into transverse and the longitudinal parts
 \be
 G^{ij}_R(\omega,\bfk)
 =
 \hatk^i \hatk^j G_L(\omega,\bfk^2) + \hat\delta^{ij} G_T(\omega,\bfk)
 \ee
 where $\hat\bfk = \bfk/|\bfk|$ is the unit vector and 
 $\hat\delta^{ij}=\delta^{ij} - \hatk^i \hatk^j$ is the transverse
 projector.
 Combining Eqs.(\ref{eq:GR001}) and (\ref{eq:GR002}) gives
 \be
 \omega^2 G_R^{00}(\omega,\bfk) 
 & = &
 \bfk^2 G_L(\omega,\bfk) 
 \label{eq:rhob_cons}
 \ee
 What we are interested in is the behaviour of $G_L(\omega,\bfk)$.
 
 Consider the small $\omega$ limit of $G_L$ first.
 From Eq.(\ref{eq:GR0}), we know that
 \be
 g_{00}(\bfk) = G_R^{00}(0,\bfk) = G_E^{00}(0,\bfk) 
 \ee
 is real and positive.
 The small $\omega$ limit of $G_L$ can be then expressed as
 \be
 G_L(\omega,\bfk) \approx {\omega^2 g_{00}(\bfk)\over \bfk^2}
 \ee
 Now consider taking the $\bfk\to 0$ limit with a fixed $\omega\ne 0$.
 The retarded density-density correlation function in this limit is
 \be
 G_R^{00}(\omega,0)
 & = &
 i\int_0^\infty dt\, e^{-i\omega t}\,
 \int d^3x\, \ave{[\hat\rho_B(t,\bfx), \hat\rho_B(0)]}
 \non
 & = &
 i\int_0^\infty dt\, e^{-i\omega t}\,
 \ave{[\hatQ_B, \hat\rho_B(0)]} = 0
 \ee
 where we used the fact that $\hatQ_B = \int d^3x\, \hat\rho_B(t,\bfx)$
 is the net baryon number operator.
 Since the net baryon number is conserved, 
 $\hatQ_B$ is independent of time. In particular,
 it can be evaluated at $t=0$. Therefore, the commutator vanishes.
 This then indicates that $G_L(\omega, \bfk)$ 
 is well behaved in the zero $|\bfk|$ limit with $\omega \ne 0$. 
 Consequently, the $\omega\to 0$ limit and the $\bfk \to 0$ limit do not
commute,
 indicating the presence of a massless pole.
 We also know that the imaginary part of $G_R^{00}$ (the spectral density
 $\rho^{00}$) has to be an odd function of $\omega$ since isotropy in space
demands that it be a function only of $\bfk^2$ (c.f.~Eq.(\ref{eq:odd_rhoAA})). 

 The most general form of $G_L$ consistent with the above conditions is
 \be
 G_L(\omega,\bfk) 
 = 
 {\omega^2 (g_{00}(\bfk) + i\omega A(\omega,\bfk))
 \over \bfk^2 - i\omega/D(\omega,\bfk) -\omega^2 B(\omega, \bfk) }
 \label{eq:GLgen}
 \ee
 The functions $D(\omega,\bfk)$, $A(\omega,\bfk)$ and $B(\omega,\bfk)$ 
 are all of the form
 \be
 D(\omega, \bfk) = D_R(\omega, \bfk) - i\omega D_I(\omega, \bfk) 
 \ee
 where $D_R(\omega,\bfk)$ and $D_I(\omega,\bfk)$ are 
 are real-valued even functions
 of $\omega$ and $\bfk$.
 The real parts $D_R$ and $B_R$
 must have a non-zero limit as $\omega\to 0$ and $\bfk \to 0$.
 All other parts of $A,B$ and $D$ must have finite limits
 as $\omega\to 0$ and $\bfk \to 0$.
 In the small $\omega$ and $\bfk$ limit, 
 the response function becomes
 \be
 G_R^{00}(\omega,\bfk) 
 \approx 
 {D \bfk^2 g_{00}(0) \over -i\omega + D\,\bfk^2}
 \ee
 where we defined the diffusion constant $D = D_R(0,0)$.
 
 The pole structure of $G_R^{00}$ dictates that in the small $\omega$ and
 $|\bfk|$ limit,
 $\delta\rho_B(t,\bfx) = \delta\langle \hatrho_B(t,\bfx) \rangle$
 obeys the diffusion equation
 \be
 \partial_t\delta\rho_B = D \nabla^2 \delta\rho_B
 \ee
 This is our first example of a dissipative hydrodynamic equation.
 The conservation law, current algebra, thermodynamic stability,
 and the general analytic structure of the  
 correlation functions are all the ingredients one needs to 
 get this diffusion equation for baryon density.
 Hence diffusion is a very general phenomenon whenever
 there is a conserved current.
 Microscopic dynamics only enters through the value of the diffusion constant.
 
 If we now go back to the conservation equation
 \be
 \partial_t \delta\rho_B = -\partial_i \deltaJ_B^i
 \ee
 we can see that the diffusion equation above is equivalent
 to the constitutive relationship
 \be
 \deltaJ^i = V_B^i = D\partial^i \delta\rho_B
 \ee
 valid in the fluid cell rest frame.
 In the more general frame boosted by $u^\mu$,
 it becomes
 \be
 V_B^\mu = D \Delta^{\mu\nu} \partial_\nu \rho_B
 \ee
 where again $\Delta^{\mu\nu} = g^{\mu\nu} - u^\mu u^\nu$ is the local 3-metric.

 The diffusion constant can be calculated
 by taking the appropriate limits of $G_L$
 \be
 \lim_{\omega\to 0}\lim_{\bfk \to 0}
 {1\over \omega} {\rm Im}\, G_L(\omega,\bfk)
 & = &
 Dg_{00}(0)
 \label{eq:Kubo_B}
 \ee
 which is our first example of the Kubo formula which relates an {\em equilibrium}
 correlation function to a dissipative coefficient. 
 As the response function is singular in the small $\omega$ and small $\bfk$ limit,
 the order of limits in the Kubo formula is important. The $\bfk\to 0$
 limit must be taken first.
 Since the transport coefficients are defined as Lorentz scalars,
 Eq.(\ref{eq:Kubo_B}) is to be evaluated using the underlying microscopic theory 
 such as thermal QCD in the rest frame of the equilibrium system.

 \subsection{Stress-Energy Tensor Correlation Functions}

 To carry out the linear response analysis of the energy-momentum currents
 and their hydrodynamic and dissipative behavior,
 one need to know the Ward identity among the correlation functions.
 Defining the retarded correlation functions of $\hatT^{\mu\nu}$ turned out
 to be not so straightforward due to the fact that 
 the conserved quantities
 $\hatP^\mu = \int d^3x\, \hatT^{0\mu}$ are also the generators of the
 space-time evolution.  Hence, in general the equal time commutators of
 $\hatT^{0\mu}$ are non-zero unlike the net baryon current case.
 
 To begin the analysis, consider the static partition function given by
 \be
 Z_E[g_E] = \int\calD\phi\, e^{-S_E[\phi,g_E]}
 \ee
 where 
 \be
 S_E[\phi,g_E] = \int d^3x\int_0^\beta d\tau\, \sqrt{g_E}\,
\calL(\phi,g_E)
 \ee
 is the Euclidean action and $\tau$ is the imaginary time.
 Here, $\phi$ denotes the collection of field variables and $g_E$ is the 
 Euclidean metric. 

 Using the Hilbert definition of the stress-energy tensor density, we have
 \be
 \ave{T^{\mu\nu}(x)}_E = -2{\delta\over \delta g^E_{\mu\nu}(x)}\ln
Z_E[g_E]
 \ee
 The two point functions are given by
 \be
 \barG^{\alpha\beta,\mu\nu}_R(x_E,y_E)
 &=& \ave{\calT_\tau T^{\mu\nu}(x_E)T^{\alpha\beta}(y_E)}_E
 \non
 &=&
 4{ \delta^2\over 
 \delta g^E_{\alpha\beta}(y_E) \delta g^E_{\mu\nu}(x_E) }\ln Z_E[g_E]
 \ee
 where $x_E = (\tau,\bfx)$ and
 $\calT_\tau$ is the time ordering operator in $\tau$.
 For the tensor density $T^{\mu\nu}$, the covariant conservation law is
 \be
 \partial_\mu \ave{T^{\mu\nu}}_E 
 + \Gamma^\nu_{\sigma\rho}\ave{T^{\sigma\rho}}_E
 =0
 \label{eq:EuclCons}
 \ee
 where
 $\Gamma^\nu_{\sigma\rho} = {1\over 2}g_E^{\mu\nu}
 (g^E_{\sigma\mu,\rho} + g^E_{\rho\mu,\sigma} - g^E_{\sigma\rho,\mu})$
 is the Christoffel symbol,
 
 By differentiating Eq.(\ref{eq:EuclCons}) once more with respect to
 $g^E_{\alpha\beta}(y)$, we obtain the Ward identity among the Euclidean
correlation
 functions in the flat space \cite{Herzog:2009xv,Kovtun:2012rj} where 
 $g^{\mu\nu}_E = \delta^{\mu\nu}$
 \be
 0 = k_\alpha^E \left(\barG_E^{\alpha\beta,\mu\nu}(k_E)
 + 
 \delta^{\beta\mu}\ave{T^{\alpha\nu}}
 +
 \delta^{\beta\nu}\ave{T^{\alpha\mu}}
 -
 \delta^{\alpha\beta}\ave{T^{\mu\nu}}
 \right)
 \label{eq:kGEabmn}
 \ee
 Here $k_E^\alpha = (\omega_n, \bfk)$ and $\omega_n = 2\pi n T$ 
 is the Matsubara frequency.
 
 To obtain the real time correlation functions, we perform the analytic 
 continuation.  From Eqs.(\ref{eq:GRAA}) and (\ref{eq:GEAA}) one can see
 that the analytically continuation
 \be
 \omega_n &\to& -ik^0 + \epsilon
 \ee
 changes the Euclidean correlation function to the
 real-time retarded correlation function.
 This also means that each time index of $T^{\mu\nu}$
 gets a factor of $(-i)$.
 Since our Minkowski metric is mostly negative,
 going from the Euclidean metric to the mostly negative Minkowski metric 
 means $\delta_{\mu\nu} \to -g_{\mu\nu}$.
 The real-time version of Eq.(\ref{eq:kGEabmn}) is then
 \be
 0 = k_\alpha
 \left(\barG_R^{\alpha\beta,\mu\nu}(k)
 -g^{\beta\mu}\ave{T^{\alpha\nu}}
 -g^{\beta\nu}\ave{T^{\alpha\mu}}
 +g^{\alpha\beta}\ave{T^{\mu\nu}}
 \right)
 \label{eq:kGRabmn}
 \ee
 where $k^\mu = (\omega, \bfk)$.
 
 The presence of the single stress-energy tensor average terms in
 Eq.(\ref{eq:kGRabmn}) implies that the correlation function
 $\barG_R^{\alpha\beta,\mu\nu}(x,x')$ 
 is not the same as
 \be
 G_R^{\mu\nu,\alpha\beta}(x,x') = 
 i\theta(t)\ave{[\hatT^{\mu\nu} (t,\bfx),
\hatT^{\alpha\beta}(t',\bfx')]}_{\rm eq}
 \ee
 but differs by terms containing $\delta(x-x')$ (as well as the contact
terms containing spatial derivatives of $\delta(x-x')$ \cite{Deser:1967zzf}).
 As the response function in the linear response theory, these delta-function
 terms do not matter since $t$ and $t' < t$ can never be the same.

 Defining the real-time correlation function by the analytic continuation
 of the Euclidean correlation function
 enables us to gain the following important relationship between the two
 \be
 \barG_E^{\mu\nu,\alpha\beta}(0,\bfk) =
\barG_R^{\mu\nu,\alpha\beta}(0,\bfk) 
 \ee
 which has a well-defined limit ($\mu$ and $\nu$ here are not summed)
 \be
 \lim_{\bfk\to 0}\barG_R^{\mu\nu,\mu\nu}(0,\bfk) 
 =
 \lim_{\bfk\to 0}\barG_E^{\mu\nu,\mu\nu}(0,\bfk) > 0
 \ee

 \subsection{Momentum Diffusion and Shear Viscosity}
 
 In this and the following section for the bulk
 viscosity, we will not consider finite net baryon density for simplicity.
 For analysis with finite $\mu_B$, see Ref.\cite{Kovtun:2012rj}.

 Suppose that we set up a system where the the flow velocity 
 at $t =0$ has a single non-zero component in the $x$-direction
 $u_x(y)$ which depend only on $y$. 
 In this situation, two layers of the fluid at $y$ and at $y+\Delta y$ have
different
 fluid velocities in the orthogonal $x$-direction. In ideal hydrodynamics,
 this difference is maintained because there is no dissipation. 
 In a normal fluid, however, particle diffusion between the two layers
 will eventually make them move with the same equilibrated speed.
 How fast two layers equilibrate depends on the size of the scattering mean free path,
 which in turn determines the diffusion constant, or the shear viscosity.
 
 To set up a shear flow, let the perturbing Hamiltonian be
 \be
 \delta\hat{H}(t) = 
 -\int d^3x\, e^{\epsilon t}\, \hatT^{x0}(t,\bfx)\beta_x(y) 
 \ee
 The corresponding linear response is
 \be
 \delta\langle T^{x0}(t,k_y) \rangle
 & = &
 \beta_x(k_y)
 \int_{-\infty}^{\infty} dt'\,
 \theta(-t') e^{\epsilon t'}
 \barG_R^{x0,x0}(t-t',k_y) 
 \label{eq:LinResTx0}
 \ee
 for $t > 0$.
 
 In Eq.(\ref{eq:kGRabmn}),
  $\barG_R^{x0,x0}$ appears 
 in the following sequences when $\bfk = (0,k_y,0)$
 \be
 \omega (\barG_R^{x0,x0}(\omega,k_y) + \varepsilon)
 &=&
 k_y \barG^{x0,xy}_R(\omega, k_y)
 \label{eq:omegaGRx0x0}
 \\
 \omega \barG_R^{x0,xy}(\omega,k_y) 
 &=&
 k_y (\barG^{xy,xy}_R(\omega, k_y) + P)
 \label{eq:omegaGRx0xy}
 \ee
 Combined, these become
 \be
 \barG^{xy,xy}_R(\omega,k_y)+P
 & = &
 {\omega^2\over k_y^2}
 \left(
 \barG_R^{x0,x0}(\omega,k_y) + \varepsilon
 \right)
 \ee
 Except the extra $P$ and $\varepsilon$, the structure of these equations
 is exactly the same as the baryon current case. The following analysis is
 therefore a repeat of that case.
 
 In the $\omega\to 0$ limit, $\barG_R^{x0,x0}(\omega,k_y)$ 
 must have a well defined limit since it is a thermodynamic quantity.
 Furthermore, the imaginary part of $\barG^{xy,xy}_R(\omega,k_y)$ must
 be an odd function of $\omega$.
 As in the baryon current case, the $k_y\to 0$ limit of 
 the correlation functions must be well-behaved.
 Thus, we can parametrize $\barG_R^{xy,xy}$ as
 \be
 \barG_R^{xy,xy}(\omega,k_y)
 = 
 {\omega^2 (\varepsilon + g_T(k_y) + i\omega A_T(\omega, k_y))
 \over
 k_y^2 - i\omega/D_T(\omega,k_y) - \omega^2 B_T(\omega,k_y) }
 - P
 \label{eq:GRxyxy}
 \ee
 and
 \be
 \barG^{x0,x0}_R(\omega, k_y)
 = 
 {k_y^2 (\varepsilon + g_T(k_y) + i\omega A_T(\omega, k_y))
 \over
 k_y^2 - i\omega/D_T(\omega,k_y) - \omega^2 B_T(\omega,k_y) }
 - \varepsilon 
 \non
 \ee
 where $g_T(k_y) = \barG_R^{x0,x0}(0,k_y)$.
 Here the functions $A_T$, $B_T$ and $D_T$ 
 all have the form
 \be
 D_T(\omega,k_y) = D_T^R(\omega, k_y) - i\omega D_T^I(\omega,k_y)
 \ee
 where $D_T^R(\omega,k_y)$ and $D_T^I(\omega,k_y)$ are 
 real-valued even functions of $\omega$ and $k_y$.
 The real parts $D_T^R$ and $B_T^R$
 must have a non-zero limit as $\omega\to 0$ and $k_y \to 0$.
 All other parts of $A_T, B_T$ and $D_T$ must have finite limits
 as $\omega\to 0$ and $k_y \to 0$.
 
 In the configuration space, the constant $-\varepsilon$ term becomes
 $-\varepsilon \delta^4(x-x')$. In Eq.(\ref{eq:LinResTx0}), 
 this $\delta$-function term does not contribute.
 Hence, in the small $\omega$ and $k_y$ limit,
 the evolution of $T^{x0}$ is determined by
 $i\omega = D_T k_y^2$ or
 \be
 \left( \partial_t - D_T\partial_y^2\right) T^{x0}(t,y) = 0
 \label{eq:Tx0eq}
 \ee
 where we defined the momentum diffusion constant $D_T = D_T^R(0,0)$.
 This is our second dissipative hydrodynamic equation.
 The diffusion equation combined with the conservation law
 implies the constitutive relationship 
 \be
 T^{xy}(t,y) = D_T\partial^y T^{x0}(t,y) 
 = \eta \partial^y u^x
 \label{eq:Txy}
 \ee
 valid in the local rest frame.
 Here $\eta = D_T (\varepsilon + P)$ is the shear viscosity.
 It is clear from Eq.(\ref{eq:Tx0eq}) that $D_T$ has the
 physical interpretation of the diffusion constant for momentum diffusion.

 Recognizing Eq.(\ref{eq:Txy}) to be a part of the spin 2 component of the
second rank tensor,
 we can generalize this result to
 \be
 \pi_{\rm NS}^{ij}(t,\bfx)
 & = &
 \eta 
 \left(\partial^i u^{j} + \partial^j u^{i}
 - {2g^{ij}\over 3}\partial_l
 u^{l}\right)
 \ee
 again in the rest frame of the fluid cell.
 In the moving frame, this becomes
 \be
 \label{eq:shear}
 \pi_{\rm NS}^{\mu\nu}(t,\bfx)
 & = &
 2\eta \Delta^{\mu\nu}_{\alpha\beta} \partial^\alpha u^\beta 
 \equiv 2\eta\sigma^{\mu\nu}
 \ee
 where $\sigma^{\mu\nu}$ is the velocity shear tensor and 
 $\Delta^{\mu\nu}_{\alpha\beta}$ is the the spin-2 projector  
 defined by
\be
 \Delta^{\mu\nu}_{\alpha\beta} =
{1\over 2}
\left(
\Delta_{\alpha}^{\mu}\Delta_{\beta}^{\nu}
+
\Delta_{\alpha}^{\nu}\Delta_{\beta}^{\mu}
-
{2\over 3}\Delta^{\mu\nu}\Delta_{\alpha\beta}
\right)
\label{eq:spin2proj}
\ee
Here the label NS indicates that this is the Navier-Stokes form of the shear
tensor.
 The Kubo formula for the shear viscosity is
 \be
 \lim_{\omega\to 0}\lim_{k_y\to 0}
 {1\over \omega} {\rm Im}\, \barG_R^{xy,xy}(\omega, k_y)
 = 
 D_T (\varepsilon + P) = \eta
 \label{eq:KuboShear}
 \ee
 where we used the fact that $g_T(0) =P$ which can be
 determined from Eq.(\ref{eq:kGRabmn}).

 The Kubo formula for the shear viscosity $\eta$ 
 can be also expressed in terms of the full shear-tensor
 correlation function\cite{forster1975,Kadanoff1963}
 \be
 \eta
 & = &
 \lim_{\omega\to 0}\lim_{\bfk\to 0}
 {1\over 10\omega} {\rm Im}\barG_R^{\pi_{ij},\pi_{ij}}(\omega,\bfk)
 \label{eq:KuboShearFull}
 \ee
 where the shear-tensor is given by
 \be
 \pi_{ij} = T_{ij} - (\delta_{ij}/3) T^k_k
 \ee

 \subsection{Sound Propagation and Bulk Viscosity}
 \label{sec:sound_bulk}

 So far, only the diffusion type of hydrodynamic flow is discussed 
 which are not the main bulk excitation.
 To get the main excitation which must also include the ideal 
 hydrodynamics part,
 one needs to look at the disturbance in the energy density.
 This bulk excitation, of course, is the sound wave.

 Suppose we perturb the energy density with
 \be
 \delta\hat{H}(t) = 
 -\int d^3x\, e^{\epsilon t}\, \hatT^{00}(t,\bfx)\beta_0(\bfx) 
 \ee
 The linear response is then
 \be
 \delta\langle T^{00}(t,\bfk) \rangle
 & = &
 \beta_0(\bfk)
 \int_{-\infty}^{\infty} dt'\,
 \theta(-t') e^{\epsilon t'}
 \barG_R^{00,00}(t-t',\bfk) 
 \label{eq:LinResT00}
 \ee

 Applying the conservation law to each index of 
 $\barG^{\mu\nu,\alpha\beta}_R$ in Eq.(\ref{eq:kGRabmn}), we get
 \be
 \omega^4  \barG^{00,00}_R(\omega,\bfk) 
  &=& 
 \omega^4 \varepsilon 
 - \omega^2 \bfk^2 (\varepsilon + P)
 + 
 \bfk^4 \barG_L(\omega,\bfk)
 \label{eq:EEcorr}
 \ee
 where 
 \be
 \bfk^4
 \barG_L(\omega,\bfk)
 =
 k_i k_j k_l k_m 
 \left(\barG^{ij,lm}_R(\omega,\bfk) 
 + P(\delta^{il}\delta^{jm} + \delta^{im}\delta^{jl} -
\delta^{ij}\delta^{lm})
 \right)
 \ee
 In the small $\omega$ limit, Eq.(\ref{eq:EEcorr}) gives
 \be
 \barG_L(\omega,\bfk) \approx  
 {\omega^2\over \bfk^2}(\varepsilon + P)
 + 
 {\omega^4\over \bfk^4}\left( \barG_R^{00,00}(0,\bfk) - \varepsilon\right)
 \label{eq:GLomegalim}
 \ee
 We also know that the imaginary part of $\barG_L$ must be an odd function 
 of $\omega$ and the correlation functions are well-behaved in the $\bfk\to
 0$ limit.
 The most general form consistent with these conditions
 is
 \be
 \barG_L(\omega,\bfk)
 & = &
 {\omega^2 \left( \varepsilon + P  + i\omega^3 Q(\omega, \bfk) \right)
 \over \bfk^2 - \omega^2/Z(\omega,\bfk) + i\omega^3 R(\omega,\bfk)
 }
 \label{eq:GLbulk}
 \ee
 Here $Z(\omega,\bfk)$, $Q(\omega, \bfk)$ and $R(\omega,\bfk)$
 all have the form
 \be
 Z(\omega,\bfk) = Z_R(\omega, \bfk) - i\omega Z_I(\omega,\bfk)
 \ee
 where $Z_R(\omega,\bfk)$ and $Z_I(\omega,\bfk)$ are 
 real-valued even functions of $\omega$ and $\bfk$.
 The real parts $Z_R$ and $R_R$
 must have non-zero limits as $\omega\to 0$ and $\bfk \to 0$.
 All other parts of $Z, Q$ and $R$ must have finite limits
 as $\omega\to 0$ and $\bfk \to 0$.
 Matching the small $\omega$ limit (\ref{eq:GLomegalim}) demands that
 \be
 Z_R(0,\bfk) = 
{ \varepsilon + P\over  \barG^{00,00}_R(0,\bfk) - \varepsilon }
 \ee
 
Up to the quadratic terms in $\omega$ and $\bfk$, 
the poles of $\barG_L(\omega,\bfk)$ for small $\omega$
and $|\bfk|$ are determined by
\be
\omega^2 - Z_R(0,0)\bfk^2 + i\omega Z_I(0,0)\bfk^2 = 0 
\label{eq:sound_dispersion}
\ee
This has the structure of the dispersion relationship of a damped sound
wave.
Hence, in the small $\omega$ and the small $|\bfk|$ limit,
$ Z_R(0,0) = v_s^2 $ is the speed of sound squared and $Z_I(0,0)$
is the sound damping coefficient.

To relate $Z_I(0,0)$ to the shear and the bulk viscosities,
let us consider the constitutive relationships once again.
From the shear part, we already have the spin-2 part of the stress
tensor in the fluid cell rest frame
 \be
 \pi_{\rm NS}^{ij}(t,\bfx)
 & = &
 D_T \left(\partial^i T^{j0} + \partial^j T^{i0} 
 - {2g^{ij}\over 3}\partial_l T^{l0}\right)
 \label{eq:piij_consti}
 \ee
 To this we add a spin-0 part
 $-\gamma g^{ij} \partial_t\varepsilon = \gamma g^{ij}\partial_l T^{l0}$
 to get
 \be
 \delta T^{ij}(t,\bfx)
 & = &
 D_T \left(\partial^i T^{j0} + \partial^j T^{i0} 
 - {2g^{ij}\over 3}\partial_l T^{l0}\right)
 + \gamma g^{ij} \partial_l T^{l0}
 \label{eq:FirstConstRel}
 \ee
 The energy conservation law in the local rest frame\footnote{
 In the local rest frame, $\bfu(t,\bfx) = 0$, but $\partial_i u_j \ne 0$.
}
 now becomes
 using Eq.(\ref{eq:Tmunu2}) with the dissipative part given
 by Eq.(\ref{eq:FirstConstRel})
 \be
 0 & = &
 \partial_\mu \partial_\nu T^{\mu\nu}
 \non
 & = &
 \partial_t^2 \varepsilon
 - \nabla^2 P 
 -D_T {4\over 3}\nabla^2 \partial_t\varepsilon 
 -\gamma \nabla^2 \partial_t\varepsilon
 \non
 & \to &
 \left(-\omega^2 + v_s^2 \bfk^2 
 - i(4D_T/3 + \gamma)\bfk^2\omega\right)\delta\varepsilon
 \label{eq:sound_eq}
 \ee
 where we used $g^{ij}\partial_i \partial_j  = -\nabla^2$
 and $ \partial_t\varepsilon = -\partial_l T^{l0} $.
 Comparing with Eq.(\ref{eq:sound_dispersion}), one can identify
 $v_s^2 = \partial P/\partial\varepsilon = Z_R(0,0)$,
 and 
 \be
 Z_I(0,0) = \Gamma = 4D_T/3 + \gamma
 \label{eq:BulkGamma}
 \ee
 as the sound attenuation constant.
 Since we have already identified $D_T = \eta/(\varepsilon + P)$, 
 this allows us to identify $\gamma = \zeta/(\varepsilon + P)$ where 
 $\zeta$ is the bulk viscosity.
 In the fluid cell rest frame, the added term corresponds to 
 the constitutive relationship $\Pi_{\rm NS} = -\zeta \partial_i u^i$.
 In the general frame, this becomes
 \be
 \Pi_{\rm NS} = -\zeta \partial_\mu u^\mu
 \label{eq:Pi_consti}
 \ee
 Again, the label NS indicates that this is the Navier-Stokes form of the bulk
pressure.
 The minus sign in Eq.(\ref{eq:Pi_consti}) makes sense
 since the effective pressure $P + \Pi$ 
 should be less than the equilibrium pressure
 when the fluid is expanding (positive $\partial_i u^i$).
 
 We have so far identified $Z_R(0,0)$ and $Z_I(0,0)$ as the speed of sound squared
 and the sound attenuation coefficient.  The role of $R(\omega,\bfk)$
 is still to be identified.
 In the Kubo formula for the attenuation coefficient, $R_R(0,0)$ appears as
 \be
 \lim_{\omega\to 0}\lim_{\bfk\to 0}{\rm Im}\, {\barG_L(\omega,\bfk)\over \omega}
 = 
 \left(\varepsilon + P)(Z_I(0,0) - Z_R(0,0)^2\, R_R(0,0)\right)
 \label{eq:KuboBulk}
 \ee
 which does not allow one to identify the right hand side with $\Gamma$
 if $R_R(0,0)\ne 0$.
 Actually, the left hand side of Eq.(\ref{eq:KuboBulk}) does yield $\Gamma$.
 It is just that we have not been consistent in power counting
 since the wave equation
 Eq.(\ref{eq:sound_eq}) does not contain $O(\omega^3)$ term while 
 the pole of $\barG_L$ does.
 One may consider this discrepancy
 as the first sign of the trouble with the first order constitutive
relationship
 Eq.(\ref{eq:FirstConstRel}).

\subsection{Second Order Viscous Hydrodynamics}
\label{sec:2nd_order_visc}

 Let us consider the consequence of having the first order constitutive
relationship more
 closely. 
The diffusion equation with a source $S$
\be
\left(\partial_t - D_T\nabla^2\right)\deltan(t,\bfx) = S(t,\bfx)
\ee
has the solution 
\be
\deltan(x) = 
\int d^4x'\, G_R(x-x')\, S(x')
\ee
Here the retarded Green function is
\be
G_R(x-x')
= 
\theta(t-t')
{e^{-{|\bfx-\bfx'|^2\over 4 D (t-t')}}\over 8 (\pi D(t-t'))^{3/2}}
\label{eq:G_diff}
\ee
If one has a point source, $S(x') = N_0\delta(x')$, 
then $\deltan(t,\bfx) = N_0G_R(t,\bfx)$.
At $t=0$, 
the space is empty except at the origin. But at any time after
that, there is non-zero $\deltan$ everywhere. 
This is clearly acausal.

On the other hand, the solution of the sound equation
\be
-(\partial_t^2 - v_s^2 \nabla^2)\delta\epsilon(x) = S(x)
\ee
for a point source $S(x) = \Lambda_0 \delta(x)$ is
\be
\delta\epsilon(t,\bfx) = \Lambda_0\theta(t){1\over 4\pi} {\delta(|\bfx|-v_st)\over
|\bfx|} 
\ee
This is causal since the disturbance only moves with the speed of sound.

The origin of acausality in diffusion is the mismatch between the number of
time derivatives
and the number of spatial derivatives in the diffusion equation.
The diffusive dispersion relationship
$ \omega = -iD\bfk^2 $ gives the group velocity
\be
{\partial\omega\over \partial \bfk}
=
-2iD\bfk
\ee
which becomes large in the large $\bfk$ limit.
This problem can be remedied if one replaces the constitutive equation, 
$J^i = D\partial^i n$ with a relaxation type equation
\be
\partial_t J^i = -{1\over \tau_R} (J^i - D\partial^i n)
\label{eq:JB_relax}
\ee
then the conservation law becomes
\be
\partial_t^2 n 
=
-\partial_i \partial_t J^i
=
-{1\over \tau_R}\partial_t n
+
{D\over \tau_R}\nabla^2 n
\ee
For large $k$ where we previously had a problem, we now have
\be
\omega^2 \approx v_R^2 k^2
\ee
with the
propagation speed $v_R = \sqrt{D/\tau_R}$.

This type of relaxation equation was actually anticipated already: 
Up to the second order in $\omega$,
the poles of the the density-density correlator in Eq.(\ref{eq:GLgen})
are determined by
\be
D\bfk^2 - i\omega - \omega^2 D B = 0
\ee
Comparing, we see that 
\be
B =\tau_R/D
\ee
For the viscous stress-energy tensor components,
the following relaxation equations apply in the local rest frame
\be
\left(\partial_t + {1\over\tau_\pi}\right) \pi^{ij} 
&=& {1\over \tau_\pi} \pi_{\rm NS}^{ij}
\label{eq:shear_relax}
\\
\left(\partial_t + {1\over\tau_\Pi}\right) \Pi
&=& {1\over\tau_\Pi}\Pi_{\rm NS} 
\label{eq:bulk_relax}
\ee
 where
 \be
 \pi^{lm} = T^{lm} -  {g^{lm}\over 3}T^k_k
 \ee
 is the traceless part of the stress tensor and
 \be
 \Pi = -\left( {1\over 3}T^k_k + P\right)
 \ee
 is the bulk pressure. They are not to be identified with the 
 Navier-Stokes forms (\ref{eq:piij_consti}) and (\ref{eq:Pi_consti}).
 Rather, they will relax to the Navier-Stokes forms. 
 
 To see if the acausality in the momentum diffusion
 is cured, we start with the momentum conservation 
 \be
 \partial_t T^{k0} = -\partial_l T^{kl}
 \ee
 Applying the curl gives
 \be
 \partial_t \pi^i_T
 & = &
 -\epsilon_{ijk} \partial_j \partial_l \pi^{kl}
\ee
where we defined $\pi^i_T = \epsilon_{ijk}\partial_j T^{k0}$.
Applying $(\partial_t + 1/\tau_\pi)$ and using Eq.(\ref{eq:piij_consti})
yields
 \be
 0 = \left(\tau_\pi \partial_t^2 + \partial_t - D_T\nabla^2\right)\pi^i_T
\ee
As long as $D_T/\tau_\pi < 1$, this is now causal.
 
For the sound modes, we start with the conservation law in the local rest
frame
\be
\partial_t^2\varepsilon 
= \nabla^2 P + \partial_l\partial_m \pi^{lm}
- \nabla^2 \Pi
\label{eq:varepsilon_eq}
\ee
 Applying
 $(\tau_\pi\partial_t + 1)(\tau_\Pi\partial_t + 1)$
 to Eq.(\ref{eq:varepsilon_eq}) and using Eqs.(\ref{eq:shear_relax}) and
 (\ref{eq:bulk_relax}), one obtains 
 the following dispersion relation for the bulk mode propagation (in this
 case $\delta\varepsilon$)
\be
\lefteqn{
0 = 
\tau_\pi\tau_\Pi \omega^4 
- \tau_\pi\tau_\Pi v_s^2 \omega^2\bfk^2 
- \tau_\Pi{4D_T\over 3}\bfk^2\omega^2
- \tau_\pi\gamma \bfk^2\omega^2
}&&
\non &&
- \omega^2
+ v_s^2\bfk^2
-
i\left({4 D_T\over 3} + \gamma
+ v_s^2 \left(\tau_\pi + \tau_\Pi\right)
\right)
\bfk^2\omega
+ i\omega^3\left(\tau_\pi + \tau_\Pi\right)
\label{eq:bulk_dispersion}
\non
\ee
Comparing with the small $\omega$ and small $|\bfk|$ expansion
of the denominator in Eq.(\ref{eq:GLbulk}), we can identify
$Z_R(0,0) = v_s^2$,
 \be
 Z_I(0,0) = {4 D_T\over 3} + \gamma + v_s^2 \left(\tau_\pi + \tau_\Pi\right)
 \ee
 and
 \be
 R_R(0,0) = (\tau_\pi + \tau_\Pi)/v_s^2
 \ee
The Kubo formula for the damping constant now makes more sense
 \be
 \lim_{\omega\to 0}\lim_{\bfk\to 0}{\rm Im}\, {\barG_L(\omega,\bfk)\over \omega}
 &=& 
 (\varepsilon + P)(Z_I(0,0) - Z_R(0,0)^2\, R_R(0,0))
 \non
 & = &
 (\varepsilon + P)\left({4D_T\over 3} + \gamma \right)
 \non
 & = &
 {4\eta\over 3} + \zeta
 \ee
 and for the bulk viscosity only,
 \be
 \zeta = 
 \lim_{\omega\to 0}\lim_{\bfk\to 0}\,
 {1\over\omega}
 \left( {\rm Im}\, {\barG_L(\omega,\bfk)}
 -{4\over 3} 
 {\rm Im}\, {\barG_R^{xy,xy}(\omega, k_y)}
 \right)
 \ee
 The Kubo formula for the bulk viscosity $\zeta$ is also available
 in terms of the pressure-pressure correlation function
\cite{forster1975,Kadanoff1963}
\be
\zeta =  \lim_{\omega\to 0}\lim_{\bfk\to 0}
{1\over \omega}
{\rm Im}\, G_R^{PP}(\omega, \bfk)
\ee 
Using the fact that the correlation functions of $T^{00}$ vanishes in the $\bfk\to 0$ limit,
one can use in place of $P$ 
the trace $T^{\mu}_\mu/3$ or the combination $P - v_s^2\varepsilon$
to make it more explicit that the bulk viscosity 
is non-zero only if the conformal symmetry is broken.
The Kubo formulas for the relaxation times $\tau_\pi$ and $\tau_\Pi$ 
are not simple to determine in this analysis.
Simple Kubo formulas for $\tau_\pi$ has been worked out in 
Refs.\cite{Baier:2007ix,Moore:2010bu} as
\be
\eta\tau_\pi = -\lim_{\omega\to 0}\lim_{\bfk\to 0}
{1\over 2}
{\rm Re}\,\partial_\omega^2 G^{xy,xy}_R(\omega, \bfk)
\ee
although a simple Kubo formula for $\tau_\Pi$ is still to be found.
 
 In the dispersion relation Eq.(\ref{eq:bulk_dispersion}), the 4-th order
 terms of $O(\omega^4)$ and $O(\omega^2\bfk^2)$ are present but
$O(\bfk^4)$
 terms are not. This may seem unsatisfactory since there is no reason why
 this term should be small compared to the other two 4-th order terms when
 $\omega \sim v_s |\bfk|$.
 However, one should recall that hydrodynamics is valid only in the long
 wavelength and small frequency limits. From this point of view, 
 the 4-th order terms are not so important. They become, however,
significant when the equations are solved numerically that can include short
wavelength excitations.
Fortunately, this Israel-Stewart form of second order 
hydrodynamics\cite{Israel:1979wp} 
(comprising of
Eqs.(\ref{eq:JB_relax}), (\ref{eq:shear_relax}) and (\ref{eq:bulk_relax}))
is shown to be stable in Refs.\cite{Denicol:2008ha,Pu:2009fj}.

 If one wants to include $O(\bfk^4)$ terms, then one can modify the
 relaxation equations (\ref{eq:shear_relax}) and (\ref{eq:bulk_relax})
 to include second derivatives of $T^{ij}$. However, doing so not only
 generates $O(\bfk^4)$ terms, but it also generates (incomplete)
 terms involving 5 and 6 factors of $\omega$ and $\bfk$. 
 Since higher order terms begin to matter at large $\omega$ 
 and $|\bfk|$, having higher and higher order of frequency and momentum (or derivatives
 in the configuration space) does not guarantee that the numerical solution
 in this limit becomes more and more faithful to the real spectrum.
 One just needs to be careful not to interpret high frequency and momentum modes as physical.

As for the calculation of the viscosities, full leading order perturbative QCD 
results for both the shear viscosity and the bulk viscosity have been obtained 
in Refs.\cite{Arnold:2001ba,Arnold:2003zc,Arnold:2006fz} using the Kubo formulas
illustrated above.
QCD is an asymptotically free theory
\cite{Gross:1973ju,Politzer:1974fr,Bethke:2012zza}.
In principle, it admits a perturbative expansion
only when the energy scale exceeds at least a few GeV.
Since the typical QGP energy scale is less than $1\,\hbox{GeV}$, the strong
coupling is not small.
Phenomenologically, we must have $\alpha_S \approx 0.3$ 
\cite{Schenke:2009gb,Tribedy:2011aa}.
This value may look weak, but the gauge coupling itself 
$g = \sqrt{4\pi \alpha_S} \approx 2$ is not so small.
In the perturbative many-body QCD,
$g$ (or $g/2\pi$) is the expansion parameter not
$\alpha_S$~\cite{Gross:1980br,Andersen:2010ct,Andersen:2011sf,Haque:2012my}.
Therefore although the analysis performed
in Refs.\cite{Arnold:2001ba,Arnold:2003zc,Arnold:2006fz} are nothing short
of tour de force, having $g \approx 2$ makes 
numerical values obtained in perturbation theory not too reliable.
At this point, reliable first principle calculations at large $g$ can
only be performed on numerical lattice in Euclidean space.
Lattice QCD can straightforwardly compute static properties 
such as the equation of state. However, calculations of dynamic properties
such as the viscosities become more complicated
as they involve estimating real continuous functions 
from a finite set of discrete Euclidean data.
Nonetheless, a great deal has been accomplished in computing the
properties
of QGP through lattice QCD calculations\cite{AliKhan:2001ek,Cheng:2009zi,
Borsanyi:2013bia,Meyer:2007dy,Meyer:2007ic}
as well as through effective models 
such as the hadron resonance gas model (HR) 
\cite{NoronhaHostler:2008ju,FernandezFraile:2009mi}
and the AdS/CFT correspondence
\cite{Kovtun:2004de,Buchel:2007mf,Gubser:2008sz,Springer:2010mw}.

The purpose of this section has been to show that the
hydrodynamics is very general. No matter what the system is, there usually
is a regime where hydrodynamics is in some way applicable as long as there
exist ``macroscopically small but microscopically large'' length and time scales.
For more detailed analysis of the length and time scales and also for
demonstrating more general structure
of hydrodynamic equations, we now turn to the kinetic theory. 

\section{Hydrodynamics from kinetic theory}
\label{sec:kinetic_hydro}

\subsection{Length scales and validity of hydrodynamic approximations}
\label{sec:kinetic}

Kinetic theory describes a medium microscopically, by following the
evolution of the phase-space distribution function $f(x,p)$, a Lorentz
scalar that describes the probability of finding a particle with
four-momentum $p^\mu$ at space-time position $x^\nu$. Classical kinetic
theory assumes that the particle momenta are on-shell, $p^2=m^2$, which
requires the system to be sufficiently dilute and the mean free paths
sufficiently long to ignore collisional broadening effects on the spectral
function $\rho(p)=2\pi\delta(p^2{-}m^2)$ that defines the particles'\
propagator. The defining equation of classical kinetic theory is the
Boltzmann equation,
\be
\label{k:eq1}
  p^\mu\partial_\mu f(x,p) = C(x,p),
\ee
where $C(x,p)$ is the collision term in which the strength of the
interaction enters through their scattering cross sections. Especially for
massless degrees of freedom, its detailed form can be quite complicated
\cite{Arnold:2002zm}. A popular simplification of the collision term is
the
relaxation time approximation (RTA)
\footnote{
The Boltzmann equation with this RTA-approximated collision term is known
as the Anderson-Witting equation.}
\begin{equation}
\label{k:eq1a}
    C(x,p) =
\frac{p^\mu u_\mu(x)}{\tau_\mathrm{rel}(x,p)}\,\Bigl[f_\mathrm{eq}(x,p){-}f(x,p)\Bigr].
\end{equation}
where the relaxation time $\tau_\mathrm{rel}$ in general depends on
position through the local density and can also depend on the local rest
frame energy of the particles (indicated by the $p$-dependence). Classical
kinetic theory is valid if this relaxation time $\tau_\mathrm{rel}$, and
the associated mean free path $\lambda_\mathrm{mfp}=\langle (p/E)
\tau_\mathrm{rel}\rangle$, are sufficiently large. In other words,
interactions among the constituents must be weak.

Hydrodynamics is valid if the system is close enough to thermal
equilibrium
that its local momentum distribution (and therefore its macroscopic
fields,
such as particle and energy density and pressure, which can all be
expressed as moments of the local momentum distribution) can be
characterized by a small number of thermodynamic and transport parameters,
such as temperature, chemical potential, shear and bulk viscosity, etc.
This requires efficient interactions among the constituents of the medium
because otherwise any kind of macroscopic dynamics involving local
expansion or
compression or shear of the fluid will drive its local momentum
distribution away from its near-equilibrium form. Hydrodynamics works best
for systems made of strongly interacting constituents.

Does this mean that the validity of kinetic theory and hydrodynamics are
mutually exclusive? Not necessarily. To gain clarity consider a
relativistic system of (almost) massless degrees of freedom. It can be
characterized by three length scales, two microscopic and one macroscopic
one:
\begin{itemize}
\item the thermal wavelength $\lambda_\mathrm{th}\sim 1/T$
\item the mean free path $\lambda_\mathrm{mfp}\sim 1/(\langle \sigma
v\rangle n)$ where
        $\langle \sigma v\rangle$ is the momentum-averaged transport cross
section times
        the relative speed ($\approx 1$ in units of $c$) of the colliding
objects, and $n$ is the
        density of scatterers
\item the length scale $L_\mathrm{hydro}$ over which macroscopic fluid
dynamical variables vary;
         it can be defined in many ways that give quantitatively different
but similar order of
magnitude results: 
$L^{-1}_\mathrm{hydro}\sim \partial_\mu u^\mu \sim |\partial_\mu e|/e$ etc.
\end{itemize}
The ratio between the two microscopic scales characterizes the magnitude
of
the transport coefficients $\eta$ (shear viscosity), $\zeta$ (bulk
viscosity), and $\kappa$ (heat conductivity):
\be
\label{k:eq1b}
\frac{\lambda_\mathrm{mfp}}{\lambda_\mathrm{th}} \sim
\frac{1}{\langle \sigma \rangle n}\,\frac{1}{\lambda_\mathrm{th}} \sim
\frac{1}{\langle\sigma\rangle \lambda_\mathrm{th}}\frac{1}{s} \sim
\frac{\eta}{s},\ \frac{\zeta}{s},\ \frac{T\kappa}{s},
\ee
where we used $\eta,\ \zeta,\ T\kappa \sim
1/(\langle\sigma\rangle\lambda_\mathrm{th}) \sim \lambda_\mathrm{mfp}T^4$
and the entropy density $s\simeq 4n\sim T^3$ for a near-thermalized system
of particle number density $n$ for massless degrees of freedom.

In terms of the two microscopic length scales we can define three regimes
of microscopic dynamics:
\begin{enumerate}
\item {\bf Dilute gas regime}:
\be
\label{k:eq1c}
\frac{\lambda_\mathrm{mfp}}{\lambda_\mathrm{th}} \sim \frac{\eta}{s}\gg 1
\quad \Longleftrightarrow \quad \langle\sigma\rangle \ll
\lambda_\mathrm{th}^2\sim\frac{1}{T^2}
\ee
This is the {\em weak-coupling regime} where the microscopic system
dynamics can be described in terms of on-shell quasi-particles and
many-body correlations are suppressed. In this regime the Boltzmann
equation applies.
\item {\bf Dense gas regime:}
\be
\label{k:eq1d}
\frac{\lambda_\mathrm{mfp}}{\lambda_\mathrm{th}} \sim \frac{\eta}{s}\sim 1
\quad \Longleftrightarrow \quad \langle\sigma\rangle \sim
\lambda_\mathrm{th}^2
\ee
In this case interactions happen on the scale $\lambda_\mathrm{th}$. We
call this the {\em moderate coupling regime} where the microscopic system
dynamics must be described by off-shell quasiparticles (whose spectral
functions have a finite collisional width) and many-body correlation
effects are non-negligible. Here the Boltzmann equation must be replaced
by
a quantum kinetic approach based on Wigner distributions, and the BBGKY
hierarchy of coupled equations for the $N$-body distribution functions can
no longer be efficiently truncated.
\item {\bf Liquid regime:}
\be
\label{k:eq1e}
\frac{\lambda_\mathrm{mfp}}{\lambda_\mathrm{th}} \sim \frac{\eta}{s}\ll 1
\quad \Longleftrightarrow \quad \langle\sigma\rangle \gg
\lambda_\mathrm{th}^2
\ee
This is the {\em strong-coupling regime} where the system has no
well-defined quasiparticles and no valid kinetic theory description.
\end{enumerate}

\noindent
To judge the validity of a macroscopic hydrodynamic approach we compare
the
microscopic to the macroscopic length scales. To simplify the discussion,
let us agree on using the inverse of the scalar expansion rate
$\theta=\partial_\mu u^\mu$ to represent the macroscopic length scale
$L_\mathrm{hydro}$.
\footnote{%
The shorter $L_\mathrm{hydro}$, the faster the system is driven away from
local equilibrium. The scalar expansion rate directly drives the bulk
viscous pressure $\Pi$. It is parametrically of the same order as the
shear
tensor $\sigma^{\mu\nu}=\Delta^{\mu\nu}_{\alpha\beta}\partial^\alpha
u^\beta\equiv \nabla^{\langle\mu} u^{\nu\rangle}$ defined in
Eq.~(\ref{eq:shear}) that drives the shear viscous pressure $\pi^{\mu\nu}$
and as the diffusion force $I^\mu\EQ\nabla^\mu(\mu/T)$ associated with
space-time gradients of conserved charge densities that drives the heat
flow $V^\mu$.
}
The figure of merit controlling the validity of a fluid dynamic picture is
the {\em Knudsen number}:
\be
\label{k:eq1f}
\mathrm{Kn} = \lambda_\mathrm{mfp}\cdot\theta \sim \frac{\eta}{s}
\lambda_\mathrm{th}\cdot\theta
\sim \frac{\eta}{sT}\cdot\theta \sim \theta\tau_\mathrm{rel}.
\ee
The Knudsen number is the small parameter that controls the convergence of
the expansion in gradients of thermodynamic quantities that underlies the
derivation of hydrodynamics as an effective theory for the long-distance
dynamics of a general quantum field theory \cite{Dubovsky:2011sj}. Again,
we can use it to define three regimes:
\begin{enumerate}
\item{\bf Ideal fluid dynamics:}
\be
\label{k:eq1g}
\mathrm{Kn}\approx 0 \quad \Longleftrightarrow \quad
\frac{\eta}{s}\approx0\ \text{or}\ \theta\approx 0 \ \text{such that}\
\theta\tau_\mathrm{rel}\approx 0
\ee
\item{\bf Viscous fluid dynamics:}
\be
\label{k:eq1h}
\mathrm{Kn}\lesssim1 \quad \Longleftrightarrow \quad \frac{\eta}{s}\
\text{or}\ \theta\ \text{small} \
\text{such that}\ \theta\tau_\mathrm{rel}\lesssim 1
\ee
\item{\bf Hydrodynamics breaks down:}
\be
\label{k:eq1i}
\mathrm{Kn}\gg1 \quad \Longleftrightarrow \quad \frac{\eta}{s}\ \text{or}\
\theta\ \text{large} \
\text{such that}\ \theta\tau_\mathrm{rel}\gg 1
\ee
\end{enumerate}

In high-energy heavy-ion collisions, the initial energy deposition occurs
in an approximately boost-invariant fashion along the beam direction,
leading to an expansion rate $\theta$ that diverges like $1/\tau$ for
small time $\tau$ after impact.%
\footnote{Here, $\tau = \sqrt{t^2 - z^2}$ is the longitudinal proper time and 
the boost-invariance refers to independence of the space-time rapidity 
$\eta = \tanh^{-1}(z/t)$. For more details, see section \ref{sec:taueta}.}
On the other hand we now know that the quantum
mechanical uncertainty relation places a lower bound $\eta/s \gtrsim
1/(4\pi)$ on any system, even at infinitely strong coupling
\cite{Danielewicz:1984ww,Policastro:2001yc,Kovtun:2004de}. Therefore,
hydrodynamics is inapplicable during the earliest stage of a heavy-ion
collision. At the end of a heavy-ion collision, the mean free paths of
hadrons become large compared to the Hubble radius $\sim 1/\theta$ of the
expanding fireball, and hydrodynamics breaks down again. This process is
called {\em kinetic decoupling}. Between the early pre-equilibrium and the
final decoupling stage stretches an extended period of applicability of
viscous fluid dynamics. The most important factor ensuring this is the
strong collective coupling of the quark-gluon plasma (QGP) phase which is
characterized by a small specific shear viscosity $\eta/s\sim
(2{-}3)/(4\pi)$
\cite{Romatschke:2007mq,Song:2010mg,Gale:2012rq,Heinz:2013th}.

Note that the validity of hydrodynamics does not rely directly on $\eta/s$
being small, only on $(\eta/s)\cdot(\theta/T)$ being small. So, strictly
speaking, strong coupling is not required for hydrodynamics to be valid.
Only in extreme situations, such as heavy-ion collisions which are
characterized by extreme expansion rates, does hydrodynamics require very
strong coupling. In this case, hydrodynamics is applicable even though
classical kinetic theory is not, because very strongly coupled quantum
field theories do not allow a description in terms of on-shell
quasi-particles. It is generally believed that the very earliest stage of
a
heavy-ion collision has no well-defined quasiparticles at all and is
better
described by a theory of classical or quantum fields than by a (quantum)
kinetic approach. On the other hand, weakly coupled system, with very
large
values of $\eta/s$ in which the applicability of (even classical) kinetic
theory is ensured, can still be describable macroscopically through fluid
dynamics if they are sufficiently homogeneous and expand slowly. In this
case the smallness of $\theta$ can compensate for the largeness of
$\eta/s$, resulting in a small Knudsen number. Systems with a large
$\eta/s$ but a small product $(\eta/s)\cdot(\theta/T)$ admit simultaneous
microscopic classical kinetic and macroscopic hydrodynamic descriptions.
In
the following subsections we will study such systems to derive the
macroscopic hydrodynamic equations from the microscopic kinetic theory.
Hydrodynamics being an effective long-distance theory, the form of the
resulting equations does not rely on the validity of the underlying
kinetic
theory (although the values for the transport coefficients do); they can
therefore also be applied to a strongly-coupled liquid such as the QGP.

\subsection{Ideal fluid dynamics}

We define $p$-moments of the distribution function weighted with some
momentum observable $O(p)$ by
\begin{equation}
\label{eq2}
 \langle O(p)\rangle \equiv \int_p O(p)\, f(x,p) \equiv g
\int \frac{d^3p}{(2\pi)^3p^0} O(p)\, f(x,p)
\end{equation}
($g$ is a degeneracy factor) and $p^0 = E_p = \sqrt{m^2 + \bfp^2}$.
The particle number current and energy
momentum tensor are then written as
\be
\label{eq3}
j^\mu \EQ \langle p^\mu\rangle,\quad
T^{\mu\nu}\EQ\langle p^\mu p^\nu\rangle.
\ee
Usually there is more than one particle species in the system, and the
conserved baryon charge current $J^\mu_B$ and energy-momentum tensor
$T^{\mu\nu}$ are given in terms of linear combinations of $\langle
p^\mu\rangle_i$ and $\langle p^\mu p^\nu\rangle_i$ where the subscript $i$
labels the particle species whose distribution function is $f_i(x,p)$:
\begin{equation}
\label{eq3a}
J^\mu_B = \sum_i b_i j^\mu_i = \sum_i b_i \langle p^\mu\rangle_i,\quad
T^{\mu\nu} = \sum_i T^{\mu\nu}_i = \sum_i \langle p^\mu p^\nu\rangle_i;
\end{equation}
here $b_i$ is the baryon charge carried by each particle of species $i$.
For simplicity, we restrict the following discussion to a single particle
species.

The particle number current and energy-momentum tensor take their ideal
fluid dynamical form
\be
\label{eq3b}
j^\mu_\mathrm{id}\EQ{n}u^\mu, \qquad
T_\mathrm{id}^{\mu\nu}\EQ{\varepsilon} u^\mu u^\nu -
P\Delta^{\mu\nu},
\ee
where the spatial projector in the local rest frame (LRF)
$\Delta^{\mu\nu}$
is given in Eq.~(\ref{eq:Deltamunu}), if we assume that the system is locally
momentum isotropic:
\begin{equation}
\label{eq4}
f(x,p)=f_\mathrm{iso}(x,p)\equiv
f_\mathrm{iso}\left(\frac{p^\mu u_\mu(x)-\mu(x)}{T(x)}\right).
\end{equation}
The local equilibrium distribution
\be
\label{eq4a}
f_\mathrm{eq}(\zeta)=\frac{1}{e^\zeta+a},
\ee
where $\zeta\equiv(p^\mu u_\mu(x){-}\mu(x))/T(x)$ and $a=1,-1,0$ for
Fermi-Dirac, Bose-Einstein, and classical Boltzmann statistics,
respectively, is a special form of $f_\mathrm{iso}(x,p)$. It is defined as
the distribution for which the collision term $C(x,p)$ in the Boltzmann
equation (\ref{k:eq1}) vanishes. Note that the ideal fluid decomposition
(\ref{eq3b}) does not require chemical equilibrium, i.e.~it holds for
arbitrary values of the chemical potential $\mu(x)$, nor does it require
complete thermal equilibrium, i.e.~$f_\mathrm{iso}$ is not required to
depend on its argument exponentially as is the case for the equilibrium
distribution (\ref{eq4a}). If the dependence is non-exponential, the
collision term in the Boltzmann equation is non-zero, but its
$p^\mu$-moment still vanishes, $\int_p p^\mu C\EQ0$, due to
energy-momentum
conservation.

The ideal hydrodynamic equations follow by inserting the ideal fluid
decomposition (\ref{eq3b}) into the conservation laws Eq.~(\ref{eq:1}):
\be
\label{eq:ideal}
 \dot{n} =-n\theta,\qquad
 \dot{\varepsilon} = -(\varepsilon{+}P)\theta,\qquad
 \dot{u}^\nu = \frac{\nabla^\mu P}{\varepsilon{+}P}
  = \frac{c_s^2}{1+c_s^2}\,\frac{\nabla^\mu \varepsilon}{\varepsilon},
\ee
where the very last expression assumes an EOS of type $P=c_s^2
\varepsilon$. $\dot F$ denotes the LRF time derivative of a function $F$,
$\dot F{\,\equiv\,}D_\tau F{\,\equiv\,}u^\mu \partial_\mu F$, and
$\nabla^\mu=\Delta^{\mu\nu}\partial_\nu$ the spatial gradient in the LRF. 
Thus, $\partial_\mu = u^\mu D_\tau +\nabla^\mu$.%
\footnote{%
Note that in curvilinear coordinates or curved space-times, the partial
derivative $\partial_\mu$ must be replaced by the covariant derivative
$d_\mu$.
}

Equations~(\ref{eq:ideal}) can be solved numerically for the local
particle
density $n(x)$, energy density $\varepsilon(x)$, and flow velocity
$u^\mu(x)$, with the temperature $T(x)$, chemical potential $\mu(x)$ and
pressure $P(x)$ following from the equation of state (EOS) of
the
fluid. Local deviations from chemical equilibrium result in a
non-equilibrium value of the local chemical potential $\mu(x)$ and a
non-zero right hand side in the current conservation equation for $j^\mu$.
Deviations from thermal equilibrium (while preserving local isotropy) must
be accounted for by a non-equilibrium pressure in the EOS\,
$P(\varepsilon,n)$. In both cases, the conservation laws, Eqs.~(\ref{eq:1}),
lead to a non-vanishing entropy production rate $\partial_\mu
S^\mu{\,\sim\,}1/\tau_\mathrm{rel}{\,\ne\,}0$.

\subsection{Viscous fluid dynamics}
\subsubsection{Navier-Stokes (NS), Israel-Stewart (IS) and
Denicol-Niemi-Molnar-Rischke (DNMR) theory ({\sc vHydro})}

Israel-Stewart (IS) and Denicol-Niemi-Molnar-Rischke (DNMR) second-order
viscous fluid dynamics \cite{Israel:1979wp} are obtained by using in
(\ref{eq3}) for $f(x,p)$ the ansatz
\begin{equation}
\label{eq6}
 f(x,p)= f_\mathrm{iso}\left(\frac{p^\mu u_\mu(x)-\mu(x)}{T(x)}\right)
+\delta f(x,p).
\end{equation}
The correction $\delta f$ describes the deviation of the solution $f(x,p)$
of the Boltzmann equation from local momentum isotropy. It is supposed to
be ``small'', in a sense that will become clearer below, and will thus be
treated perturbatively.

Most authors set $f_\mathrm{iso}=f_\mathrm{eq}$, i.e. they expand around a
local equilibrium state. To obtain the correct form of the hydrodynamic
equations this is not necessary; only the form of the equation of state
$P(\varepsilon,n)$ and the values of the transport coefficients
depend on this choice. We, too, will make this choice for simplicity, but
emphasize that under certain conditions the perturbative treatment of
$\delta f$ may be better justified if the leading-order distribution
$f_\mathrm{iso}$ is not assumed to be thermal.

For later convenience we decompose $p^\mu$ into its temporal and spatial
components in the LRF:
\be
\label{eq6a}
p^\mu\EQ(u^\mu u^\nu{+}\Delta^{\mu\nu})p^\mu={\barE_p}u^\mu {+}
p^{\langle\mu\rangle}
\ee
where $\barE_p{\,\equiv\,}u_\mu p^\mu$ and
$p^{\langle\mu\rangle}{\,\equiv\,}\Delta^{\mu\nu}p_\nu$ are the energy and
spatial momentum components in the LRF. Then
\be
\label{eq6b}
n = \langle \barE_p\rangle,\qquad
\varepsilon = \langle \barE_p^2\rangle.
\ee

The decomposition (\ref{eq6}) is made unique by Landau matching: First,
define the LRF
by solving the eigenvalue equation (\ref{EVeq}) with the
constraint $u^\mu u_\mu\EQ1$ which selects among the four eigenvectors of
$T^{\mu\nu}$ the timelike one. Eq.~(\ref{EVeq}) fixes the flow vector
$u^\mu(x)$ and the LRF energy density. Next, we fix $T(x)$ and $\mu(x)$ by
demanding that $\delta f$ gives no contribution to the local energy and
baryon density:
\be
\label{eq6c}
\langle \barE_p\rangle_\delta = \langle \barE_p^2\rangle_\delta=0.
\ee
Inserting (\ref{eq6}) into (\ref{eq3}) we find the general decomposition
\begin{equation}
\label{eq7}
j^\mu = j^\mu_\mathrm{id}  + V^\mu, \quad
T^{\mu\nu}=T^{\mu\nu}_\mathrm{id} - \Pi\Delta^{\mu\nu} + \pi^{\mu\nu},
\end{equation}
with a non-zero number flow in the LRF,
\be
\label{eq7a}
V^\mu\EQ\bigl\langle p^{\langle\mu\rangle}\bigr\rangle_\delta, 
\ee
a bulk viscous pressure
\be
\label{eq7b}
\Pi\EQ-\frac{1}{3}\bigl\langle p^{\langle\alpha\rangle}
p_{\langle\alpha\rangle}\bigr\rangle_\delta,
\ee
and a shear stress
\be
\label{eq7c}
\pi^{\mu\nu}\EQ\bigl\langle p^{\langle\mu} p^{\nu\rangle}
\bigr\rangle_\delta,
\ee
where $\langle\dots\rangle_\delta$ indicates moments taken with the
deviation $\delta f$ from $f_\mathrm{iso}$. In the last equation we
introduced the notation
\be
\label{eq7d}
A^{\langle\mu\nu\rangle}\equiv \Delta^{\mu\nu}_{\alpha\beta}
A^{\alpha\beta},
\ee
where $\Delta^{\mu\nu}_{\alpha\beta}$ is the
spin-2 projector introduced in Eq.(\ref{eq:spin2proj})
denoting the traceless and transverse (to $u^\mu$) part of a tensor
$A^{\mu\nu}$. The shear stress tensor
$\pi^{\mu\nu}\EQ{T}^{\langle\mu\nu\rangle}$ thus has 5 independent
components while $V^\mu$, which is also orthogonal to $u^\mu$ by
construction, has 3 independent components.

Using the viscous hydrodynamic decomposition (\ref{eq7}) in the
conservation laws $\partial_\mu T^{\mu\nu} = 0$ and $\partial_\mu j^\mu = 0$,
we obtain the {\sc vHydro} viscous
hydrodynamic evolution equations 
\begin{align}
\label{eq:vhydro}
&\dot n = -n\theta -\nabla_\mu V^\mu,
\nonumber\\
&\dot{\varepsilon}=-(\varepsilon{+}P{+}\Pi)\theta +
\pi_{\mu\nu}\sigma^{\mu\nu},
\nonumber\\
&(\varepsilon{+}P{+}\Pi)\dot{u}^\mu=\nabla^\mu(P{+}\Pi)
    - \Delta^{\mu\nu} \nabla^\sigma\pi_{\nu\sigma} +
\pi^{\mu\nu}\dot{u}_\nu.
\end{align}
where $\sigma^{\mu\nu} = \nabla^{\langle\mu}u^{\nu\rangle}$ is the velocity shear tensor
introduced in Eq.(\ref{eq:shear}).
They differ from the ideal fluid dynamical equations (\ref{eq:ideal}) by
additional source terms  arising from the dissipative flows. Altogether,
the deviation $\delta f$ has introduced (3+1+5)=9 additional dissipative
flow degrees of freedom for which additional evolution equations are
needed. These cannot be obtained from the macroscopic conservation laws
but
require input from the microscopic dynamics. In a system that is initially
in local equilibrium, the deviation $\delta f$ is caused by the dynamical
response of the system to gradients in the thermodynamic and flow
variables. The forces that drive this deviation can be classified by their
Lorentz structure as a scalar, a vector and a tensor force:
\begin{align}
\label{eq7e}
\text{scalar force:}\quad \theta &= \partial_\mu u^\mu\ \text{(scalar expansion rate)};
\nonumber\\
\text{vector force:}\quad I^\mu &= \nabla^\mu\left(\frac{\mu}{T}\right)\
\text{(fugacity gradient)};
\nonumber\\
\text{symmetric tensor force:}\quad \sigma^{\mu\nu} &=
\nabla^{\langle\mu}u^{\nu\rangle}\ \text{(velocity shear tensor)}.
\nonumber\\
\text{antisymmetric tensor force:}\quad \omega^{\mu\nu} &=
\frac{1}{2}\left(\nabla^\mu u^\nu{-}\nabla^\nu u^\mu\right)\
\text{(vorticity tensor)}.
\end{align}
These forces generate dissipative flows, the scalar bulk viscous pressure
$\Pi$, the heat flow vector $V^\mu$, and the shear stress tensor
$\pi^{\mu\nu}$.
\footnote{
The energy-momentum tensor is symmetric, so dissipative flows have no
antisymmetric tensor contribution, but the antisymmetric vorticity tensor
couples to the other dissipative forces and flows at second order in the
Knudsen and inverse Reynolds numbers.
}
The strength of the forces (\ref{eq7e}) driving the system away from local
equilibrium is characterized by the Knudsen number. The system response
can
be characterized by inverse Reynolds numbers associated with the
dissipative flows \cite{Denicol:2012cn}:
\be
\label{eq7f}
\mathrm{R}_\Pi^{-1} = \frac{|\Pi|}{P},\quad
\mathrm{R}_V^{-1} = \frac{\sqrt{-I_\mu I^\mu}}{P},\quad
\mathrm{R}_\pi^{-1} = \frac{\sqrt{\pi_{\mu\nu}\pi^{\mu\nu}}}{P}.
\ee
Due to the time delay $\tau_\mathrm{rel}$ between the action of the force
and the system response, built into the collision term of the Boltzmann
equation, the inverse Reynolds numbers are not necessarily of the same
order as the Knudsen number: For example, an initially small bulk viscous
pressure can remain small, due to critical slowing down, as the system
passes through a phase transition, even though the bulk viscosity becomes
large during the transition \cite{Song:2009rh}. Conversely, strong
deviations from local equilibrium during a rapidly expanding
pre-equilibrium stage in heavy-ion collisions can lead to large initial
values for the dissipative flows, and a slow equilibration rate may cause
them to to stay large for a while even though the expansion rate decreases
with longitudinal proper time as $1/\tau$. Deviations from equilibrium,
and
the accuracy of their description by viscous fluid dynamics, are therefore
controlled by a combination of Knudsen and inverse Reynolds numbers
\cite{Denicol:2012cn}.

The 9 equations of motion describing the relaxation of the 9 dissipative
flow components are controlled by microscopic physics, encoded in the
collision term on the right hand side of the Boltzmann equation, and can
be
derived from approximate solutions of that equation. This was first done
almost 50 years ago by Israel and Stewart in Ref.\cite{Israel:1979wp}, but
when
the problem was recently revisited it was found \cite{Baier:2007ix,
Betz:2009zz, Denicol:2012cn, Denicol:2014vaa} that the relaxation
equations
take a much more general form than originally derived. Specifically,
Denicol {\it et al.} \cite{Denicol:2012cn} found the following general structure
\begin{eqnarray}
\label{eq:DNMR}
&\tau_\Pi \dot{\Pi} + \Pi = -\zeta \theta + \mathcal{J} + \mathcal{K} +
\mathcal{R},
\nonumber\\
&\tau_V \Delta^{\mu\nu}\dot{V}_\nu + V^\mu = \kappa I^\mu +
\mathcal{J}^\mu
+ \mathcal{K}^\mu
             + \mathcal{R}^\mu,
\nonumber\\
&\tau_\pi \Delta^{\mu\nu}_{\alpha\beta}\dot{\pi}^{\alpha\beta} +
\pi^{\mu\nu} = 2\eta \sigma^{\mu\nu} + \mathcal{J}^{\mu\nu}
            + \mathcal{K}^{\mu\nu} + \mathcal{R}^{\mu\nu}.
\end{eqnarray}
Here all calligraphic terms are of second order in combined powers of the
Knudsen and inverse Reynolds numbers. $\mathcal{J}$ terms contain products
of factors that are each of first order in the Knudsen and inverse
Reynolds
numbers; $\mathcal{K}$ terms are second order in Knudsen number, and
$\mathrm{R}$ terms are second order in inverse Reynolds numbers.

In relaxation time approximation, with an energy-independent relaxation
time $\tau_\mathrm{rel}$, the relaxation times for the dissipative flows
all agree with each other:
$\tau_\Pi\EQ\tau_V\EQ\tau_\pi\EQ\tau_\mathrm{rel}$ \cite{Denicol:2014vaa}.
The same does not hold for more general forms of the collision term. If we
set in Eqs.~(\ref{eq:DNMR}) the relaxation times and all other second
order
terms to zero, we obtain the equations of {\bf relativistic Navier-Stokes
theory}:
\be
\label{eq:NS}
\Pi= -\zeta\theta,\quad
V^\mu = \kappa I^\mu,\quad
\pi^{\mu\nu}=2\eta\sigma^{\mu\nu}.
\ee
The relaxation equations (\ref{eq:DNMR}) have solutions that, for
sufficiently small expansion rates (see below), approach asymptotically
(at
times $\tau\gg\tau_{\Pi,V,\pi}$) the Navier-Stokes values (\ref{eq:NS}).
However, plugging the Navier-Stokes solutions (\ref{eq:NS}) directly back
into the decompositions (\ref{eq7}) and using them in the conservation
laws
(\ref{eq:1}) leads to viscous hydrodynamic equations of motion that are
acausal and numerically unstable
\cite{Hiscock:1983zz,Hiscock:1985zz}. The physical reason for this is the
instantaneous response of the dissipative flows to the dissipative forces
encoded in Eqs.~(\ref{eq:NS}) which violates causality. A causal and
numerically stable implementation of viscous fluid dynamics must account
for the time delay between cause and effect of dissipative phenomena and
therefore be by necessity of second order in Knudsen and Reynolds numbers.

The first relativistic causal second-order theory of viscous fluid
dynamics
was Israel-Stewart (IS) theory \cite{Israel:1979wp}. It amounts to
dropping
the $\mathcal{K}$ and $\mathcal{R}$ terms in (\ref{eq:DNMR}) and replacing
(for massless particles) $\mathcal{J}\to{-}\frac{4}{3}\tau_\Pi\theta\Pi$,
$\mathcal{J}^\mu\to{-}\tau_V\theta V^\mu$, and
$\mathcal{J}^{\mu\nu}\to{-}\frac{4}{3}\tau_\pi\theta\pi^{\mu\nu}$. The
importance of keeping specifically these second-order $\mathcal{J}$-terms
for the preservation of conformal invariance in a system of massless
degrees of freedom was stressed by Baier {\it et al.}.\cite{Baier:2007ix}

For conformal systems the resulting Israel-Stewart relaxation equations
can
be written in the form \cite{Song:2008si}
\begin{align}
\label{eq:IS}
\dot{\Pi} &= -\frac{1}{\tau'_\Pi}\Bigl(\Pi{+}\zeta'\theta\Bigr),
\nonumber\\
\Delta^{\mu\nu}\dot{V}_\nu &= -\frac{1}{\tau'_V}\Bigl(V^\mu{-}\kappa'
I^\mu\Bigr),
\nonumber\\
\Delta^{\mu\nu}_{\alpha\beta}\dot{\pi}^{\alpha\beta} &=
-\frac{1}{\tau'_\pi}\Bigl(\pi^{\mu\nu}{-}2\eta' \sigma^{\mu\nu}\Bigr),
\end{align}
with effective transport coefficients and relaxation times that are
modified by the scalar expansion rate as follows:
\be
\label{eq7h}
\zeta' = \frac{\zeta}{1{+}\gamma_\Pi},\quad
\kappa' = \frac{\kappa}{1{+}\gamma_V},\quad
\eta' = \frac{\eta}{1{+}\gamma_\pi},\quad
\tau'_i = \frac{\tau_i}{1{+}\gamma_i} \ (i=\Pi,\,V,\,\pi),
\ee
where 
$\gamma_i=\frac{4}{3}\theta\tau_i$
for $i=\Pi,\pi$ and $\gamma_i=\theta\tau_i$ for $i=V$.

These relaxation equations describe an asymptotic approach of the
dissipative flows to effective Navier-Stokes values that, for a positive
scalar expansion rate $\theta$, are reduced relative to their first-order
values (\ref{eq:NS}) by a factor $1{+}\gamma_i$, while the effective rate
of approach to this effective Navier-Stokes limit is sped up by the same
factor. $\gamma_i$ involves the product of the scalar expansion rate and
the respective relaxation time. So, compared to a more slowly expanding
system, a rapidly expanding system with the same microscopic scattering
cross sections is characterized by lower effective viscosities and shorter
effective relaxation times.\cite{Heinz:2005bw,Song:2008si}

When all the second order terms are kept, the DNMR equations become quite
complicated. The $\mathcal{J}$, $\mathcal{K}$ and $\mathcal{R}$ terms
listed by DNMR \cite{Denicol:2012cn} add up to 16 terms for $\Pi$, 18
terms
for $V^\mu$, and 19 terms for $\pi^{\mu\nu}$, each with its own transport
coefficient. While many of these transport coefficients have been computed
for massless theories at both weak and strong coupling (a subject too rich
to be fully reviewed here, see papers by Denicol {\it et
al.}\cite{Denicol:2012cn,Molnar:2013lta,Denicol:2014vaa}, Moore {\it et
al.}\cite{York:2008rr,Moore:2012tc} and Noronha {\it et
al.}\cite{Finazzo:2014cna} for lists of references to relevant work),
quite
a few are still unknown. Furthermore, the QGP is neither weakly nor
strongly enough coupled, nor is it sufficiently conformally symmetric for
any of these calculations to be quantitatively reliable for heavy-ion
collisions. For these reasons, many practical applications of viscous
fluid
dynamics employ phenomenological values for the transport coefficients,
and
work studying which terms need to be kept and which might be of lesser
importance is still ongoing \cite{Molnar:2013lta}.

One important non-linear coupling mechanism that enters at second order
are
bulk-shear couplings where shear stress drives a bulk viscous pressure and
vice versa
\cite{Denicol:2012cn,Molnar:2013lta,Denicol:2014vaa,Denicol:2014mca}.
Heavy-ion collisions are characterized by initially very large differences
between the longitudinal and transverse expansion rates that cause large
shear stress. The latter, in turn, creates a bulk viscous pressure via
bulk-shear coupling that can dominate over the one generated \`a la
Navier-Stokes by the scalar expansion rate \cite{Denicol:2014mca} and may
actually be able to flip its sign. This should be taken into account in
phonemenological applications of viscous hydrodynamics to heavy-ion
collisions.

\subsubsection{Anisotropic hydrodynamics ({\sc aHydro})}

The dissipative flows are given by moments of the deviation $\delta
f(x,p)$
of the distribution function from local equilibrium, and their relaxation
equations are derived from the Boltzmann equation using approximations
that, in one way or another, assume that $\delta f$ is small. However,
systems featuring strongly anisotropic expansion, such as the early
evolution stage of the fireballs created in ultra-relativistic heavy-ion
collisions, generate strong local momentum anisotropies: the width of the
LRF momentum distribution along a certain direction is inversely
proportional to the local expansion rate in that direction, and this
momentum-space distortion growth with the magnitude of the shear
viscosity.
In viscous hydrodynamics, where we expand $f$ around a locally isotropic
LO
distribution (see Eq.~(\ref{eq6})), this local momentum anisotropy must be
absorbed entirely by $\delta f$, making $\delta f$ large and rendering the
approximations used for calculating the evolution of the dissipative flow
generated by $\delta f$ unreliable. Indeed, even for moderate specific
shear viscosities $\eta/s\sim 5{-}10$ the (negative) longitudinal
component
of the viscous shear pressure can become so large in Israel-Stewart theory
that it overwhelms the thermal pressure, resulting in a negative total
pressure along the beam direction -- which, according to the kinetic
definition $P_L=\frac{1}{3}\langle p_z^2\rangle$, should never
happen.

Anisotropic hydrodynamics \cite{Martinez:2010sc,Florkowski:2010cf} is
based
on the idea to account already in the LO distribution for the local
momentum anisotropy resulting from anisotropic expansion, by parametrizing
\cite{Romatschke:2003ms}
\begin{equation}
\label{eq8}
  f(x,p) = f_\mathrm{RS}(x,p)\equiv
  f_\mathrm{iso}\left(\frac{\sqrt{p_\mu\Xi^{\mu\nu}(x)
p_\nu}-\tilde\mu(x)}{\Lambda(x)}\right),
\end{equation}
where $\Xi^{\mu\nu}(x) = u^\mu(x) u^\nu(x) +\xi(x) z^\mu(x) z^\nu(x)$,
$z^\mu(x)$ being a unit vector in longitudinal $z$ direction in the LRF.
The subscript {\it RS} refers to Romatschke and Strickland who 
are the authors of Ref.\cite{Romatschke:2003ms}.
This distribution is characterized by 3 flow parameters $u^\mu(x)$ and
three ``thermodynamic'' parameters: the ``transverse temperature''
$\Lambda(x)$, the effective chemical potential $\tilde{\mu}(x)$, and the
momentum-anisotropy parameter $\xi(x)$. Inserting (\ref{eq8}) into
(\ref{eq3}) yields the {\sc aHydro} decomposition
\begin{eqnarray}
\label{eq9}
\!\!\!\!\!\!\!\!\!\!\!\!\!\!
  && j^\mu_\mathrm{RS} = n_\mathrm{RS} u^\mu, \quad
  T^{\mu\nu}_\mathrm{RS} = \varepsilon_\mathrm{RS} u^\mu u^\nu -
P_T \Delta^{\mu\nu}
  + (P_L-P_T)z^\mu z^\nu,
\\\label{eq10}
\!\!\!\!\!\!\!\!\!\!\!\!\!\!
  &&n_\mathrm{RS} = \langle E\rangle_\mathrm{RS} = {\cal R}_0(\xi)\,
n_\mathrm{iso}(\Lambda,\tilde\mu),
      \quad
       \varepsilon_\mathrm{RS} = \langle E^2\rangle_\mathrm{RS} = {\cal
R}(\xi)\,
       \varepsilon_\mathrm{iso}(\Lambda,\tilde\mu),
\\\nonumber
    &&P_{T,L} = \langle p^2_{T,L}\rangle_\mathrm{RS} = {\cal
R}_{T,L}(\xi) \,
      P_\mathrm{iso}(\Lambda,\tilde\mu).
\end{eqnarray}
For massless systems, the local momentum anisotropy effects factor out via
the $\mathcal{R}(\xi)$-functions:\cite{Martinez:2010sc}
\begin{align}
\label{eq:R}
&{\cal R}_0(\xi)=\frac{1}{\sqrt{1+\xi}},
&{\cal R}(\xi) = \frac{1}{2}\left(\frac{1}{1+\xi}
+\frac{\arctan\sqrt{\xi}}{\sqrt{\xi}} \right) \, ,
\nonumber\\
&{\cal R}_\perp(\xi) = \frac{3}{2 \xi} \left( \frac{1+(\xi^2{-}1){\cal
R}(\xi)}{\xi + 1}\right) \, ,
&{\cal R}_L(\xi) = \frac{3}{\xi} \left( \frac{(\xi{+}1){\cal
R}(\xi)-1}{\xi+1}\right) \, .
\end{align}
The isotropic pressure is obtained from a locally isotropic equation of
state
$P_\mathrm{iso}(\Lambda,\tilde\mu)\EQ 
P_\mathrm{iso}(\varepsilon_\mathrm{iso}(\Lambda,\tilde\mu),
n_\mathrm{iso}(\Lambda,\tilde\mu))$. 
For massless noninteracting partons,
$P_\mathrm{iso}(\Lambda,\tilde\mu)=
\frac{1}{3}\varepsilon_\mathrm{iso}(\Lambda,\tilde\mu)$ 
independent of chemical composition. 
To compare with ideal and IS viscous hydrodynamics, 
we need to assign the locally anisotropic system an appropriate temperature 
$T(x)\EQ{T}\bigl(\xi(x),\Lambda(x),\tilde{\mu}(x)\bigr)$ 
and chemical potential
$\mu(x)\EQ\mu\bigl(\xi(x),\Lambda(x),\tilde\mu(x)\bigr)$,
thinking of $f_\mathrm{RS}(\xi,\Lambda)$ as an expansion around the locally 
isotropic distribution $f_\mathrm{iso}(T)$. 
For this we impose the generalized Landau matching conditions 
$\varepsilon_\mathrm{RS}(\xi,\Lambda,\tilde\mu)
\EQ{\varepsilon}_\mathrm{iso}(T,\mu)$ and 
For example, using an exponential (Boltzmann) function for 
$f_\mathrm{iso}$ with $\mu=\tilde\mu=0$,
one finds $T\EQ\Lambda {\cal R}^{1/4}(\xi)$. With this matching we can write
\begin{eqnarray}
\label{eq11}
\!\!\!\!\!\!\!\!\!\!\!\!\!\!
   &&T^{\mu\nu}_\mathrm{RS} =  T^{\mu\nu}_\mathrm{id}
   - (\Delta P+\Pi_\mathrm{RS})\Delta^{\mu\nu} +
\pi^{\mu\nu}_\mathrm{RS},
\\\label{eq12}
 \!\!\!\!\!\!\!\!\!\!\!\!\!\!
  &&\Delta P + \Pi_\mathrm{RS} = -\frac{1}{3}\int_p p_\alpha
\Delta^{\alpha\beta} p_\beta
      (f_\mathrm{RS}-f_\mathrm{iso}) \qquad (= 0\ \mathrm{for}\ m=0),
\\\label{eq13}
\!\!\!\!\!\!\!\!\!\!\!\!\!\!
  &&\pi^{\mu\nu}_\mathrm{RS} = \int_p p^{\langle\mu} p^{\nu\rangle}
(f_\mathrm{RS}{-}f_\mathrm{iso})
       = (P_T{-}P_L) \,\frac{x^\mu x^\nu+y^\mu y^\nu-2
z^\mu z^\nu}{3}.
\end{eqnarray}
We see that $\pi^{\mu\nu}_\mathrm{RS}$ has only one independent component,
$P_T{-}P_L$, so {\sc aHydro} leaves 4 of the 5
components of $\pi^{\mu\nu}$ unaccounted for. For massless particles we
have
$(P_T{-}P_L)/P_\mathrm{iso}(\varepsilon)\EQ
{\cal R}_T(\xi){-}{\cal R}_L(\xi)$,
so the equation of motion for $\pi^{\mu\nu}_\mathrm{RS}$ can be replaced by 
one for $\xi$. For (2+1)-dimensional expansion with longitudinal 
boost-invariance these equations can be found and were solved numerically by Martinez
{\it et al.} \cite{Martinez:2012tu}. For $m\ne0$ we need an additional  
``anisotropic EOS'' for 
$(\Delta P/P_\mathrm{iso}){\,\equiv\,}
(2P_T{+}P_L)/(3P_\mathrm{iso}) - 1$, 
in order to separate $\Delta P$ from the viscous bulk pressure $\Pi$.\\

\subsubsection{Viscous anisotropic hydrodynamics ({\sc vaHydro})}

As explained above, {\sc aHydro} \cite{Martinez:2010sc,Florkowski:2010cf}
accounts only for one (albeit largest) of the five independent components
of the shear stress tensor $\pi^{\mu\nu}$. It can therefore not be used to
compute the viscous suppression of elliptic flow which is sensitive to
e.g.
$\pi^{xx}{-}\pi^{yy}$. On the other hand, since the four remaining
components of the shear stress tensor never become as large as the
longitudinal/transverse pressure difference (with smooth initial density
profiles they start out as zero, and with fluctuating initial conditions
they are initially small), they can be treated ``perturbatively'' \`a la
Israel and Stewart, without running into problems even at early times.
Combining the non-perturbative dynamics of $P_L{-}P_T$
via {\sc aHydro} with a perturbative treatment of the remaining viscous
stress terms $\tilde{\pi}^{\mu\nu}$ \`a la Israel-Stewart defines the {\sc
vaHydro} scheme \cite{Bazow:2013ifa}. {\sc vaHydro} is expected to perform
better than both IS theory and {\sc aHydro} during all evolution stages.

The {\sc vaHydro} equations are obtained by generalizing the ansatz
(\ref{eq8}) to include arbitrary (but small) corrections to the
spheroidally deformed $f_\mathrm{RS}(x,p)$:
\begin{equation}
\label{eq14}
  f(x,p) = f_\mathrm{RS}(x,p) + \delta\tilde f(x,p) =
  f_\mathrm{iso}\left(\frac{\sqrt{p_\mu\Xi^{\mu\nu}(x)
p_\nu}-\tilde\mu(x)}{\Lambda(x)}\right)
  + \delta\tilde f(x,p) .
\end{equation}
The parameters $\Lambda$ and $\tilde{\mu}$ in (\ref{eq14}) are
Landau-matched as before, {\it i.e.} by requiring $\langle
E\rangle_{\tilde\delta}\EQ\langle E^2\rangle_{\tilde\delta}\EQ0$. To fix
the value of the deformation parameter $\xi$ one demands that
$\delta\tilde{f}$ does not contribute to the pressure anisotropy
$P_T{-}P_L$; this requires $(x_\mu x_\nu{+} y_\mu
y_\nu{-}2 z_\mu z_\nu)\langle p^{\langle\mu}
p^{\nu\rangle}\rangle_{\tilde\delta}\EQ0$. Then, upon inserting
(\ref{eq14}) into (\ref{eq3}), we obtain the {\sc vaHydro} decomposition
\be
\label{eq15}
       j^\mu = j^\mu_\mathrm{RS} +\tilde{V}^\mu,\quad
       T^{\mu\nu}=T^{\mu\nu}_\mathrm{RS} - \tilde\Pi \Delta^{\mu\nu} +
\tilde\pi^{\mu\nu},
\ee
with
\be
\label{eq15a}
       \tilde{V}^\mu = \bigl\langle
p^{\langle\mu\rangle}\bigr\rangle_{\tilde\delta},\quad
       \tilde\Pi = -{\textstyle\frac{1}{3}} \bigl\langle
p^{\langle\alpha\rangle}
       p_{\langle\alpha\rangle}\bigr\rangle_{\tilde\delta},\quad
       \tilde{\pi}^{\mu\nu} = \bigl\langle p^{\langle\mu}
p^{\nu\rangle}\bigr\rangle_{\tilde\delta},
\ee
subject to the constraints
\be
\label{eq15b}
u_\mu \tilde\pi^{\mu\nu}\EQ\tilde\pi^{\mu\nu} u_\nu\EQ(x_\mu x_\nu{+}y_\mu
y_\nu{-}2 z_\mu z_\nu)\tilde\pi^{\mu\nu}\EQ\tilde\pi^\mu_\mu\EQ0
\ee
Clearly, the additional shear stress $\tilde\pi^{\mu\nu}$ arising from
$\delta\tilde{f}$ has only 4 degrees of freedom.

The strategy in {\sc vaHydro} is now to solve hydrodynamic equations for
{\sc aHydro} \cite{Martinez:2012tu} (which treat
$P_T{-}P_L$ nonperturbatively) with added source terms
describing the residual viscous flows arising from $\delta\tilde f$,
together with IS-like ``perturbative'' equations of motion for
$\tilde\Pi,\,\tilde V^\mu$, and $\tilde\pi^{\mu\nu}$. The hydrodynamic
equations are obtained by using the decomposition (\ref{eq15}) in the
conservation laws (\ref{eq:1}). The evolution equations for the
dissipative
flows $\tilde \Pi,\, \tilde V^\mu$, and $\tilde\pi^{\mu\nu}$ are derived
by
generalizing the DNMR\cite{Denicol:2012cn} procedure to an expansion of
the
distribution function around the spheroidally deformed $f_\mathrm{RS}$ in
(\ref{eq8}), using the 14-moment approximation. These equations are
lengthy; for massless systems undergoing (2+1)-dimensional expansion with
longitudinal boost invariance they were derived by Bazow\,{\it et al.}
\cite{Bazow:2013ifa}. Generalizations to massive systems and full
(3+1)-dimensional expansion are in progress. We give their simplified form
for (0+1)-d expansion in the next subsection. Especially at early times
$\delta\tilde f$ is much smaller than $\delta f$, since the largest part
of
$\delta f$ is already accounted for by the momentum deformation in
(\ref{eq8}). The inverse Reynolds number
$\tilde{\mathrm{R}}_\pi^{-1}=\sqrt{\tilde\pi^{\mu\nu}\tilde\pi_{\mu\nu}}/{\cal P}_\mathrm{iso}$ associated with the residual shear stress $\tilde\pi^{\mu\nu}$ is therefore strongly reduced compared to the one associated with $\pi^{\mu\nu}$, significantly improving the range of applicability of {\sc vaHydro} relative to standard second-order viscous hydrodynamics.

\subsection{Testing different hydrodynamic approximations}

For (0+1)-d longitudinally boost-invariant expansion of a transversally
homogeneous system, the Boltzmann equation can be solved exactly in the
relaxation time approximation (RTA), both for massless
\cite{Baym:1984np,Florkowski:2013lza,Florkowski:2013lya} and massive
particles\cite{Florkowski:2014sfa,Florkowski:2014bba}. More recently, an
exact solution of this equation was also found for massless systems
undergoing (1+1)-dimensional
expansion,\cite{Denicol:2014xca,Denicol:2014tha} with a boost invariant
longitudinal and azimuthally symmetric transverse velocity profile
discovered by Gubser (``Gubser flow'') \cite{Gubser:2010ze,Gubser:2010ui}
as the result of imposing a particular conformal symmetry (``Gubser
symmetry'') on the flow. These exact solutions of the kinetic theory can
be
used to test various hydrodynamic approximation schemes, by imposing the
symmetry of the exact solution also on the hydrodynamic solution, solving
both with identical initial conditions, and comparing the predictions of
both approaches for the evolution of macroscopic observables
\cite{Florkowski:2013lza,Florkowski:2013lya,Bazow:2013ifa,Denicol:2014xca,Denicol:2014tha,Nopoush:2014qba}.

We will here use the (0+1)-d case to test the {\sc vaHydro} approach
\cite{Bazow:2013ifa}. This illustrates the procedure and the kind of
conclusions one can draw from such a comparison. Setting homogeneous
initial conditions in $r$ and space-time rapidity $\eta_s$ and zero
transverse flow, $\tilde\pi^{\mu\nu}$ reduces to a single non-vanishing
component $\tilde\pi$:
$\tilde\pi^{\mu\nu}=\mathrm{diag}(0,-\tilde\pi/2,-\tilde\pi/2,\tilde\pi)$
at $z=0$.  The factorization $n_\mathrm{RS}(\xi,\Lambda)\EQ{\cal
R}_0(\xi)\,n_\mathrm{iso}(\Lambda)$ etc. are used to obtain equations of
motion for $\dot\xi,\, \dot\Lambda,\,\dot{\tilde\pi}$
\cite{Bazow:2013ifa}:
\begin{align}
\label{eq:vahydro0+1}
&\frac{\dot\xi}{1{+}\xi}-6\frac{\dot\Lambda}{\Lambda}=
\frac{2}{\tau}+\frac{2}{\tau_\mathrm{rel}}\left(1-\sqrt{1{+}\xi}\,{\cal
R}^{3/4}(\xi)\right)\;,
\nonumber\\
&{\cal R}'(\xi)\, \dot\xi + 4 {\cal R}(\xi) \frac{\dot\Lambda}{\Lambda} =
- \Bigl({\cal R}(\xi) + {\textstyle\frac{1}{3}} {\cal R}_L(\xi)\Bigr)
\frac{1}{\tau}
+\frac{\tilde\pi}{\varepsilon_\mathrm{iso}(\Lambda)\tau},
\nonumber\\
&\dot{\tilde\pi}=
-\frac{1}{\tau_\mathrm{rel}}\Bigl[\bigl({\cal R}(\xi){\,-\,}{\cal R}_{\rm
L}(\xi)\bigr)P_\mathrm{iso}(\Lambda)+\tilde\pi\Bigr]
-\lambda(\xi)\frac{\tilde\pi}{\tau}
\\\nonumber
&\qquad  +12\biggl[
   \frac{\dot{\Lambda}}{3\Lambda}\Bigl({\cal R}_{\rm L}(\xi){\,-\,}{\cal
R}(\xi)\Bigr)
  +\Bigl(\frac{1{+}\xi}{\tau}-\frac{\dot{\xi}}{2}\Bigr)
  \Bigl({\cal R}^{zzzz}_{-1}(\xi){\,-\,}\frac{1}{3}{\cal
R}^{zz}_{1}(\xi)\Bigr)
  \biggr] P_\mathrm{iso}(\Lambda).
\end{align}
$\tau_\mathrm{rel}$ and the ratio of shear viscosity $\eta$ to entropy
density $s$, $\eta/s$, are related by
$\tau_\mathrm{rel}\EQ5\eta/(sT)\EQ5\eta/(\mathcal{R}^{1/4}(\xi)s\Lambda)$.
The numerical solution of these equations \cite{Bazow:2013ifa} can be
compared with the exact solution of the Boltzmann equation
\cite{Florkowski:2013lza}, and also with the other hydrodynamic
approximation schemes discussed above, plus a 3rd-order viscous
hydrodynamic approximation \cite{Jaiswal:2013vta}.
\begin{figure}[hbt!]
\begin{center}
\includegraphics[width=0.8\linewidth]{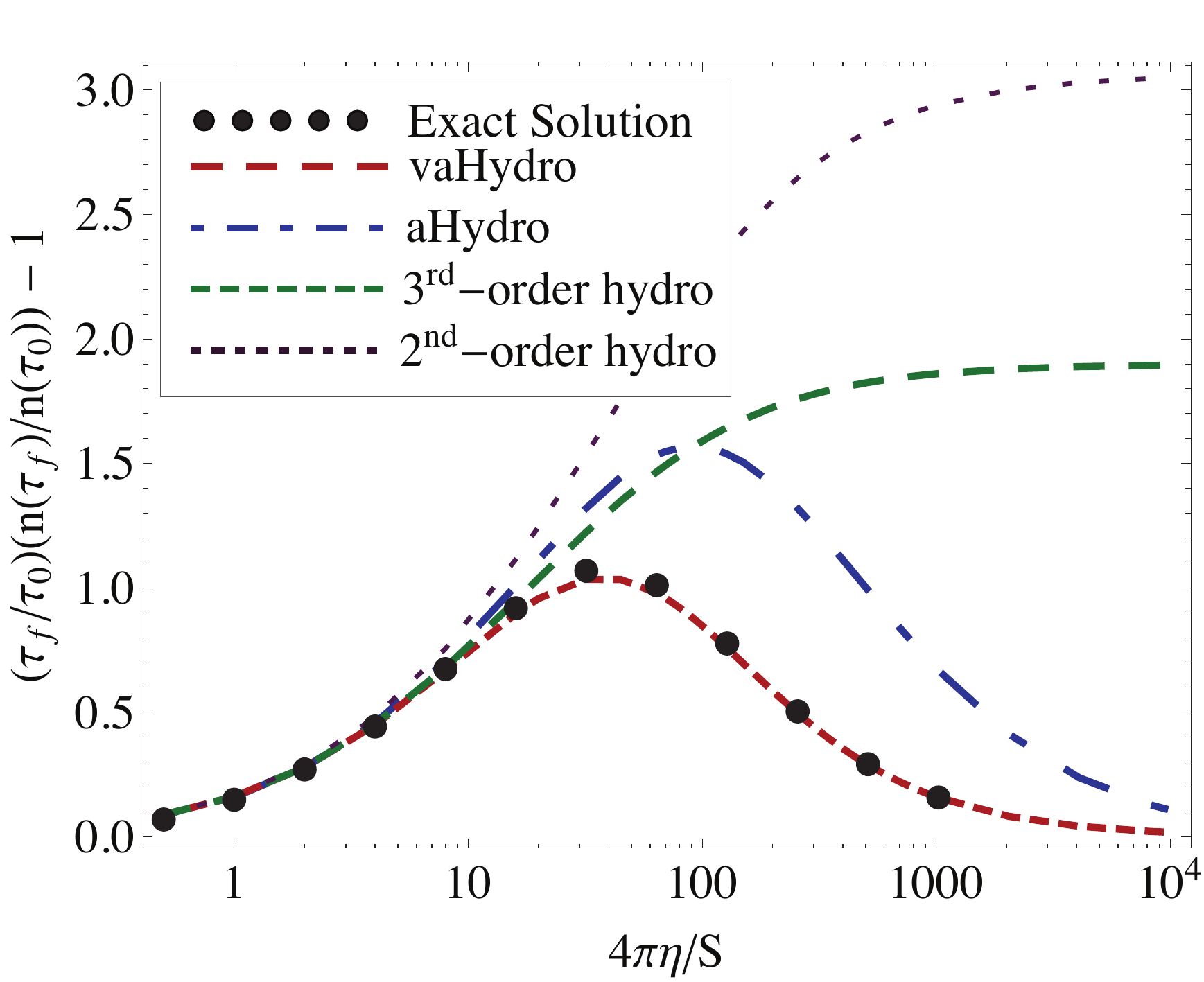}
\end{center}
\caption{(Color online) The particle production measure $(\tau_f
n(\tau_f))/(\tau_0 n(\tau_0)) -1$ as a function of $4\pi\eta/s$. The black
points, red dashed line, blue dashed-dotted line, green dashed line, and
purple dotted line correspond to the exact solution of the Boltzmann
equation, {\sc vaHydro}, {\sc aHydro}, third-order viscous hydrodynamics
\cite{Jaiswal:2013vta}, and DNMR second-order viscous hydrodynamics
\cite{Denicol:2012cn}, respectively. The initial conditions are
$T_0\EQ600$\,MeV, $\xi_0\EQ0$, and $\tilde\pi_0\EQ0$ at
$\tau_0\EQ0.25$\,fm/$c$. The freeze-out temperature was taken to be
$T_f\EQ150$\,MeV.
\label{F1}
}
\end{figure}
As an example, we show in Fig.~\ref{F1} the entropy production (measured
by
the increase in particle number $\tau n(\tau)$) between start and end of
the dynamical evolution from an initial temperature of 600\,MeV to a final
one of 150\,MeV. For this extreme (0+1)-d scenario, where the difference
between longitudinal and transverse expansion rates is maximal, {\sc
vaHydro} is seen to reproduce the exact solution almost perfectly,
dramatically outperforming all other hydrodynamic approximations. In
particular, it should be noted that only the {\sc aHydro} and {\sc
vaHydro}
approximations are able to correctly reproduce entropy production (or
rather the lack thereof) in both the extreme strong coupling (ideal fluid
dynamics, $\tau_\mathrm{real}\EQ\eta/s\EQ0$) and the extreme weak coupling
(free-streaming particles, no collisions,
$\tau_\mathrm{real},\,\eta/s\EQ\to\infty$) limits of the microscopic
dynamics. {\sc vHydro} schemes based on an expansion around a locally
isotropic equilibrium distribution cannot reproduce the constraint that
entropy production should vanish as collisions cease; these schemes break
down for large $\eta/s$ values.

Similar comparisons have been done for the massive (0+1)-dimensional case
\cite{Florkowski:2014sfa,Florkowski:2014bba,BHM} and for the
(1+1)-dimensional Gubser flow. In all cases one finds the following
hierarchy of hydrodynamic approximations, when listed in order of
improving
accuracy in their descriptions of the moments of the exactly known
microscopic dynamics: first-order viscous hydrodynamics (Navier-Stokes
theory), second-order Israel-Stewart theory, second-order DNMR theory,
third-order viscous hydrodynamics \`a la Jaiswal, {\sc aHydro}, and {\sc
vaHydro}. In view of the increasing sophistication of these approximation
schemes, as discussed in the preceding subsections, this ordering is not
surprising, and some variant of {\sc vaHydro} is likely to become the
standard hydrodynamic modelling framework in the future. At the moment,
however, only {\sc vHydro} and {\sc aHydro} have been implemented
numerically for (2+1)-d and (3+1)-d expansion which do not rely on
simplifying assumptions such as longitudinal boost-invariance and
azimuthal
symmetry. The fireballs created in heavy-ion collisions are not
azimuthally
symmetric, and experiments tell us that they feature characteristic
anisotropic flow patterns that could never arise from an azimuthally
symmetric initial condition. Longitudinal boost-invariance is not a good
approximation either for particles emitted at large forward and backward
rapidities, and it becomes worse when going to lower energies. Therefore,
much effort is presently being expended into developing (2+1)-d and
(3+1)-d
implementations of the {\sc aHydro}  and {\sc vaHydro} schemes.

\section{Numerical Implementation of Hydrodynamics}

\subsection{Need for $\tau$ and $\eta$}
\label{sec:taueta}

The hydrodynamic simulations of the ultra-relativistic
heavy ion collisions is best implemented in the 
hyperbolic coordinate system\cite{Bjorken:1982qr}
(also known as the Milne coordinate system)
where instead of $t$ (the laboratory time) and $z$ (the beam direction),
one uses
the longitudinal proper time
\be
\tau = \sqrt{t^2 - z^2}
\ee
and the space-time rapidity
\be
\eta = \tanh^{-1}(z/t)
\ee
Equivalently, $t = \tau\cosh\eta$ and $z = \tau\sinh\eta$.
\begin{figure}[th]
    \centerline{\includegraphics[width=0.4\tw]{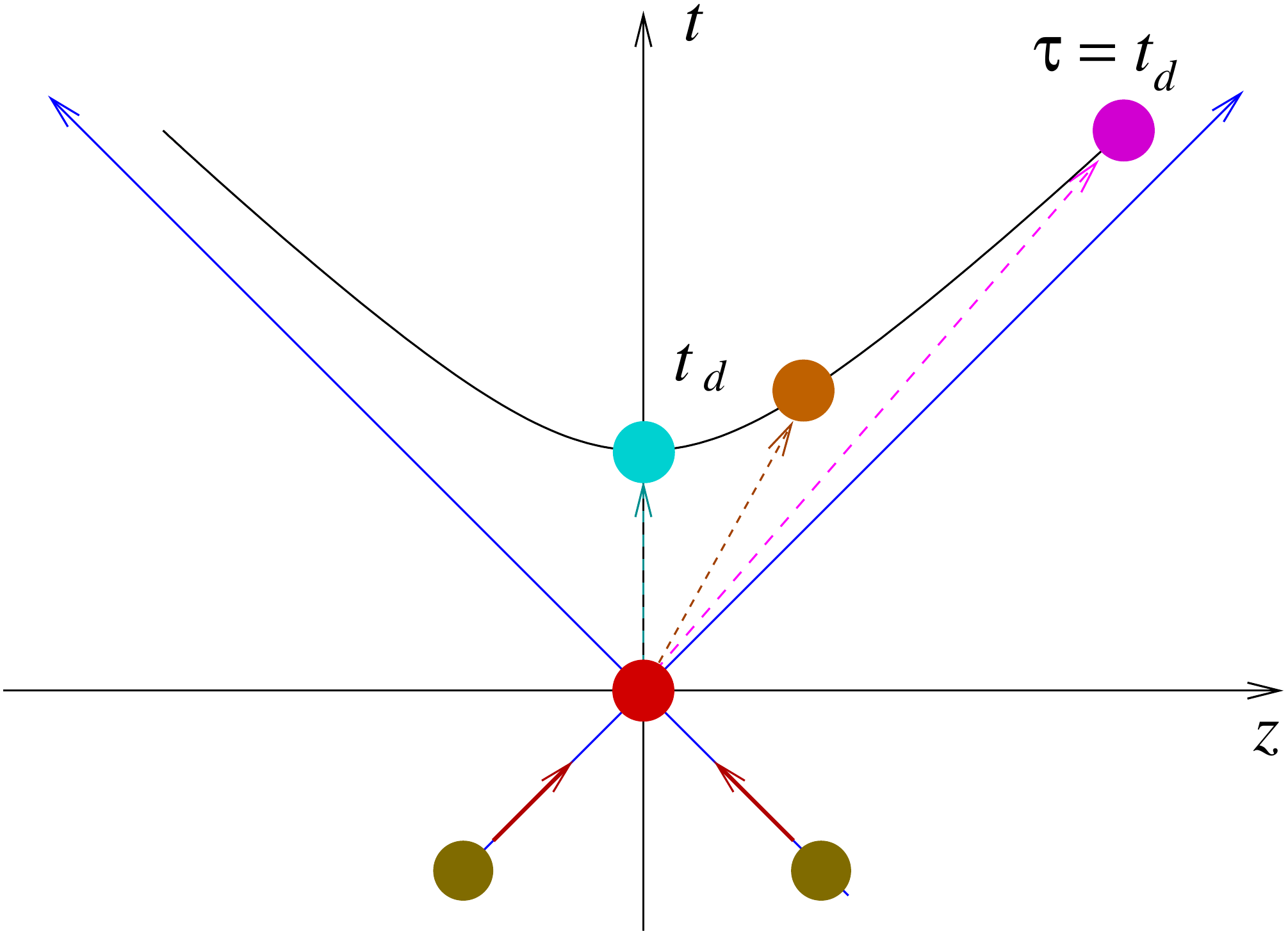}}
    \caption{A schematic diagram showing evolution of fireballs
    with differing $v_z$ after the collision of two heavy ions at 
    $t = 0$ and $z = 0$.}
    \label{fig:why_tau}
\end{figure}

One reason this is useful is that as shown in Fig.~\ref{fig:why_tau}, 
the evolution of the ultra-relativistic
heavy ion collisions can occur only in the forward light cone
bounded by the light cone axes, equivalently $\tau = 0$.
More physically,
suppose two identical systems were created at $t = 0$ and $z = 0$ by the
initial collision
of the two nuclei. Further suppose that
one of the two has the collective velocity $v$ in the $z$ direction, but
the other one is at
rest. In this case, due to the time dilation, the same stage of the
evolution will be
reached when the lab time is at $t_d$ for the system at rest and at
$t_d/\sqrt{1 - v^2}$ for
the moving system. In relativistic systems, these two Minkowski
times can be very much different even
though the two system are at the same stage in their respective evolution.
However, in terms of the proper time both are at the same $\tau = t_d$
since $\tau$ is nothing but the local rest frame time.
Hence, it is very natural that we should use the $\tau-\eta$ coordinate
system when there is a strong longitudinal flow of matter. 
In the ultra-relativistic heavy ion collisions, this is the case
due to the original beam momenta of the projectile and the target.

For numerical implementation of hydrodynamics,
one first needs to
formulate the conservation laws in this coordinate system.
For this,
we need to know the transformation law between $\tau-\eta$ and $t-z$.
We can start with the derivatives
 \be
 \partial_\tau
 & = &
 {\partial t\over \partial\tau}\partial_t 
 +
 {\partial z\over \partial\tau}\partial_z 
 \non
 & = &
 \cosh\eta \partial_t
 +
 \sinh\eta \partial_z
 \ee
and 
 \be
 \partial_{\eta}
 & = &
 {\partial t\over \partial\eta}\partial_t 
 +
 {\partial z\over \partial\eta}\partial_z 
 \non
 & = &
 \tau\sinh\eta \partial_t
 +
 \tau\cosh\eta \partial_z
 \ee
 which can be summarized as Lorentz transformations
 \be
 \tilde{\partial}_a = \Lambda_a^\mu \partial_\mu
 \ \ \hbox{and}\ \ 
 \partial_\mu
 = \Lambda^a_\mu 
 \tilde{\partial}_a 
 \label{eq:L_transform}
 \ee
 where $\tilde{\partial}_a = (\partial_\tau, \nabla_\perp,
(1/\tau)\partial_\eta)$
 and
 \be
 \Lambda_\mu^a
 =
 \bmat{cccc}
 \cosh\eta & 0 & 0 & -\sinh\eta\\
 0 & 1 & 0 & 0\\
 0 & 0 & 1 & 0\\
 -\sinh\eta & 0 & 0 & \cosh\eta
 \emat
 \label{eq:Lambda_mua}
 \ee
 The inverse transform $\Lambda^\mu_a$ is obtained by substituting $\eta$
with $-\eta$.
 From now on, we will use the first letters of the roman alphabet ($a, b,
\cdots$)
 to represent the components in the Milne space as above.
 
 To figure out the transformation of the conservation law, we apply
 Eq.(\ref{eq:L_transform}) to
 \be
 \partial_t j^0 + \partial_z j^3
 & = &
 \cosh\eta \partial_\tau j^0 - {\sinh\eta\over \tau}\partial_{\eta} j^0
 +
 {\cosh\eta\over\tau} \partial_{\eta} j^3 - {\sinh\eta}\partial_{\tau} j^3
 \non
 & = &
 {1\over \tau}\partial_\tau(\tau \cosh\eta j^0 - \tau \sinh\eta j^3)
 -
 \partial_\eta( {1\over \tau}\sinh\eta j^0 - {1\over \tau}\cosh\eta j^3)
 \non
 \ee
 Defining
 \be
 q^\tau &=& \cosh\eta\, q^0 - \sinh\eta\, q^3
 \non
 \tildeq^\eta &=& \cosh\eta\, q^3 - \sinh\eta\, q^0
 \ee
 and 
 \be
 \tildeq^a = (q^\tau, \bfq_\perp, \tildeq^\eta),
 \ee
 for any 4-vector $q^\mu$, the conservation law becomes
 \be
 0 & = &
 \partial_\tau (\tau j^\tau)
 + \nabla_\perp{\cdot}(\tau\bfj_\perp)
 + \partial_\eta \tildej^\eta
 = 
 \tilde\partial_a (\tau \tildej^a)
 \label{eq:consJ}
 \ee
 where $\bfj_\perp$ only has the $x,y$ components.
 One can also show
 \be
 D_\tau
 & = &
 u^\mu \partial_\mu
 = \tildeu^a \tilde\partial_a
 \ee
 Note that our definition of the $\eta$ component is different from
 the curvilinear definition of the $\eta$ component by a
 factor of $\tau$.  We do this to keep the dimension of the $\eta$ component
 the same as the other components.
If one uses the curvilinear definition
 \be
 j^\eta = {1\over \tau}\tildej^\eta
 \ee
 then the conservation law becomes
 \be
 0 & = &
 \partial_\tau (\tau j^\tau)
 + \nabla_\perp{\cdot}(\tau\bfj_\perp)
 + \partial_\eta (\tau j^\eta)
 = \partial_a (\tau j^a)
 \ee
 where $\partial_a = (\partial_\tau, \nabla_\perp, \partial_\eta)$.

 For the energy momentum conservation, one needs to apply the
transformation law 3 times for
 the each index in $\partial_\mu T^{\mu\nu} = 0$.
 The algebra is a bit tedious but
straightforward\cite{Kolb:2000sd,Schenke:2010nt}.
 The result is
\be
\partial_\tau \calT^{\tau\tau}
+ {1\over \tau}\partial_\eta \calT^{\eta\tau} + \partial_v \calT^{v\tau}
=
- {1\over \tau} \calT^{\eta\eta} 
\label{eq:calTtautau}
\ee
where we defined
\be
\calT^{ab} = \tau \Lambda^a_\mu \Lambda^b_\nu\, T^{\mu\nu}
\ee
and the index $v = x, y$.
For the $\tau\eta$ component,
\be
\partial_\tau \calT^{\tau\eta} 
+ {1\over \tau} \partial_\eta \calT^{\eta\eta}
+ \partial_v \calT^{v\eta}
= 
- {1\over \tau} \calT^{\eta\tau}
\label{eq:calTtaueta}
\ee
Both $\calT^{\tau\tau}$ and $\calT^{\tau\eta}$ conservation laws contain
the geometrical source term.
Transverse momentum conservation is simpler:
\be
\partial_\tau \calT^{\tau v}
+ 
{1\over \tau}\partial_\eta \calT^{\eta v}
+
\partial_w \calT^{w v}
& = &
0
\ee
where $v$ and $w$ are transverse coordinate indices.
When testing one's code for the energy-momentum conservation,
the following form may be more convenient
\be
{d\over d\tau} \int dx dy d\eta\, \tildeT^{\tau\mu} = 0
\ee
The half-transformed
$\tildeT^{a\mu} = \tau \Lambda^a_\nu T^{\nu\mu}$ satisfies the conservation
law
$ \partial_\tau \tildeT^{\tau \mu}
+ \tilde\partial_\eta \tildeT^{\eta \mu} + \partial_v \tildeT^{v
\mu} = 0$
without the geometrical source terms exactly like Eq.(\ref{eq:consJ}).

For the shear and the bulk evolution equations 
in Eq.(\ref{eq:DNMR}) or (\ref{eq:IS}),
transformation from the Minkowski space to the $\tau-\eta$ coordinate
space is
straightforward, but it is a lot more involved algebraically. For more
details
see Refs.\cite{Muronga:2003ta,Heinz:2005bw,Schenke:2010rr}.

\subsection{Numerical solution of conservation equations}
\label{sec:numerical}

In this section, we will first discuss conservation laws in the Minkowski
coordinates where the conservation laws are
\be
\partial_\mu T^{\mu\nu} = 0,\ \ \partial_\mu J^\mu_B = 0
\label{eq:vhydro_eq}
\ee
In the Milne coordinate system, the energy-momentum conservation
takes a slightly different form
\be
\tilde\partial_a \calT^{ab} = \calS^b,
\ee
where $\calS^b$ is the geometric source term in 
Eqs.(\ref{eq:calTtautau}) and (\ref{eq:calTtaueta}).
However, as will be shown shortly, the methods illustrated 
below can be easily adapted to this case.

For {\sc vHydro}, the energy-momentum tensor is given in a general
reference frame by the decomposition
\be
\label{eq:T}
T^{\mu\nu} = T_{\rm id}^{\mu\nu}
+ \pi^{\mu\nu} - \Pi \Delta^{\mu\nu}
\ee
where 
\be
T_{\rm id}^{\mu\nu} = \varepsilon u^\mu u^\nu
- P(\varepsilon, \rho_B) \Delta^{\mu\nu}
\label{eq:Tmunu_id}
\ee
is the ideal fluid part of the tensor. 
The net baryon current has the form 
$J^\mu_B = \rho_B u^\mu + V_B^\mu$.
The equations that need to be solved are given in Eq.(\ref{eq:vhydro_eq}), 
together with the relaxation equations for the dissipative flows, 
for example the Israel-Stewart equations (\ref{eq:IS}). 
The first step in solving the hydrodynamic equations is their
initialization. 
Let us assume that some microscopic pre-equilibrium dynamical theory
provides us with 
a baryon current $J^\mu_B(x)$ and energy-momentum tensor $T^{\mu\nu}(x)$ 
for points $x^\mu$ on some Cauchy surface on which we want to initialize
the hydrodynamic 
evolution stage. The following projection steps, to be taken at each point
$x$ 
on that surface, yield the required initial value fields:%
\footnote{%
The following paragraph refers to the {\sc vHydro} decomposition 
(\ref{eq:JB1},\ref{eq:T}). A slightly modified projection method
applies for the {\sc aHydro} and {\sc vaHydro} decompositions
(\ref{eq9}), (\ref{eq10}) and (\ref{eq15}) \cite{Bazow:2013ifa}.
}

First we define the local fluid rest frame by solving the eigenvalue
equation $T^{\mu}_{\ \nu} u^\nu = \varepsilon u^\mu$
for its normalized timelike eigenvector $u^\mu$. The
associated eigenvalue gives us the LRF energy density $\varepsilon$. The
LRF baryon density is obtained from $J^\mu_B$ by projecting onto $u_\mu$:
$\rho_B\EQ{u}_\mu J_B^\mu$. The initial heat flow $V_B^\mu$ is the component
of $J_B^\mu$ perpendicular to $u^\mu$:
$V^\mu\EQ\Delta^{\mu\nu}J_{B,\nu}$. Now that we know the LRF energy and
baryon densities we can compute the LRF pressure $P$ from the
equation of state of the fluid, $P(\varepsilon,\rho_B)$. Next,
the bulk viscous pressure is obtained from
$\Pi\EQ-\frac{1}{3}\Delta_{\mu\nu}T^{\mu\nu}-P$. Finally, the
shear stress is obtained as $\pi^{\mu\nu}\EQ{T}^{\mu\nu}-\varepsilon u^\mu
u^\nu + (P{+}\Pi)\Delta^{\mu\nu}$ or, equivalently, as
$\pi^{\mu\nu}\EQ{T}^{\langle\mu\nu\rangle}{\,\equiv\,}\Delta^{\mu\nu}_{\alpha\beta} T^{\alpha\beta}$.

For a simple illustration of numerical methods that can solve the
hydrodynamic equations,
let us first consider a single conservation law in 1-D.
There is no difficult conceptual obstacle in extending this case to the
multi-component, multi-dimension cases such as the Israel-Stewart
equations.  The conservation equation is
\be
\partial_t u = -\partial_x j
\ee
We need to supplement this equation with a relationship between 
the density $u$ and the current $j$.
Simplest example is $j = vu$ with a constant speed of propagation $v$. 
But in general $j$ is a non-linear function(al) of $u$.
For instance, the ideal part, Eq.(\ref{eq:Tmunu_id}), is certainly not in
this simple
form due to the normalization condition $u_0^2 = \sqrt{1 - \bfu^2}$ and
also to the presence
of the pressure term.
In the dissipative cases, the relaxation equation 
\be
(\partial_t + 1/\tau_R)j = -(D/\tau_R)\partial_x u
\ee
determines the relationship between $j$ and $u$.
In such cases, the numerical methods discussed in this section 
needs to be applied in two steps. In the first step, the conservation laws
are used to advance the time component of the currents using the methods
that
will be discussed here. In the second step,
the spatial part of the currents needs to be reconstructed from the time
components.
The relaxation equations also need to be solved separately, although the
techniques discussed
here can be easily adapted to handle the relaxation equation as well.

For the simple $j=vu$ case with a constant $v$, the equation becomes
\be
\partial_t u = -v\partial_x u
\ee
This is an advection equation
and has a simple solution
\be
u(t,x) = f(x-vt)
\label{eq:sol_advec}
\ee
That is, at any given time, the solution is just
the translation of the initial profile by $vt$. 
Analytically, this is trivial. However, it is remarkable how difficult
it can be to maintain the initial profile in numerical solutions.
We will often use this as the simplest test case for our algorithms.

To solve the conservation equation numerically, one first needs to
discretize
time and space. We define
\be
u^{n}_{i} = u(t_n, x_i)
\ee
where $t_n = t_0 + n h$ and $x_i = x_0 + ia$
for any function $u(t,x)$ of time and space. Here $h$ is the time step
size
and $a$ is the spatial cell size.
Physically, it is important to have 
the following properties in the discretization method.
First, the method should conserve total $u$ explicitly, that is,
$\sum_i u^n_i = \sum_i u^{n+1}_i$ modulo the boundary terms.
For this, one requires that the discretized form of the divergence to take
the
form
\be
(\partial_x j)^n_i \to {j_i(u^n) - j_{i-1}(u^n)\over a} 
\label{eq:discrete_j}
\ee
where $j_i(u^n)$ is the discretized representation of the current $j$ at
$x_i$ ant $t_n$.
One can easily see that in the sum $\sum_{i=0}^N (\partial_x j)_i^n$  only
the boundary
terms would survive. The details of the boundary terms depend on the
method. However, as
long as the boundaries are far away from the physical region, the boundary
terms should be
vanishingly small. If the boundary of the space is not too far away from
the physical
region, then some suitable discrete boundary conditions should be imposed.

The second requirement is simple, yet quite demanding: If $u^0_{i} \ge 0$ for all $i$,
then we would like this property to be
maintained for any future time. For instance, if $u$ represents the energy
density,
then it should never become negative. 

To illustrate some of these issues, consider again the advection equation
$\partial_t u = -v\partial_x u$ with $v > 0$. Let the initial condition be
a rectangle: $u^0_i = u_c$ for $b \le i \le f$ and $u_i^0 = 0$ otherwise.
This is a prototype of many situations where two smoothly varying regions
are joined by a
stiff gradient.
The simplest discretization method of $\partial_t u = -v\partial_x u$
is the forward-time centered-space (FTCS) method
\be
u_i^{n+1} =  u_i^{n} -{vh\over 2a}(u^{n}_{i+1} - u^{n}_{i-1})
\label{eq:ftcs1}
\ee
which is correct up to $O(h^2)$ and $O(a^2)$ errors.
Since the second term in the right hand side is supposed to be
a correction, we require $ |vh/a| < 1$.
This is certainly in the form of Eq.(\ref{eq:discrete_j}) and hence conserves
the total $u$.

According to the analytic solution Eq.(\ref{eq:sol_advec}), 
the space behind the back edge ($x_b$) of the rectangle should always have
$u^n = 0$.
However, according to Eq.(\ref{eq:ftcs1}) the cell right behind the back
edge
becomes non-zero and
negative after the first time step
\be
u_{b-1}^1 = -{vh\over 2a}u_c
\ee
At the same time, the front edge starts to get distorted by the same
amount
\be
u_{f}^{1} = u_c - u_{b_1}^1 > u_c
\ee
At the next time step, 
$u_{b-1}^2$ becomes even more negative
\be
u_{b-1}^{2} 
= 
-{vh\over a}u_c + \left(vh\over 2a\right)^2 u_c
\ee
while $u_f$ deviates even more from $u_c$
\be
u_f^{2} = u_c - u_{b-1}^2 
\ee
Furthermore at $t_2$,
the next cell in the empty region becomes non-zero
$ u_{b-2}^2 =  \left({vh\over 2a}\right)^2 u_c $.
Clearly, both the profile preservation and the positivity of the
solution are grossly violated by this method even though 
the total $u$ is conserved.  In addition,
one can easily see that the growth at $x_{b-1}$ will continue,
indicating that this method is unconditionally unstable.

In order to cure the negativity problem,
one may note that the trouble above mainly
comes from the centered nature of the
$O(a^2)$ numerical derivative in Eq.(\ref{eq:ftcs1}).
Instead, one may try the first order approximation\footnote{
This is in fact trivially exact when $a = vh$ since in that
case $u_i^{n+1} = u_{i-1}^n$. That is, the whole profile moves by
one spatial cell at each time step. However, 
this wouldn't work for more general currents.}
\be
u_i^{n+1} =  u_i^{n} -{vh\over a}(u^{n}_{i} - u^{n}_{i-1})
\label{eq:upwind}
\ee
In this ``up-wind scheme'', 
$u_{b-1}^1$ trivially vanishes at $t_1$.
In fact, all $u_{b-i}$ for $i \ge 1$ will remain zero for all times.
Similarly, one can show that the numerical solution is positive and
bounded everywhere as long as $|vh/a| < 1$.
If $v<0$, mirror-image conclusions can be reached
if one uses $(\partial_x u)_i \approx (u_{i+1}-u_i)/a$.

Positivity is thus maintained in this up-wind scheme.
However, as the system evolves in time, the shape
of the solution gets more and more distorted.
This is because the first order difference
\be
(u_i^n - u^n_{i-1})/a
= \partial_x u(t_n, x_i) - {a\over 2}\partial_x^2 u(t_n,x_i)
+ O(a^2)
\label{eq:approx_ux}
\ee
is too crude an approximation of the first order derivative. The second
derivative term
in Eq.(\ref{eq:approx_ux}) in fact
introduces too much artificial (numerical) damping to preserve the shape
for long.
In effect, the profile at $t$ is the convolution of the initial profile and
the 
Gaussian Green function of the diffusion equation (Eq.(\ref{eq:G_diff}))
with the
diffusion constant given by $D = {a\over 2h}$. As one can see in
Eq.(\ref{eq:G_diff}), the
width of the Gaussian grows linearly with time. Hence, the initial profile
will be smeared
out quickly.

\begin{figure}[th]
\centerline{
\includegraphics[width=0.7\tw]{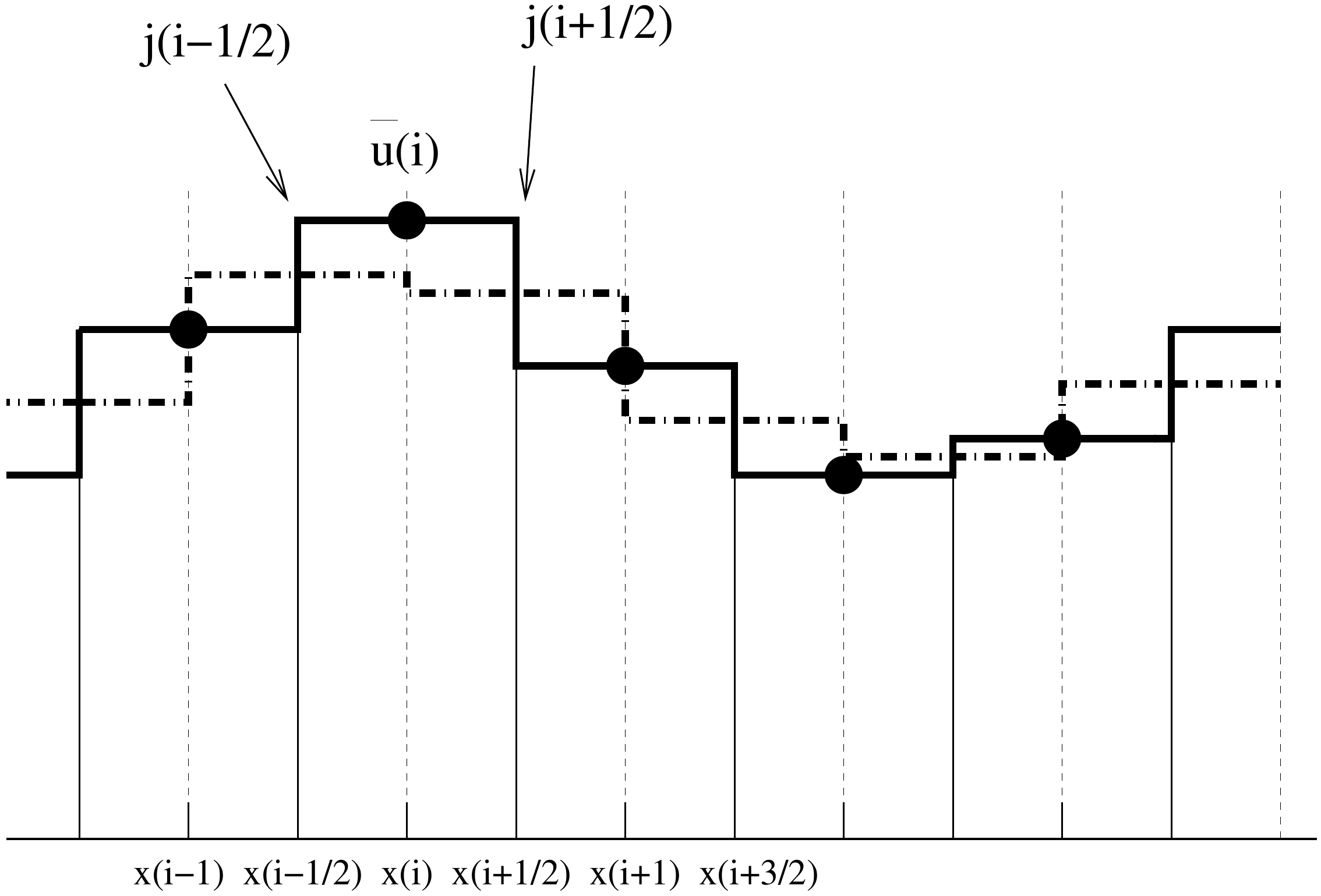}
}
\caption{
Spatial grid used in deriving finite volume methods.
The solid line is 
the lowest order histogram function approximation of $u(t,x)$ 
in the $x_i$-centered grid.
The dot-dash line is the projection of the $x_i$-centered grid onto
the $x_{i+1/2}$-centered grid.
}
\label{fig:histo}
\end{figure}
Better discretization methods must keep the positivity
preserving
nature of the up-wind method and at the same time
must have a better approximation for the 
spatial derivative than the simple first order difference for shape
preservation.
To devise better discretization methods more systematically,
consider first dividing the space into $N+1$ cells of size $a$
labelled by integers $0$ through $N$.
The $i$-th cell starts at  
$x_{i-1/2} = x_i-a/2$ and ends at $x_{i+1/2} = x_i+a/2$. 
(See Fig.~\ref{fig:histo}.) 
Averaging over one spatial cell and integrating over one time step,
the conservation law $\partial_t u = -\partial_x j$ becomes
\be
\baru_{i}^{n+1}
&=&
\baru_{i}^n
-{1\over a}\int_{t_n}^{t_{n+1}}dt\, \Big( j(t,x_{i+1/2}) - j(t,x_{i-1/2})
\Big)
\label{eq:exact1}
\ee
where $\baru_{i}^n = \baru_i(t_n)$ is the cell average.
This is an exact expression which can be used
as the basis for further approximation. This type of methods are known as
the finite volume methods.

Approximating the time integral using the midpoint rule, we get
\be
\baru_{i}^{n+1}
&=&
\baru_{i}^n
-{h\over a} \Big( j(t_{n+1/2},x_{i+1/2}) - j(t_{n+1/2},x_{i-1/2}) \Big)
+
O(h^3)
\label{eq:mid1}
\ee
where $t_{n+1/2} = t_n+h/2$. 
There are few things that should be mentioned here.
First, the basic quantities to calculate are the cell-averaged values 
$\baru_i^n$.
Second, we need to approximate the function $u(t,x)$ itself from
$\baru_i^n$
because the right
hand side contains $u(t,x)$ evaluated at $t_{n+1/2}$ and $x_{i\pm 1/2}$.
An approximate form of $u(t_n,x)$ is also
needed for obtaining $\baru_i^{n+1}$ for the next time step (see below).
Therefore, how we approximate $u(t,x)$ and evaluate $j^{n+1}_{i\pm 1/2}$
determine different numerical schemes and the accuracy of the given scheme.

\begin{figure}[th]
\bcent
 \includegraphics[width=0.8\tw]{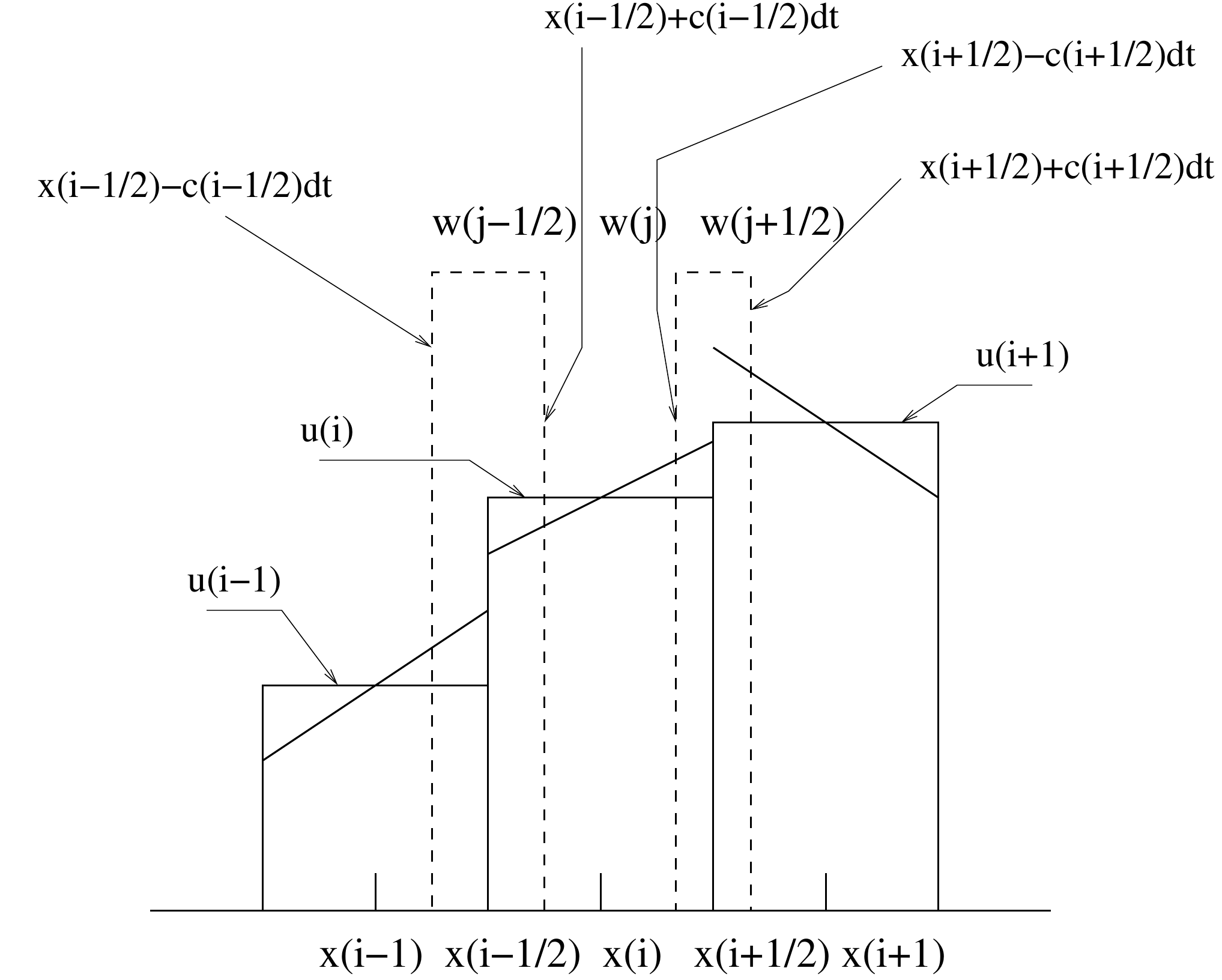}
\ecent
\caption{A schematic view of the cell division
used in the Kurganov-Tadmor scheme.}
\label{fig:KTScheme}
\end{figure}

In many schemes, the values at spatial half points are not unique because
the interpolating function approximating $u(t,x)$ is usually only
piece-wise
continuous.
For instance, see Fig.~\ref{fig:histo} and Fig.~\ref{fig:KTScheme}.
One way to deal with the ambiguity in evaluating $j(t_{n+1/2}, x_{i\pm
1/2})$ 
is to just avoid evaluating it at the boundaries. The staggered nature
of the space and time indices in Eq.(\ref{eq:mid1}) suggests an obvious way
to do so.
Suppose that the initial data is given for $\baru_{i}^n$ where the $i$-th
grid 
is centered at $x_i$ and has the spatial interval $[x_{i-1/2},
x_{i+1/2}]$.
For the next time step, instead of updating the values of $\baru_i$ within
the intervals $[x_{i-1/2}, x_{i+1/2}]$, we update the values of
$\baru_{i+1/2}$
within the shifted intervals $[x_i, x_{i+1}]$. 
Using the lowest order approximation for $u(t,x)$
(the histogram functions in Fig.~\ref{fig:histo}),
this yields
\be
\baru_{i+1/2}^{n+1} 
= 
{\baru_{i+1}^n + \baru_{i}^n\over 2}
- {h\over a}\left( j(u^{n+1/2}_{i+1}) - j(u^{n+1/2}_{i}) \right)
\label{eq:NT1}
\ee
where ${(\baru_{i}^n+\baru_{i+1}^n)/2}$ is the 
average value of $u(t_n,x)$ in the shifted interval $[x_{i}, x_{i+1}]$
using the histogram function in Fig.~\ref{fig:histo}.
For the values of $u_{i+1}^{n+1/2}$ at $t_{n+1/2}$,
the forward Euler method 
\be
u^{n+1/2}_{i} = 
\baru^{n}_{i} - {h\over 2} (\partial_x j)^{n}_i + O(h^2)
\label{eq:half_step}
\ee
is enough since the overall error in the midpoint rule is $O(h^3)$.
In this way, we avoid evaluating the current at the cell boundaries.
For the next time step, we shift back to the original grid.
This staggered method 
is a slightly generalized form of the Lax-Friedrichs scheme
which is second order accurate in space and third order accurate in time.
The original Lax-Friedrichs scheme is only second order in time
because the currents are evaluated at $t_n$ instead of at $t_{n+1/2}$.

Eq.(\ref{eq:half_step}) involves evaluating numerical derivative of the
current 
$j$ at $x_i$.  This cannot be exactly determined
when all one has is data on the average values $\baru_i^n$.
For instance, $(j_x)^n_i$ could be the first order approximations 
$(j^n_{i+1} - j^n_{i})/a$ or $(j^n_{i} - j^n_{i-1})/a$,
or the second order approximation
$(j^n_{i+1} - j^n_{i-1})/2a$, or any other approximate form.
In normal situations, choosing a higher order formula
should be better than the first order ones.
But this is not always the case when the gradient is stiff.

We have already shown that 
using the central difference $(j^n_{i+1} - j^n_{i-1})/2a$
in the forward Euler method (the 
forward-time-centered-space method in Eq.(\ref{eq:ftcs1})) can be
disastrous when the gradient is stiff, although it can be safely
(and preferably) used in smooth regions.
When gradient is stiff, one should instead use the up-wind method
Eq.(\ref{eq:upwind})
to maintain the positivity.
Therefore, an intelligent scheme would choose the derivative according to
some approximate measure of the true gradient.
One way to do this is to choose the gradient according to the following
scheme
\be
(\partial_x u)_{i}^n
=
\left\{
\begin{array}{l}
0 \hbox{ if $\baru^n_i < \baru^n_{i\pm 1}$ or $\baru^n_i > \baru^n_{i\pm
1}$}
\\
\hbox{else }
{\rm sign}(\baru^n_{i+1} - \baru^n_i)\,
{\rm min}(\theta  {\left| \baru^n_{i+1}-\baru^n_i\right|\over a},
{\left| \baru^n_{i+1}-\baru^n_{i-1} \right|\over 2},
\theta {\left| \baru^n_i - \baru^n_{i-1} \right| \over a})
\end{array}
\right.
\ee
The first line indicates that the function has either a maximum or a
minimum within the interval.
Therefore the slope at $x_i$ can be best approximated by 0.
The second line applies when the function is changing monotonically near
$x_i$.
The parameter $1 \le \theta < 2$ is there to be slightly more general.
This choice of the derivative is called the ``generalized minmod flux
limiter''.

The Lax-Friedrichs scheme represented by Eq.(\ref{eq:NT1})
is only $O(a^2)$ accurate because 
we have used the histogram function as an approximation for $u(t_n, x)$.
Needless to say, this is the lowest order approximation.
As a result, the Lax-Friedrichs scheme
contains too much numerical diffusion to be practically useful.
Both of these facts can be easily seen from the Taylor expansion
\be
{\baru_{i}^n+\baru_{i+1}^n\over 2} =
\baru_{i+1/2}^n + {a^2\over 8} \partial_x^2 u_{i+1/2}^n + O(a^4)
\label{eq:Taylor}
\ee
where the second derivative term is the $O(a^2)$ error term that also
causes strong diffusion. 
In time, this diffusion distorts the solution too much
just as in the first order up-wind method.

To obtain a better approximation,
one needs to evaluate $u(t,x)$ more accurately using the average values
from 
the nearest neighbor cells.
If one uses the linear interpolant (see Fig.~\ref{fig:KTScheme})
\be
\hatu^n_i(x) = \baru^n_i + (\partial_x u)^n_i (x-x_i)
\ee
for each interval $[x_{i-1/2}, x_{i+1/2}]$, then the scheme should become
at least $O(a^3)$. 
Using this to improve the estimate of $\baru_{i+1/2}^n$ adds a correction
term 
to the lowest order result Eq.(\ref{eq:NT1})
\be
\baru_{i+1/2}^{n+1} 
& = &
{\baru_{i}^n+\baru_{i+1}^n\over 2}
-{h\over a}
\Big( j(t_{n+1/2},x_{i+1}) - j(t_{n+1/2},x_{i}) \Big)
\non & & {}
- {a\over 8}\left((\partial_x u)^n_{i+1} - (\partial_x u)^n_{i}\right)
\label{eq:NT2}
\ee
This staggered approach is the second order NT (Nessyahu-Tadmor) 
scheme and works reasonably well \cite{NT90}. 
The last term 
in the right hand side of Eq.(\ref{eq:NT2}) 
cancels the second derivative term in Eq.(\ref{eq:Taylor})
so that together they represent $\baru_{i+1/2}^n$ with the $O(a^4)$ error.
In other words, the last term in Eq.(\ref{eq:NT2}) is the anti-diffusion
term
that is correcting the large second order diffusion
introduced by the symmetric combination 
$ (\baru_{i}^n+\baru_{i+1}^n)/2 $.
Therefore, this scheme is $O(a^4)$ in the smooth region.

Why don't we then just replace these terms with $\baru_{i+1/2}^{n}$? 
If one does that, it just becomes the forward-time-centered-space (FTCS)
scheme
in Eq.(\ref{eq:ftcs1}). The difference, the $O(a^4 \partial_x^4 u(x))$ term, 
provides just enough numerical diffusion
so that spurious oscillations do not propagate
from the stiff gradient.

The second order NT algorithm represented by Eq.(\ref{eq:NT2}) 
is related to another often-used finite element 
method called the SHASTA algorithm.
We start with Eq.(\ref{eq:NT2}) but replace
\be
{\baru_i^n + \baru_{i+1}^b\over 2}
& \approx &
\baru_{i+1/2}^n + {1\over 8}(u_{i+1} - 2u_{i+1/2} + u_{i})
+ O(a^4)
\ee
so that
\be
\baru_{i+1/2}^{n+1} 
& = &
\baru_{i+1/2}^n 
-{h\over a} \Big( j(t_{n+1/2},x_{i+1}) - j(t_{n+1/2},x_{i}) \Big)
\non & & {}
+ {1\over 8}(\baru^n_{i+1} - 2\baru^n_{i+1/2} + \baru^n_{i})
- {a\over 8}\left((\partial_x u)^n_{i+1} - (\partial_x u)^n_{i}\right)
\label{eq:SHASTA1}
\ee
At this point, this is no longer a finite volume method even though we will
keep using
the notation $\baru_i^n$ to represent the value of $u$ at $t_n$ and $x_i$.
It is also not a staggered method any more as the right hand side contains
$\baru_{i+1/2}^n$.
Re-labelling the spaces indices $i+1/2 \to i$ and $i\to i-1$
so that the new grid size is $a' = a/2$, we get
\be
\baru_{i}^{n+1} 
& = &
\baru_{i}^n
-{h\over 2a}
\Big( j(t_{n+1/2},x_{i+1}) - j(t_{n+1/2},x_{i-1}) \Big)
\non & & {}
+ {1\over 8}(\baru^n_{i+1} - 2\baru^n_{i} + \baru^n_{i-1})
- {a\over 16}\left((\partial_x u)^n_{i+1} - (\partial_x u)^n_{i-1}\right)
\label{eq:SHASTA2}
\ee
with the appropriate scaling of the derivatives in the last term
and after renaming $a' \to a$.
With $u(t_{n+1/2}, x_{i\pm 1})$ given by Eq.(\ref{eq:half_step}),
Eq.(\ref{eq:SHASTA2}) represents the basic SHASTA algorithm.

In practice,
Eq.(\ref{eq:SHASTA2}) is broken up into two stages to ensure positivity.
Specifying $j = vu$, the first transport step is\footnote{
In the original SHASTA algorithm,  
$(\baru^n_{i+1} - 2\baru^n_{i} + \baru^n_{i-1})$
is used in place of $\left((\partial_x u)^n_{i+1} 
- (\partial_x u)^n_{i-1}\right)/2$ in this stage.
In practice, as long as $|vh/a|$ is small, this difference does not matter
much.
But it is crucial that a flux limiter is used in the second anti-diffusion
step.
}
\be
w_{i}^{n+1} 
& = &
\baru_{i}^n 
+ {1\over 8}(\baru^n_{i+1} - 2\baru^n_{i} + \baru^n_{i-1})
\non & & {}
-{vh\over 2a} \Big( 
\baru^{n}_{i+1} - {vh\over 2}(\partial_x u)^{n}_{i+1}
-
\baru^{n}_{i-1} + {vh\over 2}(\partial_x u)^{n}_{i-1}
\Big)
\ee
again using Eq.(\ref{eq:half_step}).
If $\baru_l^{n} \ge 0$ for all $l$, then
as long as $|vh/a| < 1/4$, 
$w_i^{n+1}$ is positive.

The second stage is the anti-diffusion step
\be
\baru_{i}^{n+1} 
=
w_{i}^{n+1} 
- {a^2\over 8}
(\partial_x^2 w)^{n+1}_{i}
\ee
where 
$ (\partial_x^2 w)^{n+1}_{i} $
represents numerical estimate of the second derivative at $x_i$
that preserves the positivity. The numerical approximation suggested
by the last term in Eq.(\ref{eq:SHASTA2})
turned out not to preserve the positivity.
The original formulation of the SHASTA algorithm by Boris and Book uses
a conservative form
\be
(\partial_x^2 w)^{n+1}_{i}
= 
{1\over a}
\left(
(w_x)^{n+1}_{i+1} - (w_x)^{n+1}_{i}  
\right)
\ee
where
\be
(w_x)^{n+1}_{i+1} 
=
\left\{
\begin{array}{l}
0 \hbox{  if $\Delta^{n+1}_{i}$, $\Delta^{n+1}_{i+1}$ and
$\Delta^{n+1}_{i+2}$
do not all have the same sign.}
\\
\hbox{else }
{\rm sign}(\Delta^{n+1}_{i+1})\,
{\rm min}( 
{8|\Delta^{n+1}_{i}|},
{|\Delta^{n+1}_{i+1}|},
{8|\Delta^{n+1}_{i+2}|})
\end{array}
\right.
\ee
with
\be
\Delta^{n+1}_{i+1} = (w_{i+1}^{n+1} - w_i^{n+1})/a
\ee
This is similar to the minmod flux limiter and maintains the positivity.
The SHASTA algorithm is used in many hydrodynamics simulation programs
for ultra-relativistic heavy ion collisions\cite{Schneider:1993gd}
including  pioneering works in 
Refs.\cite{Sollfrank:1996hd,Kolb:2000sd}
and also later works in 
Refs.\cite{Heinz:2005bw,Petersen:2010cw,Karpenko:2010te,Holopainen:2010gz,
Pang:2012he,Roy:2011pk}.

The second order NT scheme and the SHASTA scheme
in practice work fairly well\cite{Akamatsu:2013wyk}.
However, in these schemes the $h\to 0$ limit
cannot be taken
since the numerical viscosity behaves like $\sim 1/h$. 
It would be convenient to be able to take this limit because one can 
then formulate the discretized problem as a
system of coupled ordinary differential equations in time. 
Many techniques for the ordinary differential equations  such as the
Runge-Kutta
methods then become available to control the accuracy of the time
evolution.
So far different time integration techniques were not available other than
the midpoint
rule in Eq.(\ref{eq:mid1}).

One way to achieve this is to subdivide the cells as shown in
Fig.~\ref{fig:KTScheme} with the piece-wise linear approximation
for $u(t,x)$.
The size of the cell containing the discontinuity at the half integer point
$x_{i+1/2}$
is controlled by the local propagation speed $c_{i+1/2}$. That is, the
subcell
surrounding $x_{i+1/2}$ is between the left boundary
$x_{i+1/2}^l = x_{i+1/2} - c_{i+1/2}h$ and the right boundary
$x_{i+1/2}^r = x_{i+1/2} + c_{i+1/2}h$.
Then the cells containing the boundaries and the cells not containing the
boundaries
(between $x_{i-1/2}^r$ to $x_{i+1/2}^l$)
are independently evolved.
For the subcell containing $x_{i+1/2}$
\be
w_{i+1/2}^{n+1}
=
{\baru_{i+1/2}^l + \baru_{i+1/2}^r\over 2} 
- {1\over 2c_{i+1/2}}
\left(j(\baru_{i+1/2}^r)-j(\baru_{i+1/2}^l)\right)
+ O(h)
\ee
which is basically Eq.(\ref{eq:NT1}).
Here the superscripts
$r$ and $l$ means the value of the approximate $u(t_n,x)$
at the boundary points $x_{i+1/2}^{r}$ and $x_{i+1/2}^l$, respectively.
Within the smooth region between $x_{i-1/2}^r$
and $x_{i+1/2}^l$, we get using Eq.(\ref{eq:mid1})
\be
w_{i}^{n+1}
=
{\baru^l_{i+1/2} + \baru^r_{i-1/2}\over 2}
- {h\over a}
\left(j(\baru_{i+1/2}^l)-j(\baru_{i-1/2}^r)\right)
+ O(h^2)
\ee
where the first term in the right hand side is obtained by applying
the trapezoid rule.
The divided cells are then projected onto the original grid using the size
of the cells as the weight to get
\be
\baru_{i}^{n+1}
& = &
{c_{i-1/2}h\over a} w_{i-1/2}^{n+1} 
+
{c_{i+1/2}h\over a} w_{i+1/2}^{n+1} 
+
\left( 1 - {(c_{i-1/2}+c_{i+1/2})h\over a} \right) w_i^{n+1}
\non
\ee
When the $h\to 0$ limit is taken, this procedure yields 
\be
{d\baru_i\over dt}
& = &
-{H_{i+1/2} - H_{i-1/2}\over a}
\label{eq:KT_generic}
\ee
where
\be
H_{i+1/2} = {j(\baru_{i+1/2}^+)+j(\baru_{i+1/2}^-) \over 2} 
- c_{i+1/2}{\baru_{i+1/2}^+-\baru_{i+1/2}^-\over 2}
\label{eq:KT_H_generic}
\ee
where now 
$\baru^{+}_{i+1/2}$ and $\baru^{-}_{i+1/2}$ 
are the values of the piece-wise linear $u(t, x)$ when approaching
$x_{i+1/2}$
from the right and from the left, respectively.
They are given by 
\be
\baru_{i+1/2}^+ &=& \baru_{i+1} - (a/2) (\partial_x u)_{i+1}
\\
\baru_{i+1/2}^- &=& \baru_{i} + (a/2) (\partial_x u)_{i}
\ee
again using the minmod flux limiter for the derivatives.
This is known as the second order Kurganov-Tadmor (KT) scheme
\cite{KT00} and it is 
implemented in the 3+1D event-by-event viscous hydrodynamics
simulation program {\sc Music} \cite{Schenke:2010nt,Schenke:2010rr}.
The numerical viscosity in the smooth regions is known to be
$O(a^3\partial_x^4 u)$.

The structure of the KT algorithm, Eqs.(\ref{eq:KT_generic}) and
(\ref{eq:KT_H_generic}),
is the same for the lowest order,
second order and the third order algorithms. All one needs to do to improve
accuracy is
to get a better estimate of $\baru^{\pm}_{i+1/2}$.
Actually, it is instructive to consider the lowest order result
which uses the histogram function as the approximation of $u(x)$.
One then has
$\baru^{+}_{i+1/2} = \baru_{i+1}$
and $\baru^{-}_{i+1/2} = \baru_{i}$. 
It is easy to see in this case that if $j = vu$ and hence
$c_{i+1/2} = |v|$,
Eq.(\ref{eq:KT_generic}) automatically becomes the up-wind method.

We now have a set of ODE's. What is a good choice of the ODE solver?
One of the physical requirement is again the positivity. 
For instance, suppose $u$ represents particle density. One knows that $u$ can never be
negative. 
One of the schemes that preserves the positivity is the Heun's method which
is one of the
second order Runge-Kutta schemes.
The equation $du/dt = f$ is numerically solved by following these steps
\be
\baru_j^* &=& \baru_j^n + h f(t_n, \baru^n) 
\\
\baru_j^{**} &=& \baru_j^* + h f(t_{n+1}, \baru^*)
\\
\baru_j^{n+1} &=& {1\over 2}(u_j^n + u_j^{**})
\ee
Positivity is maintained at each stage with a suitable choice
of the flux limiter such as the minmod flux limiter.

For 3-D (and similarly for 2-D), a simple extension
\be
{d\over dt}\baru_{ijk} = 
-{H^x_{i+1/2,j,k} - H^x_{i-1/2,j,k}\over a_x}
-{H^y_{i,j+1/2,k} - H^x_{i,j-1/2,k}\over a_y}
-{H^z_{i,j,k+1/2} - H^z_{i,j,k-1/2}\over a_z}
\non
\ee
works well.
In curvilinear coordinate systems such as the Milne coordinates,
one may not have the conservation law
in the form $\partial_\mu J^\mu = 0$. Instead, it may take the form
\be
\partial_a J^a = S
\ee
where $S$ is the geometrical source term that does not involve
derivatives.
The KT algorithm can be easily adapted to this case by simply adding the
source term 
on the right hand side. Namely,
\be
{d\baru_i\over dt}
& = &
-{H_{i+1/2} - H_{i-1/2}\over a} + S(\baru_i)
\ee

In implementing the KT scheme, one needs the maximum speed of propagation
at $x_{i\pm 1/2}$.
If there is only one variable $u$ and one current $j$,
then the speed of propagation in the $i$-th direction is
\be
c_i = 
\left| {\partial j_i\over \partial u} \right| 
\ee
If there are more than one current, 
then we first define the Jacobian matrix
\be
J_i^{ab} = {\partial j_i^a \over \partial u_b}
\ee
where $i = 1, 2, 3$ is the space index and $a = 1, \cdots, M$ labels
the conserved quantities. Therefore we have 3 $M\times M$ matrices.
The maximum propagation speed in the $i$-th direction is 
\be
c_i = {\rm max}(|\lambda_1|,\cdots,|\lambda_M|)
\ee
where $\lambda$'s are the eigenvalues of the Jacobian $J_i^{ab}$.
When discretizing in time, the original authors of the KT algorithm
recommended
the time step $h$ to be small enough so that $|{\rm max}(c_i) h/a| < 1/8$.

For the explicit form of the $c_i$ for the 3+1d ideal hydrodynamics including
the net baryon currently used in {\sc Music},
see Ref.\cite{Schenke:2010nt}. For implementation
of the event-by-event 3+1d viscous hydrodynamics in {\sc Music},
see Refs.\cite{Schenke:2010rr,Gale:2012rq}. 
Additional valuable information on the
algorithms used in solving the {\sc vHydro} equations can be found in
Ref.\cite{Shen:2014vra}.
For some comparisons of various schemes discussed in the section,
see Ref.\cite{Akamatsu:2013wyk}.

There are many other numerical schemes that are currently in use
but we unfortunately did not have space to discuss.
These include, but not limited to,
a PPM (Piece-wise Parabolic Method) scheme\cite{Hirano:2012yy},
a Lagrangian scheme where the grid points follow the movement of
fluid cells\cite{Nonaka:1999et},
Riemann solvers \cite{DelZanna:2013eua,Akamatsu:2013wyk},
and a SPH (smoothed particle hydrodynamics) method
\cite{Aguiar:2000hw}.
The simple FTCS scheme has also been employed for the viscous hydrodynamics 
\cite{Luzum:2008cw,Bozek:2011ua} for smooth initial conditions.

\subsection{Freeze-out Hypersurface and Cooper-Frye Formula for Particle Production}

\begin{figure}[t]
\includegraphics[width=6cm,height=5cm]{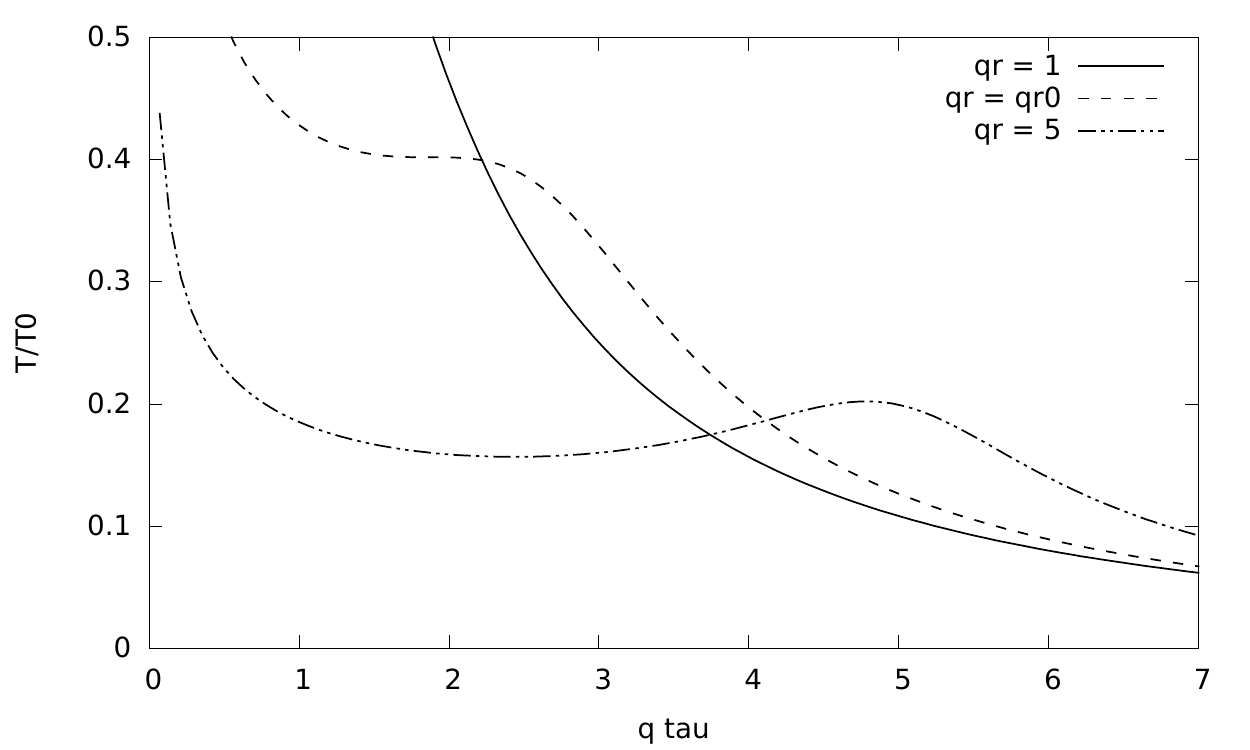}
\includegraphics[width=6cm,height=5cm]{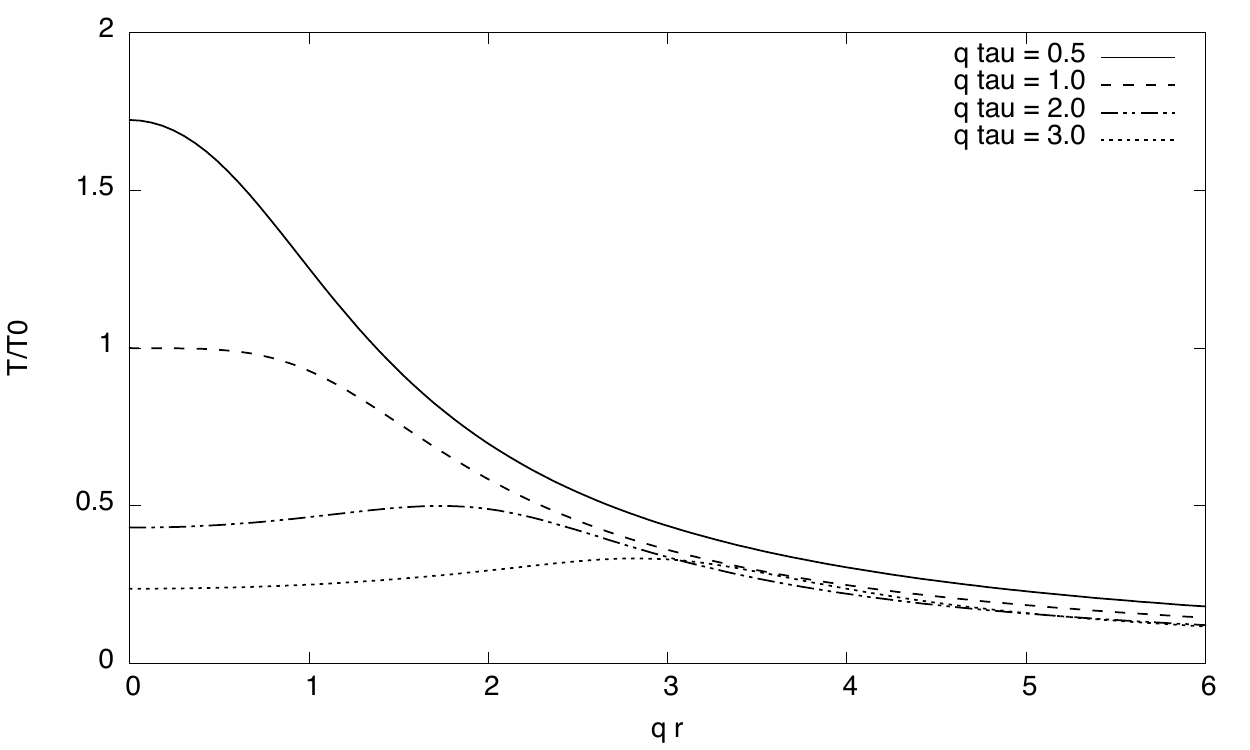}
\caption{Plot of $T(\tau,r)/T_0$ for the Gubser solution. Here $qr_0 =
(3+\sqrt{5})/2$.
The left panel shows $T(\tau,r)/T_0$ at fixed $r$'s and the right panel
shows $T(\tau,r)$
at fixed $\tau$'s.}
\label{fig:gubserT}
\end{figure}

\begin{figure}[t]
\centerline{\includegraphics[width=7cm,height=5cm]{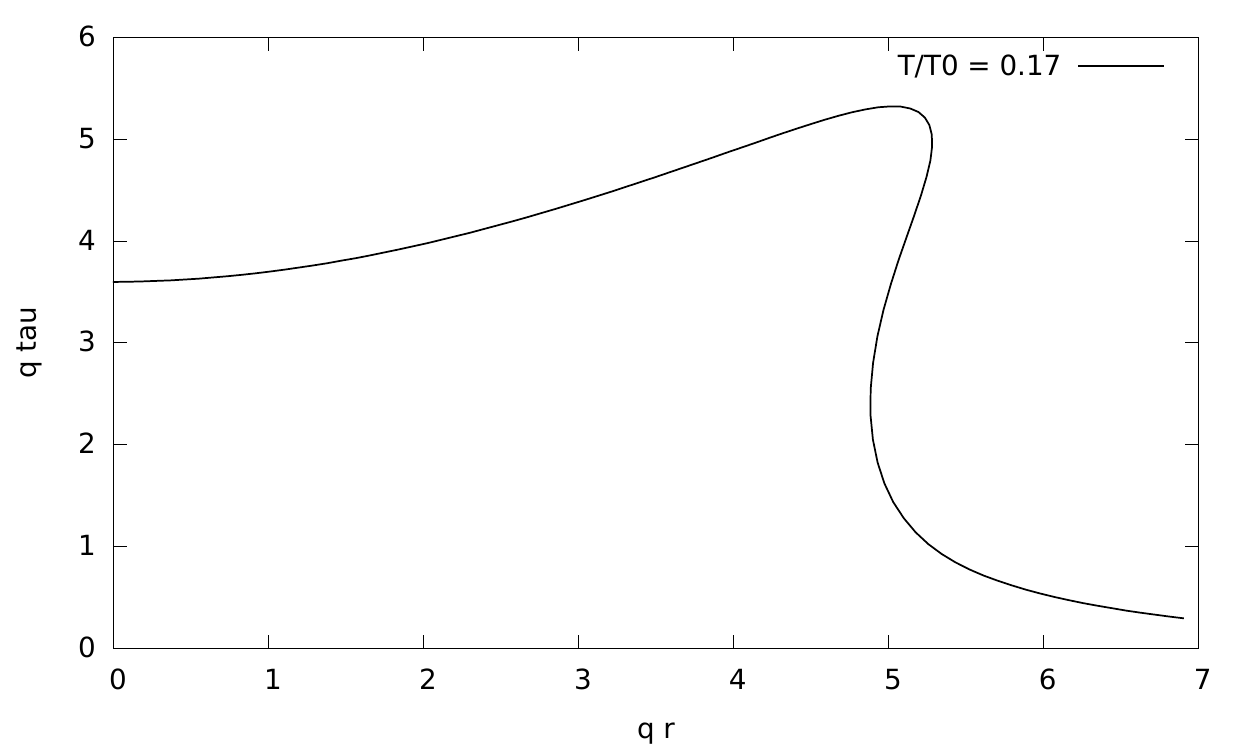}}
\caption{Plot of the hypersurface with $T(\tau,r)/T_0 = 0.17$ for the
Gubser solution.}
\label{fig:fo_curve}
\end{figure}

When a hydrodynamic system is expanding, there comes a time when the
system is too dilute to be treated with hydrodynamics 
(c.f.~section \ref{sec:kinetic}).
From this point on, the
system is basically
a collection of non-interacting particles. In realistic systems,
this time is not the same for all
fluid cells. As the system starts to expand at $\tau_0$,
there are cells at the periphery of the system that are dilute enough 
to ``freezes out'' in a very short time. As time elapses, expansion of the
system can cause
hot and dense matter to flow into the location of those frozen cells. These
will eventually
freeze-out, too. Therefore, the freeze-out surface volume cannot be a
simple 3-dimensional
volume. It is a complicated 3-d volume in the 4-d space-time.

To illustrate this point, 
consider the following Gubser solution of the boost-invariant and
azimuthally symmetric
ideal-hydrodynamics\cite{Gubser:2010ze}.\footnote{
Analytic solutions of the ideal hydrodynamics do exist for special cases,
but not for general
cases.}
\be
T(\tau,r) = 
{2 T_0 
\over
\big(2q\tau \left[1 + 2q^2(\tau^2 + r^2) + q^4(\tau^2 - r^2)^2\right]
\big)^{1/3}
}
\ee
where $r^2 = x^2 + y^2$ and $\tau = \sqrt{t^2 - z^2}$. 
It is a simple matter of taking the $\tau$ derivative of $T(\tau, r)$
to see that
for $r > r_0$ where $r_0= {3 + \sqrt{5}\over 2 q}$, 
there are two values of $\tau$ where $\partial_\tau T$ vanishes.
Since $\partial_\tau T$ is negative for small $\tau$, this means that
$T(\tau, r)$ at fixed $r > r_0$ will have a minimum and then a maximum.
This is illustrated in the left panel in Fig.~\ref{fig:gubserT}.
The solid line is for $qr = 1$ which is near
the center of the system. The temperature at that position decreases
monotonically.
At $r = r_0$, there is an inflection point but the behavior is till
monotonic.
In the $qr = 5$ case, one can clearly see that the temperature decreases
at first but a some point it starts to rise again as the pressure pushes
hot 
matter from the central region towards the periphery of the system.
Assuming that the freeze-out temperature is between the minimum and the
maximum of 
$T/T_0$, the position $qr = 5$ will contribute to the final particle
spectrum (freeze-out)
three times.

The plot of  isothermal curve with $T(\tau,r)/T_0 = 0.17$ is shown in
Fig.~\ref{fig:fo_curve}. One may think of this as representing the
``freeze-out'' hypersurface
in the Gubser solution. The long tail that can be seen above  $qr \gsim 5$
is unrealistic. 
In more realistic simulations, the simulation starts at times above the
long tail.
Nevertheless, the Gubser solution contains much of the features
that the more realistic numerical solutions 
exhibit \cite{Heinz:2002sq,Song:2010aq,Schenke:2010rr,Holopainen:2012id}.

At the freeze-out hypersurface, the fluid is to be 
converted to particles according to the local equilibrium condition.
In the kinetic theory, the particle number current for the $i$-th particle
is given by
\be
j_i^\mu(x) = g_i \int {d^3 p\over (2\pi)^3 p^0}\, f_i(x,p)\, p^\mu
\ee
where $f_i(x,p)$ is the phase space density and $g_i$ is the degeneracy.
The number of $i$-th particle in a 3-d hypersurface is then given by
\be
\int d\sigma_\mu 
j_i^\mu(x) = g_i \int {d^3 p\over (2\pi)^3 p^0}\, 
\int d\sigma_\mu\, p^\mu\, f_i(x,p)
\ee
where $d\sigma_\mu$ is the hypersurface element.
Hence, the momentum spectrum is
\be
E_p {dN_i\over d^3 p} = {g_i\over (2\pi)^3}
\int d\sigma_\mu\, p^\mu\, f_i(x,p)
\label{eq:CFmom}
\ee
This is the celebrated Cooper-Frye formula \cite{Cooper:1974mv}.

Hydrodynamics deals with the energy density and its flow. 
Experiments measure the momentum distribution of identified particles. 
Cooper-Frye formula provides the link between them.
In ideal hydrodynamics, local equilibrium is strictly maintained.
Hence once we find the freeze-out hypersurface (equivalently
the energy density via the equation of state), all one has to do is
to integrate Eq.(\ref{eq:CFmom}) over the freeze-out hypersurface
with $f_i(x,p) = 1/(e^{(p^\mu u_\mu - \mu_B)/T_{\rm FO}} + a_i)$ 
where $T_{\rm FO}$ is the freeze-out temperature,
$a_i = -1$ for bosonic statistics, $a_i = 1$ for fermionic statistics,
and $a_i = 0$ for classical (Boltzmann) statistics.

The freeze-out surface we need to integrate over has the shape shown 
in Fig.~\ref{fig:fo_curve} which requires a closer inspection.
Suppose all parts of the system reaches the freeze-out temperature at the
same
Minkowski time $t_{\rm FO}$ and only once. In that case, 
the freeze-out hypersurface is just the Minkowski spatial volume. The
integration
element is just then $d\sigma_0 = dx dy dz$. All other components vanish. 
This is simple, but not a realistic scenario in relativistic heavy ion
collisions as
explained in the previous section.

In the Milne coordinate system, the volume elements in each direction are
given by
\be
d\sigma_a = 
\left(
\tau dx dy d\eta,
-\tau d\tau dy d\eta,
-\tau d\tau dx d\eta,
-\tau d\tau dx dy
\right)
\label{eq:dSigma_a}
\ee
When the freeze-out hypersurface is specified by the
freeze-out proper time $\tau_f(x,y,\eta)$,
the surface element on this surface is obtained by replacing
$d\tau$ in Eq.(\ref{eq:dSigma_a}) with $(\partial\tau_f/\partial x_i)dx_i$
\be
d\sigma^f_a = 
\left(1,
-{\partial\tau_f\over \partial x},
-{\partial\tau_f\over \partial y},
-{\partial\tau_f\over \partial \eta}
\right)
\tau_f dx dy d\eta
\label{eq:hypersurface_dV}
\ee
If $d\sigma_a^f$ is time-like, then it is guaranteed that 
$p^a d\sigma^f_a > 0$. Therefore, Eq.(\ref{eq:CFmom}) has a well defined
interpretation that the particles are flowing out of the hypersurface.
If $d\sigma_a^f$ is space like, then depending on the direction of $p^a$,
$p^a d\sigma_a^f$ could be either positive or negative. 
Equivalently, particles can be flowing into or out of the hypersurface.

Before we discuss the physical situation when this can happen, first
we define the dynamic rapidity $y = \tanh^{-1}(p_z/E)$ so that
\be
E & = & m_T \cosh y
\\
p^z & = & m_T\sinh y
\ee
with $m_T = \sqrt{m^2 + \bfp_T^2}$. 
The Milne space momentum is
\be
p^a = 
\left(
m_T \cosh(y - \eta), \bfp_T, m_T\sinh(y-\eta)/\tau
\right)
\ee
which is equivalent to 
using $\Lambda^a_\mu p^\mu$ from (c.f.~Eq.(\ref{eq:Lambda_mua})) but with
an extra factor of $1/\tau$ for the $\eta$ component to conform with the
geometrical
definition of the hypersurface element Eq.(\ref{eq:hypersurface_dV}).

To illustrate what happens when $d\sigma_a^f$ is space-like, 
let's consider the Gubser solution again.
In this case,
$\tau_f(r)$ is a function of the transverse radius $r = \sqrt{x^2 + y^2}$
only.
Since the system is boost-invariant, we can also set $\eta = 0$ without
loss of generality.
The inner product of the momentum and the volume element is
\be
p^a d\sigma^f_a = 
\left( m_T\cosh y - (\bfp_T{\cdot}\hat{\bfr})\partial_r \tau_f
\right)\tau_f dx dy d\eta
\ee
where $\hat{\bfr}$ is the unit vector in the transverse direction.
This can become negative if 
$\partial\tau_f/\partial r > (m_T/p_T) \cosh y > 1$.
{}From Figs.~\ref{fig:gubserT} and \ref{fig:fo_curve},
one can see that large positive gradient $\partial\tau_f/\partial r$
exists in the region where the temperature reaches $T_{\rm FO}$ the second
time.
At this time, the temperature is going {\em up} from the minimum in the
left panel of
Fig.~\ref{fig:gubserT}. Therefore, this part of the $\tau_f$ is not about
the fluid
freezing-out. It is rather colder matter being heated
up to become the a part of the fluid again.

Before one could use the Cooper-Frye formula, one needs to know the freeze-out
hypersurface. This is not a trivial problem because hydrodynamic simulations
only provide the freeze-out {\em space-time points.}
The freeze-out hypersurface needs to be reconstructed from these points.
In 2+1d, this is not so complicated. But in 3+1d, it can become very involved.
Discussion of this important topic however is beyond the scope of this review.
An interested reader should consult Ref.\cite{Huovinen:2012is}.

How do we use the Cooper-Frye formula Eq.(\ref{eq:CFmom})?
If hydrodynamics is not coupled to the
hadronic cascade after-burner, then usually the hypersurface integral in
Eq.(\ref{eq:CFmom}) is performed as it is after the hypersurface is reconstructed. 
The rationale behind it is that when the cell crosses the freeze-out
boundary 3 times, the contributions from the first two times should largely
cancel each other.
This is physically sound
because when the fluid cell crosses the freeze-out surface the second
time,
the particles that are being heated up again are the remnants of the first
crossing.
When coupling to the hadronic after-burner, an additional condition such
as
$p^a d\sigma^f_a > 0$ is usually employed 
\cite{Grassi:2004dz,Sinyukov:2009ce,Huovinen:2012is,Oliinychenko:2014tqa}.

The presence of non-zero shear tensor $\pi^{\mu\nu}$ and the bulk pressure
$\Pi$ signals
non-equilibrium. In this case, the Cooper-Frye formula needs to be modified
to take into
account the non-equilibrium phase space density.
Let 
\be
f(x,p) = f_{\rm eq}(x,p) + \delta f(x,p)
\ee
Then the consistency between the hydrodynamic $T^{\mu\nu}$ and the
kinetic theory $T_{\rm Kin}^{\mu\nu} = \int {d^3p\over (2\pi)^3p^0} p^\mu
p^\nu f$ requires
\be
\pi^{\mu\nu} = 
\int {d^3 p\over (2\pi)^3 p^0}\, p^{\langle\mu} p^{\nu\rangle}\,
\deltaf(x,p)
\label{eq:deltaf_pimunu}
\ee
and
\be
\Pi = -{m^2\over 3}\int {d^3p\over (2\pi)^3 p^0}\,\deltaf(x,p)
\label{eq:deltaf_Pi}
\ee
with
\be
p^{\langle\mu}p^{\nu\rangle}
& = &
\Delta^{\mu\nu}_{\alpha\beta} p^\alpha p^\beta
\ee
These conditions can be satisfied by
\be
\deltaf(x,p)
= 
\left( A(\barE_p)\Pi(x,p) + B(\barE_p) p^{\langle\mu}p^{\nu\rangle}
\pi_{\mu\nu}(x,p)
+ \cdots
\right) 
f_{\rm eq}(x,p)(1+ a_i f_{\rm eq}(x,p))
\non
\label{eq:deltaf_exp}
\ee
where again $\barE_p = p^\mu u_\mu$ is the energy in the local rest frame.
Here $A(\barE_p)$ and $B(\barE_p)$ must be consistent with
Eqs.(\ref{eq:deltaf_pimunu}) and (\ref{eq:deltaf_Pi}), but otherwise
arbitrary at this point. 
In the presence of the net-baryon current 
Eq.(\ref{eq:deltaf_exp}) also includes $C(\barE_p)p^{\ave{\mu}}q_\mu$
with the constraint on $C(\barE_p)$ given by Eq.(\ref{eq7a}).
Exact forms of $A(\barE_p)$ and $B(\barE_p)$ (and $C(\barE_p)$)
depend on the form of the scattering cross-sections in the underlying 
microscopic system. In Ref.\cite{Dusling:2009df}, it is argued that for most 
theories, the $\barE_p$ dependence of 
$A$ and $B$ should be between $1$ and $1/\barE_p$.
See also Refs.\cite{Denicol:2012yr,Noronha-Hostler:2013gga}.
In most simulations, the constant ansatz is used.

\section{Summary}

In this review, we have tried to deliver a more general and pedagogic view
of the relativistic hydrodynamics currently used in the study of 
ultra-relativistic heavy ion collisions. One message we tried to
convey to the reader was that hydrodynamics is a very general framework and 
yet it can describe a vast set of complex phenomena.
Especially in QGP studies, hydrodynamics is an indispensable tool that connects
the QGP properties to the actual observables. 

Another message we tried to convey was that the theory of hydrodynamics is 
a fascinating subject by itself. As discussed in section \ref{sec:linear_response}
and section \ref{sec:kinetic_hydro}, there are still many unsolved problems such as finding
Kubo formulas for the second order transport coefficients and finding the right
anisotropic equation of state. In view of the apparent collectivity in the high-multiplicity
proton-proton and proton-nucleus collisions at the LHC,
the theory of collective motion in small systems 
is also in urgent need of development. In these systems, thermal fluctuations 
during the hydrodynamic evolution may not be completely 
ignored\cite{Kapusta:2011gt,Murase:2013tma,Hirano:2014dpa,Young:2014xda,Young:2014pka}.
We hope that this review has provided interested readers
enough starting points to pursue these and other interesting topics on their own.

In any short review, omission of some important subjects inevitably
occurs due to the lack
of space. One notable omission in this review is
the discussion of the initial state models. 
As briefly discussed in section \ref{sec:numerical},
the initial condition of the hydrodynamic evolution must be given outside of the
theory of hydrodynamics. For a meaningful comparison with the experimental data,
having the right initial condition including the right energy-momentum fluctuation spectra
is crucial. Unfortunately, it is outside the scope of this review and
must be left as the subject of a future review.

\section{Acknowledgments}

S.J.~thanks C.~Gale and G.~Denicol for many discussions on finer points of hydrodynamics
and L.G.~Yaffe for generous permission to use some materials from his
unpublished note.
S.J.~is supported in part by the Natural Sciences and Engineering Research
Council of Canada. U.H.~acknowledges support  by the U.S. Department of
Energy, Office of Science, Office of Nuclear Physics under Awards No.
\rm{DE-SC0004286} and (within the framework of the JET Collaboration)
\rm{DE-SC0004104}.

\bibliographystyle{ws-rv-van}

\blankpage
\printindex                         
\end{document}